\documentclass[11pt,aps,a4paper,eqsecnum,amsmath,amssymb
              ,nofootinbib
              ,notitlepage
              ]{revtex4-1}

\pdfoutput=1

\usepackage{amsfonts}
\usepackage{tikz}
\usepackage{pgfplots}
\usepackage{tabularx} \newcolumntype{C}{>{\centering}X}
\usepackage{graphicx}

\def\tp{\otimes}
\def\ds{\oplus}
\def\ket[#1]{\left| #1 \right>}
\def\bra[#1]{\left< #1 \right|}
\def\Z{\mathbb{Z}}
\def\R{\mathbb{R}}
\def\C{\mathbb{C}}
\def\L{\mathcal{L}}
\def\Wop[#1][#2][#3][#4][#5]{W\left(\left.\begin{array}{cc} #1&#2\\#3&#4\end{array}\right|#5\right)}
\def\WopAlt[#1][#2][#3][#4][#5][#6][#7]{W^{#6#7}\left(\left.\begin{array}{cc} #1&#2\\#3&#4\end{array}\right|#5\right)}

\def\notiz[#1]{\textbf{\color{red}#1}}
\begin{document}

\preprint{}
\title{%
  Integrable anyon chains: from fusion rules to face models\\
  to effective field theories
}

\author{Peter E. Finch}
\author{Michael Flohr}
\author{Holger Frahm}
\affiliation{%
Institut f\"ur Theoretische Physik, Leibniz Universit\"at Hannover,
Appelstra\ss{}e 2, 30167 Hannover, Germany}


\begin{abstract}
  Starting from the fusion rules for the algebra $SO(5)_2$ we construct
  one-dimensional lattice models of interacting anyons with commuting transfer
  matrices of `interactions round the face' (IRF) type.  The conserved
  topological charges of the anyon chain are recovered from the transfer
  matrices in the limit of large spectral parameter. The properties of the
  models in the thermodynamic limit and the low energy excitations are studied
  using Bethe ansatz methods.  Two of the anyon models are critical at zero
  temperature.  From the analysis of the finite size spectrum we find that
  they are effectively described by rational conformal field theories
  invariant under extensions of the Virasoro algebra, namely $\mathcal{W}B_2$
  and $\mathcal{W}D_5$, respectively.  The latter contains primaries with half
  and quarter spin.  The modular partition function and fusion rules are
  derived and found to be consistent with the results for the lattice model.
\end{abstract}

\maketitle


\section{Introduction}
Starting with Bethe's study of the spin-$\frac{1}{2}$ Heisenberg chain
integrable models in low dimensions have provided important insights into the
peculiarities of correlated many-body systems subject to strong quantum
fluctuations \cite{Bethe31,Baxter:book,VladB,HUBBARD}, e.g.\ the appearance of
quasi-particles with exotic properties such as fractional quantum numbers and
with unusual braiding statistics.  In a wider context these \emph{anyons}
appear as excitations in topologically ordered systems without a local order
parameter.  Realizations for such topological quantum liquids are the
fractional Hall states \cite{Laug83} and two-dimensional frustrated quantum
magnets \cite{MoSo01,BaFG02,Kita06}.
Non-Abelian anyons can also be realized as local modes with zero energy at
junctions of spin-orbit coupled quantum wires through the 'topological' Kondo
effect \cite{BeCo12,ABET13,ABET14}.
Additional interest in these objects arises from the fact that non-Abelian
anyons are protected by their topological charge which makes them potentially
interesting as resources for quantum computation \cite{Kita03,NSSF08}.

A class of integrable models describing interacting non-Abelian anyons in one
dimension can be obtained from two dimensional classical lattice systems with
interactions round the face (IRF) such as the restricted solid on solid (RSOS)
models \cite{AnBF84} in their Hamiltonian limit.  The critical RSOS models are
lattice realizations of the unitary minimal models \cite{FrQS84,Huse84} and
the family of commuting operators generated by their transfer matrices
contains Hamiltonians for Fibonacci or, more generally, $SU(2)_k$ anyons with
nearest neighbor interactions \cite{FTLT07}.  Recently, extensive studies of
these anyons and their higher spin variants subject to different interactions
or in different geometries have led to the identification of a variety of
critical phases and the corresponding conformal field theories from finite
size spectra obtained numerically \cite{TAFH08,GATL09,GATH13,LPTT11,PBTL12}.
%
%

Several approaches exist for the construction of integrable IRF models:
integrable generalizations of the RSOS models have been obtained by means of
the fusion procedure \cite{DJMO86}.  Alternatively, the correspondence between
integrable IRF models and vertex models \cite{Pasq88,Roch90} (or,
equivalently, anyon chains and quantum spin chains) with an underlying
symmetry of a quasi-triangular Hopf algebra can be exploited to construct
anyonic quantum chains \cite{Finch13,FiFr13}.  
Finally, new models can be defined by labeling the local height variables on
neighboring sites of the lattice by adjacent roots in the A-D-E Dynkin
diagrams or, more generally by primary fields of a general rational conformal
field theory related through the corresponding fusion algebra
\cite{Pasq87a,Gepn93a}.  For these models to be integrable one has to find a
parameterization of the Boltzmann weights for the configuration allowed around
a face of the lattice which satisfies the Yang-Baxter equation.  This can be
achieved by Baxterization of a representation of the braid group associated
to the symmetries of the underlying system.

In the following we apply this last approach to construct four integrable
quantum chains of a particular type of anyons responsible for the non-Fermi
liquid correlations predicted in the topological Kondo effect
\cite{ABET13,ABET14}.  Specifically, we consider anyons satisfying the fusion
rules of $SO(5)_2$.
Below we will see that -- as might be expected for a model based on
$SO(5)=B_2$ -- the anyonic quantum chains are related to IRF models with
height variables outside the A-D-E classification.  The integrability of the
models is established by construction of representations of the
Birman-Murakami-Wenzl (BMW) algebra in terms of anyonic projection operators.
The spectra of these models are determined by Bethe ansatz methods and we
identify the ground state and low lying excitations.  Two of the models can be
related to the six-vertex model in the ferromagnetic and anti-ferromagnetic
gapped regime, respectively, by means of a Temperley-Lieb equivalence.  From
the analysis of the finite size spectrum the other two integrable points are
found to be critical with the low energy sector described by unitary rational
conformal field theories (RCFTs) with extended symmetries associated to the
Lie algebras $B_2=SO(5)$ and $D_5=SO(10)$, respectively.
To present our results in a self-contained way, some general facts on the
classification and spectral data of the relevant minimal models of
Casimir-type $\mathcal{W}$-algebras are included in Appendix~\ref{appCFT}.
%
%
For the anti-ferromagnetic anyon chain we propose a modular invariant
partition function in terms of the Virasoro characters of the irreducible
highest weight representations of the 
$\mathcal{W}D_5(9,10)$ RCFT.  The $\mathcal{S}$-matrix and fusion rules of
this CFT are given in Appendix~\ref{app_WD5fus}.

\section{$SO(5)_2$ anyons}
Algebraically, anyonic theories can be described by braided tensor categories
\cite{Kita06,Bonderson07,Preskill04}.
A braided tensor category consists of a collection of objects $\{\psi_{i}\}$
(including an identity) equipped with a tensor product (fusion rules),
\begin{eqnarray*}
  \psi_{a} \tp \psi_{b} & \cong & \bigoplus_{c} N_{ab}^{c} \psi_{c}
\end{eqnarray*}
where $N_{ab}^{c}$ are non-negative integers.  In the special case
$N_{ab}^{c}\in\{0,1\}$ it is called multiplicity free.  Below we shall use a
graphical representation of 'fusion states' of the anyon model where vertices
\begin{eqnarray*}
\begin{tikzpicture}[scale=0.9]
	\tikzstyle{every node}=[minimum size=0pt,inner sep=0pt]
	\tikzstyle{every loop}=[]
	\node (na) at (0.0,1.0) {$a$};
	\node (nb) at (1.0,1.0) {$b$};
	\node (nm) at (0.5,0.5) {};
	\node (nc) at (0.5,0.0) {$c$};
	\foreach \from/\to in {na/nm,nb/nm,nc/nm} \draw (\from) -- (\to);
\end{tikzpicture}
\end{eqnarray*}
may occur provided that $\psi_c$ appears in the fusion of $\psi_a$ and
$\psi_b$.  We require associativity in our fusion i.e.\
\begin{eqnarray*}
  (\psi_{a} \tp \psi_{b}) \tp \psi_{c} & \cong & \psi_{a} \tp (\psi_{b} \tp
  \psi_{c}), 
\end{eqnarray*}
which is governed by $F$-moves, also referred to as generalized $6j$-symbols,
\begin{equation}
\label{Fmoves}
\begin{tikzpicture}[scale=0.7]
	\tikzstyle{every node}=[minimum size=0pt,inner sep=0pt]
	\tikzstyle{every loop}=[]
	\node (ns) at (0.2,0.0) {$a$};
	\node (ne) at (2.8,0.0) {$e$};
	\node (n1t) at (1.0,0.7) {$b$};
	\node (n2t) at (2.0,0.7) {$c$};
	\node (n1b) at (1.0,0.0) {};
	\node (n2b) at (2.0,0.0) {};
	\node (l1) at (1.6,-0.3) {$d$};
	\foreach \from/\to in {ns/ne} \draw (\from) -- (\to);
	\foreach \from/\to in {n1t/n1b,n2t/n2b} \draw (\from) -- (\to);
	\node (l2) at (4.8,0.0) {$=\sum_{d'} (F^{abc}_{e})^{d}_{d'}$}; 
	\node (vs) at (6.9,-0.2) {$a$};
	\node (ve) at (9.5,-0.2) {$e$};
	\node (v1) at (7.7,0.7) {$b$};
	\node (v2) at (8.7,0.7) {$c$};
	\node (v12) at (8.2,0.2) {};
	\node (vb) at (8.2,-0.2) {};
	\node (l3) at (8.55,0.05) {$d'$};
	\foreach \from/\to in {vs/ve} \draw (\from) -- (\to);
	\foreach \from/\to in {v1/v12,v2/v12,v12/vb} \draw (\from) -- (\to);
\end{tikzpicture}
\end{equation}
For more than three objects different decompositions of the fusion can be
related by distinct sequences of $F$-moves.  Their consistency for arbitrary
number of factors is guaranteed by the Pentagon equation satisfied by the
$F$-moves.

There also must be a mapping that braids two objects,
\begin{eqnarray*}
  R: \psi_{a} \tp \psi_{b} & \rightarrow & \psi_{b} \tp \psi_{a}.
\end{eqnarray*}
Note that while fusion is commutative, states of the system may pick up a
phase $R^{ab}_{c}$ under braiding.  Graphically this is represented by
\begin{equation}
\label{Rmoves}
\begin{tikzpicture}[scale=0.9]
	\tikzstyle{every node}=[minimum size=0pt,inner sep=0pt]
	\tikzstyle{every loop}=[]
	\node (na1) at (0.0,1.0) {$a$};
	\node (nb1) at (1.0,1.0) {$b$};
	\node (nm1) at (0.5,0.5) {};
	\node (nc1) at (0.5,0.0) {$c$};
	\foreach \from/\to in {na1/nm1,nb1/nm1,nc1/nm1} \draw (\from) -- (\to);
	\node (l2) at (2.25,0.5) {$=\quad R^{ab}_{c}\quad$}; 
	\node (nb2) at (3.0,1.0) {$b$};
	\node (na2) at (4.0,1.0) {$a$};
	\node (nm2) at (3.5,0.5) {};
	\node (nc2) at (3.5,0.0) {$c$};
	\foreach \from/\to in {na2/nm2,nb2/nm2,nc2/nm2} \draw (\from) -- (\to);
\end{tikzpicture}
\end{equation}
%
For a consistent anyon theory braids have to commute with fusion and satisfy
the Yang-Baxter relation.  Both properties follow from the Hexagon equation
for $F$- and $R$-moves.

In this paper we consider a system of particles which satisfy the fusion rules
for $SO(5)_2$ shown in Table~\ref{tabso52fus}.
\begin{table}[t]
\begin{center}
\begin{tabular}{|c||c|c|c|c|c|c|} \hline
	$\tp$ & $\psi_{1}$ & $\psi_{2}$ & $\psi_{3}$ & $\psi_{4}$ & $\psi_{5}$ & $\psi_{6}$ \\ \hline\hline
	$\psi_{1}$& $\psi_{1}$ & $\psi_{2}$ & $\psi_{3}$ & $\psi_{4}$ & $\psi_{5}$ & $\psi_{6}$ \\ \hline
	$\psi_{2}$& $\psi_{2}$ & $\psi_{1}\ds\psi_{5}\ds\psi_{6}$ & $\psi_{3}\ds\psi_{4}$ & $\psi_{3}\ds\psi_{4}$ & $\psi_{2}\ds\psi_{5}$ & $\psi_{2}$ \\ \hline
	$\psi_{3}$& $\psi_{3}$ & $\psi_{3}\ds\psi_{4}$ & $\psi_{1}\ds\psi_{2}\ds\psi_{5}$ & $\psi_{2}\ds\psi_{5}\ds\psi_{6}$ & $\psi_{3}\ds\psi_{4}$ & $\psi_{4}$ \\ \hline
	$\psi_{4}$& $\psi_{4}$ & $\psi_{3}\ds\psi_{4}$ & $\psi_{2}\ds\psi_{5}\ds\psi_{6}$ & $\psi_{1}\ds\psi_{2}\ds\psi_{5}$ & $\psi_{3}\ds\psi_{4}$ & $\psi_{3}$ \\ \hline
	$\psi_{5}$& $\psi_{5}$ & $\psi_{2}\ds\psi_{5}$ & $\psi_{3}\ds\psi_{4}$ & $\psi_{3}\ds\psi_{4}$ & $\psi_{1}\ds\psi_{2}\ds\psi_{6}$ & $\psi_{5}$ \\ \hline
	$\psi_{6}$& $\psi_{6}$ & $\psi_{2}$ & $\psi_{4}$ & $\psi_{3}$ & $\psi_{5}$ & $\psi_{1}$ \\ \hline
\end{tabular}
\end{center}
\caption{\label{tabso52fus}The fusion rules for $SO(5)_{2}$ anyons.}
\end{table}
This fusion algebra is the truncation of the category of irreducible
representations of the quantum group $U_{q}(so(5))$ where
$q=\mathrm{e}^{\frac{2i\pi}{5}}$ \cite{Bonderson07}.  For the $SO(5)_{2}$
fusion rules there exist four known sets of inequivalent unitary $F$-moves,
which can be found by solving the pentagon equations directly or by utilizing
the representation theory of $U_{q}(so(5))$.  Associated with each set of
$F$-moves is a modular $\mathcal{S}$-matrix which diagonalizes the fusion
rules (note that the latter may depend on the choice of $F$-moves
\cite{Bonderson07}).  We have selected the $F$-moves corresponding to the
$S$-matrix
\begin{equation}
\label{Smat}
  \mathcal{S}  =  \frac{1}{2\sqrt{5}}
  \left(\begin{array}{cccccc}
         1 & 2 & \sqrt{5} & \sqrt{5} & 2 & 1 \\
         2 & -2\phi & 0 & 0 & 2\phi^{-1} & 2 \\
         \sqrt{5} & 0 & -\sqrt{5} & \sqrt{5} & 0 & -\sqrt{5} \\
         \sqrt{5} & 0 & \sqrt{5} & -\sqrt{5} & 0 & -\sqrt{5} \\
         2 & -2\phi^{-1} & 0 & 0 & 2\phi & 2 \\
         1 & 2 & -\sqrt{5} & -\sqrt{5} & 2 & 1 \\
        \end{array} \right)
\end{equation}
where $\phi = \frac{1+\sqrt{5}}{2}$.  With this choice of $F$-moves we have
four possible sets (in two mirror pairs) of $R$-moves.  For the purposes of
this article, however, we do not need to explicitly choose one.

Given a consistent set of rules for the fusion and braiding we can construct a
one-dimensional chain of interacting `anyons' with topological charge
$\psi_j$.  First a suitable basis has to be built using fusion paths
\cite{MoSe89,Preskill04}:
starting with an auxiliary anyon $\psi_{a_0}$ we choose one of the objects
appearing in the fusion $\psi_{a_0}\tp\psi_{j}$, say $\psi_{a_1}$.  The latter
is fused to another $\psi_{j}$, and so forth.  Recording the irreducible
subspace of the auxiliary space and the subsequent irreducible subspaces which
appear after fusion we construct basis states \vspace{0.2cm}
\begin{equation}
  \label{fuspath}
  \begin{tikzpicture}[scale=1.0]
    \put (30,0){$\psi_{j}$}	\put (60,0){$\psi_{j}$}	\put (90,0){$\psi_{j}$}	\put (150,0){$\psi_{j}$}	\put (180,0){$\psi_{j}$}
    \put (0,-20){$\psi_{a_{0}}$}	\put (40,-27){$\psi_{a_{1}}$}	\put (70,-27){$\psi_{a_{2}}$}	\put (100,-27){$\psi_{a_{3}}$}	\put (160,-27){$\psi_{a_{\mathcal{L}-1}}$}	\put (205,-20){$\psi_{a_{\mathcal{L}}}$}
    \draw (0.5,-0.6) -- (4.0,-0.6);	\draw (4.2,-0.6) -- (4.4,-0.6);	\draw (4.6,-0.6) -- (4.8,-0.6);	\draw (5.0,-0.6) -- (7.0,-0.6);
    \draw (1.15,-0.1) -- (1.15,-0.6);	\draw (2.20,-0.1) -- (2.20,-0.6);	\draw (3.25,-0.1) -- (3.25,-0.6);	\draw (5.35,-0.1) -- (5.35,-0.6);	\draw (6.40,-0.1) -- (6.40,-0.6);
  \end{tikzpicture} \vspace{0.6cm} \hspace{1.0cm}
  \equiv \ket[a_{0}a_{1}a_{2}...a_{\mathcal{L}-1}a_{\mathcal{L}}]\,.
\end{equation}
Note that the construction implies that any two neighboring labels,
$a_{i}a_{i+1}$ must satisfy a local conditions, namely
$N_{a_i,j}^{a_{i+1}}\ne0$.  This local condition can be presented in the form
of a graph in which labels which are allowed to be neighbors correspond to
adjacent vertices.

For a system with periodic boundary conditions we have to identify $a_{0}$ and
$a_{\L}$ (which allows to remove the label $a_0$ from the basis states
(\ref{fuspath})).  In this case the Hilbert space of the model can be further
decomposed into sectors labeled by topological charges.  To measure these
charges one inserts an additional anyon of type $\psi_\ell$ into the system
which is then moved around the chain using (\ref{Fmoves}) and finally removed
again \cite{FTLT07}.  The corresponding topological operator $Y_\ell$ has
matrix elements
\begin{equation}
  \label{Ytopo}
  \bra[a_{1}'...a_{\mathcal{L}}'] Y_{\ell} \ket[a_{1}...a_{\mathcal{L}}] 
  = \prod_{i=1}^{\mathcal{L}} \left(F^{\ell
      a_{i}'j}_{a_{i+1}}\right)^{a_{i}}_{a_{i+1}'}\,,
  \quad \ell=1,\ldots,6\,.
\end{equation}
The spectrum of these operators is known: their eigenvalues are given in terms
of the matrix elements of the modular $\mathcal{S}$-matrix as
$\frac{\mathcal{S}_{k\ell}}{\mathcal{S}_{1\ell}}$ \cite{Kita06}.

Local operators acting on the space of fusion paths of the $\psi_j$ anyons can
be written in terms of projection operators.  In terms of the $F$-moves
(\ref{Fmoves}) they can be introduced in the following way \cite{FTLT07},
\vspace{0.3cm}
\begin{eqnarray*}
	p^{(b)}_{i} \left\{\hspace{0.4cm} 
\begin{tikzpicture}[scale=1.0]
	\put (26,5){$j$}	\put (56,5){$j$}
	\put (0,-27){$a_{i-1}$}	\put (40,-27){$a_{i}$}	\put (72,-27){$a_{i+1}$}
	\draw (0.35,-0.6) -- (2.85,-0.6);
	\draw (1.00,0.1) -- (1.00,-0.6);	\draw (2.05,0.1) -- (2.05,-0.6);
\end{tikzpicture} \hspace{0.4cm} \right\}
	& = & \sum_{b'} (F^{a_{i-1}jj}_{a_{i+1}})_{b'}^{a_{i}} \, \delta_{b}^{b'} \left\{\hspace{0.35cm} 
\begin{tikzpicture}[scale=1.0]
	\put (26,5){$j$}	\put (56,5){$j$}
	\put (8,-27){$a_{i-1}$}	\put (47,-15){$b'$}	\put (67,-27){$a_{i+1}$}
	\draw (0.60,-0.6) -- (2.60,-0.6);
	\draw (1.00,0.1) -- (1.525,-0.25);	\draw (2.05,0.1) -- (1.525,-0.25); 	\draw (1.525,-0.25) -- (1.525,-0.6);
\end{tikzpicture} \hspace{0.4cm} \right\} \\
	& = & \sum_{a_{i}'} \left[\left(F^{a_{i-1}jj}_{a_{i+1}}\right)^{a_{i}'}_{b}\right]^{*} \left(F^{a_{i-1}jj}_{a_{i+1}}\right)^{a_{i}}_{b}
\left\{ \hspace{0.4cm} 
\begin{tikzpicture}[scale=1.0]
	\put (26,5){$j$}	\put (56,5){$j$}
	\put (0,-27){$a_{i-1}$}	\put (40,-27){$a_{i}'$}	\put (72,-27){$a_{i+1}$}
	\draw (0.35,-0.6) -- (2.85,-0.6);
	\draw (1.00,0.1) -- (1.00,-0.6);	\draw (2.05,0.1) -- (2.05,-0.6);
\end{tikzpicture} \hspace{0.4cm} \right\}\,.
\end{eqnarray*}
In the second line we have used that the $F$-moves are unitary.  For our
choice of $SO(5)_{2}$ $F$-moves this can be ensured by choosing a suitable
gauge.  In terms of the basis states (\ref{fuspath}) the $p^{(b)}_{i}$ can be 
written alternatively as
\begin{equation} 
  \label{eqnProjOp}
  p^{(b)}_{i}
  =  \sum_{a_{i-1},a_{i},a_{i}',a_{i+1}}
  \left[\left(F^{a_{i-1}jj}_{a_{i+1}}\right)^{a_{i}'}_{b}\right]^{*}
  \left(F^{a_{i-1}jj}_{a_{i+1}}\right)^{a_{i}}_{b}
  \ket[..a_{i-1}a_{i}'a_{i+1}..]\bra[..a_{i-1}a_{i}a_{i+1}..].
\end{equation}
Note that the matrix elements of these operators depend on triples of
neighboring labels $a_{i-1}a_{i}a_{i+1}$ in the fusion path but only the
middle one may change under the action of the $p^{(b)}_{i}$.


\section{The $\psi_{3}$-anyon chain}
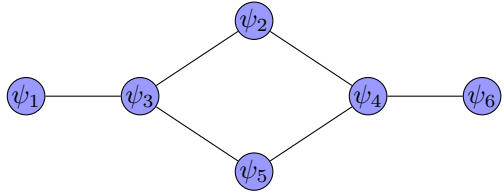
\begin{figure}[t]
\begin{center}
\begin{tikzpicture}[scale=1.0]
	\tikzstyle{every node}=[circle,draw,thin,fill=blue!40,minimum size=14pt,inner sep=0pt]
	\tikzstyle{every loop}=[]
	\node (n0) at (0.0,0.0) {$\psi_{1}$};
	\node (n1) at (3.0,1.0) {$\psi_{2}$};
	\node (n2) at (1.5,0.0) {$\psi_{3}$};
	\node (n3) at (4.5,0.0) {$\psi_{4}$};
	\node (n4) at (3.0,-1.0) {$\psi_{5}$};
	\node (n5) at (6.0,0.0) {$\psi_{6}$};
	\foreach \from/\to in {n0/n2,n1/n2,n1/n3,n2/n4,n3/n4,n3/n5} \draw (\from) -- (\to);
\end{tikzpicture}
\end{center}
\caption{A graphical representation of allowed neighboring labels in the
  fusion paths of the $\psi_{3}$ anyon chain. The vertices/nodes of the graphs
  are the labels of anyons which are connected via an edge if and only if the
  two anyon labels can appear next to each other. \label{FigPsi3Neig}}
\end{figure}

%
Based on the $SO(5)_2$ fusion rules, Table~\ref{tabso52fus}, non-trivial
models can be defined for $\psi_{2}$ (or, equivalently, $\psi_5$) and
$\psi_{3}$ ($\psi_4$) anyons.  In this paper we construct integrable chains of
$\psi_3$ anyons.  The selection rules for neighboring labels in the fusion
path (\ref{fuspath}) displayed in Figure~\ref{FigPsi3Neig}.  This defines the
Hilbert space of the anyon chains.

\subsection{Local Hamiltonians}
We shall concentrate on models with local interactions given in terms of the
projection operators (\ref{eqnProjOp}).  The resulting Hamiltonians are the
anyon chain (or face model) analogue of spin chains with nearest neighbor
interactions \cite{Finch13,Pasq88}.  There are only three non-zero linearly
independent projection operators which can be defined on the tensor product
$\psi_3\tp\psi_3 = \psi_{1}\ds\psi_{2}\ds\psi_{5}$, i.e.\
\begin{equation*}
  p^{(1)}_i+p^{(2)}_i + p^{(5)}_i=\mathbf{1}\,.
\end{equation*}
Therefore the most general local interaction has a single free coupling
parameter
\begin{equation}
  \label{eqnLocalHamth}
  h_{i}(\theta) = \cos\left(\frac{\pi}{4}+\theta\right) p^{(2)}_{i} +
  \sin\left(\frac{\pi}{4}+\theta\right) p^{(5)}_{i}  \,,
\end{equation}
and the general global Hamiltonian for the anyon chain can be written as
\begin{equation}
  \label{eqnHamTheta}
  \mathcal{H}_{\theta} = \sum_{i} h_{i}(\theta)\,.
\end{equation}
Recently, the phase diagram of this model as a function of the parameter
$\theta$ has been investigated numerically \cite{Finch.etal14}.

From Table~\ref{tabso52fus}, we see that there exists an automorphism of the
fusion rules exchanging $\psi_2$ and $\psi_5$ which allows the construction of
a non-local unitary transformation
\begin{equation}
  \label{Unonloc}
  U^{\dagger}\mathcal{H}_{\theta}U = \mathcal{H}_{-\theta}\,.
\end{equation}
We also recall that 
the Hamiltonian commutes with the topological operators (\ref{Ytopo}),
\begin{equation*}
	\left[\mathcal{H}_{\theta}, Y_{\ell} \right] = 0
\end{equation*}
for $\theta \in [0,2\pi)$ and $\ell=1,..,6$.

\subsection{Points of integrability}
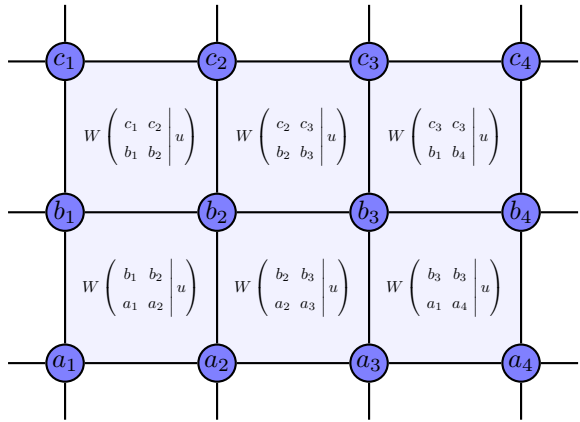
\begin{figure}[t]
  \begin{center}
    \begin{tikzpicture}[scale=2.0]
	\fill[blue!50!,opacity=0.1] (0,0) rectangle (3,2);
	\tikzstyle{every node}=[scale=0.6] 
	\node (W11) at (0.500, 0.500) {$\Wop[b_{1}][b_{2}][a_{1}][a_{2}][u]$};
	\node (W12) at (1.500, 0.500) {$\Wop[b_{2}][b_{3}][a_{2}][a_{3}][u]$};
	\node (W13) at (2.500, 0.500) {$\Wop[b_{3}][b_{3}][a_{1}][a_{4}][u]$};
	\node (W21) at (0.500, 1.500) {$\Wop[c_{1}][c_{2}][b_{1}][b_{2}][u]$};
	\node (W22) at (1.500, 1.500) {$\Wop[c_{2}][c_{3}][b_{2}][b_{3}][u]$};
	\node (W23) at (2.500, 1.500) {$\Wop[c_{3}][c_{3}][b_{1}][b_{4}][u]$};
	\tikzstyle{every node}=[scale=0.4,fill=black!00]
	\node (n01) at (0.000, -0.40) {};
	\node (n02) at (1.000, -0.40) {};
	\node (n03) at (2.000, -0.40) {};
	\node (n04) at (3.000, -0.40) {};
	\node (n10) at (-0.40, 0.000) {};
	\node (n15) at (3.400, 0.000) {};
	\node (n20) at (-0.40, 1.000) {};
	\node (n25) at (3.400, 1.000) {};
	\node (n30) at (-0.40, 2.000) {};
	\node (n35) at (3.400, 2.000) {};
	\node (n41) at (0.000, 2.400) {};
	\node (n42) at (1.000, 2.400) {};
	\node (n43) at (2.000, 2.400) {};
	\node (n44) at (3.000, 2.400) {};
	\tikzstyle{every node}=[circle,draw,thick,fill=blue!50,minimum size=14pt,inner sep=0pt] 
	\node (n11) at (0.000, 0.000) {$a_{1}$};
	\node (n12) at (1.000, 0.000) {$a_{2}$};
	\node (n13) at (2.000, 0.000) {$a_{3}$};
	\node (n14) at (3.000, 0.000) {$a_{4}$};
	\node (n21) at (0.000, 1.000) {$b_{1}$};
	\node (n22) at (1.000, 1.000) {$b_{2}$};
	\node (n23) at (2.000, 1.000) {$b_{3}$};
	\node (n24) at (3.000, 1.000) {$b_{4}$};
	\node (n31) at (0.000, 2.000) {$c_{1}$};
	\node (n32) at (1.000, 2.000) {$c_{2}$};
	\node (n33) at (2.000, 2.000) {$c_{3}$};
	\node (n34) at (3.000, 2.000) {$c_{4}$};
	\tikzstyle{every node}=[scale=0.4,fill=black!00] 
	\foreach \from/\to in {n10/n11,n11/n12,n12/n13,n13/n14,n14/n15} \draw [thick] (\from) -- (\to);
	\foreach \from/\to in {n20/n21,n21/n22,n22/n23,n23/n24,n24/n25} \draw [thick] (\from) -- (\to);
	\foreach \from/\to in {n30/n31,n31/n32,n32/n33,n33/n34,n34/n35} \draw [thick] (\from) -- (\to);
	\foreach \from/\to in {n01/n11,n11/n21,n21/n31,n31/n41} \draw [thick] (\from) -- (\to);
	\foreach \from/\to in {n02/n12,n12/n22,n22/n32,n32/n42} \draw [thick] (\from) -- (\to);
	\foreach \from/\to in {n03/n13,n13/n23,n23/n33,n33/n43} \draw [thick] (\from) -- (\to);
	\foreach \from/\to in {n04/n14,n14/n24,n24/n34,n34/n44} \draw [thick] (\from) -- (\to);
\end{tikzpicture}
  \end{center}
  \caption{\label{FigIRF}A two-dimensional lattice of heights with weights
    placed on faces.}
\end{figure}

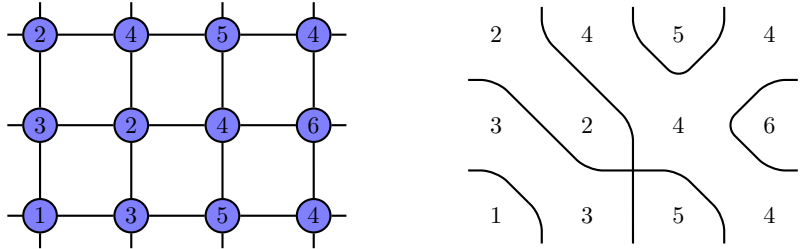
\begin{figure}[t] \begin{center}
\begin{tikzpicture}[scale=1.2]
	\tikzstyle{every node}=[scale=0.4,fill=black!00]
	\node (n01) at (0.000, -0.40) {};
	\node (n02) at (1.000, -0.40) {};
	\node (n03) at (2.000, -0.40) {};
	\node (n04) at (3.000, -0.40) {};
	\node (n10) at (-0.40, 0.000) {};
	\node (n15) at (3.400, 0.000) {};
	\node (n20) at (-0.40, 1.000) {};
	\node (n25) at (3.400, 1.000) {};
	\node (n30) at (-0.40, 2.000) {};
	\node (n35) at (3.400, 2.000) {};
	\node (n41) at (0.000, 2.400) {};
	\node (n42) at (1.000, 2.400) {};
	\node (n43) at (2.000, 2.400) {};
	\node (n44) at (3.000, 2.400) {};
	\tikzstyle{every node}=[scale=0.9,circle,draw,thick,fill=blue!50,minimum size=14pt,inner sep=0pt] 
	\node (n11) at (0.000, 0.000) {$1$};
	\node (n12) at (1.000, 0.000) {$3$};
	\node (n13) at (2.000, 0.000) {$5$};
	\node (n14) at (3.000, 0.000) {$4$};
	\node (n21) at (0.000, 1.000) {$3$};
	\node (n22) at (1.000, 1.000) {$2$};
	\node (n23) at (2.000, 1.000) {$4$};
	\node (n24) at (3.000, 1.000) {$6$};
	\node (n31) at (0.000, 2.000) {$2$};
	\node (n32) at (1.000, 2.000) {$4$};
	\node (n33) at (2.000, 2.000) {$5$};
	\node (n34) at (3.000, 2.000) {$4$};
	\tikzstyle{every node}=[scale=0.4,fill=black!00] 
	\foreach \from/\to in {n10/n11,n11/n12,n12/n13,n13/n14,n14/n15} \draw [thick] (\from) -- (\to);
	\foreach \from/\to in {n20/n21,n21/n22,n22/n23,n23/n24,n24/n25} \draw [thick] (\from) -- (\to);
	\foreach \from/\to in {n30/n31,n31/n32,n32/n33,n33/n34,n34/n35} \draw [thick] (\from) -- (\to);
	\foreach \from/\to in {n01/n11,n11/n21,n21/n31,n31/n41} \draw [thick] (\from) -- (\to);
	\foreach \from/\to in {n02/n12,n12/n22,n22/n32,n32/n42} \draw [thick] (\from) -- (\to);
	\foreach \from/\to in {n03/n13,n13/n23,n23/n33,n33/n43} \draw [thick] (\from) -- (\to);
	\foreach \from/\to in {n04/n14,n14/n24,n24/n34,n34/n44} \draw [thick] (\from) -- (\to);
	\tikzstyle{every node}=[scale=0.9]
	\node (n11) at (5.000, 0.000) {$1$};
	\node (n12) at (6.000, 0.000) {$3$};
	\node (n13) at (7.000, 0.000) {$5$};
	\node (n14) at (8.000, 0.000) {$4$};
	\node (n21) at (5.000, 1.000) {$3$};
	\node (n22) at (6.000, 1.000) {$2$};
	\node (n23) at (7.000, 1.000) {$4$};
	\node (n24) at (8.000, 1.000) {$6$};
	\node (n31) at (5.000, 2.000) {$2$};
	\node (n32) at (6.000, 2.000) {$4$};
	\node (n33) at (7.000, 2.000) {$5$};
	\node (n34) at (8.000, 2.000) {$4$};
	\tikzstyle{every node}=[scale=0.9]
	\node (nl01) at (5.500, -0.40) {};
	\node (nl02) at (6.500, -0.40) {};
	\node (nl03) at (7.500, -0.40) {};
	\node (nu10) at (4.600, 0.500) {};
	\node (nu15) at (8.400, 0.500) {};
	\node (nu20) at (4.60, 1.500) {};
	\node (nu25) at (8.400, 1.500) {};
	\node (nl41) at (5.500, 2.400) {};
	\node (nl42) at (6.500, 2.400) {};
	\node (nl43) at (7.500, 2.400) {};
	\coordinate (nl11) at (5.500, 0.000);
	\coordinate (nl12) at (6.500, 0.000);
	\coordinate (nl13) at (7.500, 0.000);
	\coordinate (nu11) at (5.000, 0.500);
	\coordinate (nu12) at (6.000, 0.500);
	\coordinate (nu13) at (7.000, 0.500);
	\coordinate (nu14) at (8.000, 0.500);
	\coordinate (nl21) at (5.500, 1.000);
	\coordinate (nl22) at (6.500, 1.000);
	\coordinate (nl23) at (7.500, 1.000);
	\coordinate (nu21) at (5.000, 1.500);
	\coordinate (nu22) at (6.000, 1.500);
	\coordinate (nu23) at (7.000, 1.500);
	\coordinate (nu24) at (8.000, 1.500);
	\coordinate (nl31) at (5.500, 2.000);
	\coordinate (nl32) at (6.500, 2.000);
	\coordinate (nl33) at (7.500, 2.000);
	\draw [thick, rounded corners=0.2cm] (nu10) -- (nu11) -- (nl11) -- (nl01);
	\draw [thick, rounded corners=0.2cm] (nu20) -- (nu21) -- (nl21) -- (nu12) -- (nu13) -- (nl13) -- (nl03);
	\draw [thick, rounded corners=0.2cm] (nl02) -- (nl12) -- (nl22) -- (nu22) -- (nl31) -- (nl41);
	\draw [thick, rounded corners=0.2cm] (nl42) -- (nl32) -- (nu23) -- (nl33) -- (nl43);
	\draw [thick, rounded corners=0.2cm] (nu15) -- (nu14) -- (nl23) -- (nu24) -- (nu25);
\end{tikzpicture}
\end{center}
    \caption{Drawing contour lines/domain walls between regions of 
    different heights allows RSOS models to be mapped onto loop models.
    \label{FigRSOSLoop}}
\end{figure}

In order to establish integrability of the model (\ref{eqnHamTheta}) for
particular choices of the parameter $\theta$ it has to be shown that the
Hamiltonian is a member of a complete set of commuting operators.
This can be done starting from $R$-matrices depending on a parameter acting on
the tensor product $\psi_3\tp\psi_3$, i.e.\
\begin{equation}
  \label{rmat}
  \begin{aligned}
  R(u) 
  & =  w_{1}(u)\,p^{(1)} + w_{2}(u)\,p^{(2)} + w_{5}(u)\,p^{(5)}\\
  & =  \sum_{a,b,c,d} \Wop[a][b][c][d][u] \ket[abd]\bra[acd]
\end{aligned}
\end{equation}
 which satisfy the Yang--Baxter equation (YBE)
\begin{equation}
  \label{eqnYBE}
  R_{j}(u)R_{j+1}(u+v)R_{j}(v) = R_{j+1}(v)R_{j}(u+v)R_{j+1}(u)\,.
\end{equation}
%
Every solution to this equation defines an integrable statistical model on a
square lattice with interactions round a face (or restricted solid-on-solid),
see Figure~\ref{FigIRF}.
%
The configurations of this model are labeled by \emph{height} variables on
each node of the lattice with neighboring heights corresponding to adjacent
nodes on the graph Figure \ref{FigPsi3Neig}.  To each face we assign the
Boltzmann weight $\Wop[a][b][c][d][u]$ from Equation~(\ref{rmat}) and
calculate the total weight of the configuration by taking the product of all
face weights.  The partition function is given as the sum of total weights
over all possible configurations of heights. It is clear from
Figure~\ref{FigPsi3Neig} that this model does not correspond to any model
within the A-D-E classification \cite{Pasq87a,Gepn93a}.
Likewise, it is known that models of this type can be mapped to loop models
\cite{Pasq87a,WaNi93}, see Figure~\ref{FigRSOSLoop}.

As a consequence of the Yang--Baxter equation the face model can be solved
using the commuting transfer matrices on the Hilbert space of the periodic
anyon chain
\begin{equation}
\label{eqTrans}
\begin{aligned}
 \bra[b_{1}...b_{\L}]t(u)\ket[a_{1}...a_{\L}] & =  \prod_{i=1}^{\L}
 \Wop[b_{i}][b_{i+1}][a_{i}][a_{i+1}][u]\,, \\ 
 \left[ t(u), t(v)\right] & =  0.
\end{aligned}
\end{equation}
Similar to the topological $Y$-operators, the transfer matrix can be viewed as
describing the process of moving an auxiliary anyon around the chain.  In
this case, however, the braiding is governed by the parameter-dependent
$R$-matrix (or, equivalently, the Boltzmann weights) rather than by constants
of the underlying category.

Just as there is a natural relationship between two-dimensional vertex models
and spin chains there exists a correspondence between this two-dimensional
interaction round the face model (IRF) and an anyon chain.
Provided that the $R$-matrix degenerates to the identity operator for a
particular value $u_0$ of the spectral parameter it follows that at $u=u_0$
the transfer matrix becomes the shift operator,
\begin{equation*}
  t(u_0) \ket[a_{1}...a_{\L-1}a_{\L}] = \alpha \ket[a_{\L}a_{1}...a_{\L-1}]\,.
\end{equation*}
This motivates the definition of a momentum operator and Hamiltonian by
expanding the transfer matrix around $u_0$:
\begin{equation}
  \label{eqnPandH}
  \begin{aligned}
    \mathcal{P} & = \frac{1}{i} \ln\left[\frac{t(u_0)}{\alpha}\right]\,,\\
    \mathcal{H} & = \alpha' \left.\frac{d}{du} \ln\left[t(u)\right]
    \right|_{u=u_0} + \alpha''\L\,.
  \end{aligned}
\end{equation}
For a proper choice of the parameters $\alpha'$ and $\alpha''$  the local Hamiltonian
\begin{equation}
  \label{eqnLocalHam}
  h = \alpha' \left.\frac{d}{du} \ln\left[R(u)\right] \right|_{u=u_0} + \alpha'',
\end{equation}
coincides with (\ref{eqnLocalHamth}).


To construct points of integrability we construct representations of the
Birman-Murakami-Wenzl (BMW) algebra \cite{{BiWe89,*Mura87}} using the local
projection operators (\ref{eqnProjOp}) occurring in the $\psi_3$-anyon model.
We define the operators
\begin{equation}
\label{repBMW}
  \begin{aligned}
    E_{j} & = \sqrt{5}\, p_{j}^{(1)}\,, \\
    G_{j} & = R^{33}_{1}\, p_{j}^{(1)} + R^{33}_{2}\, p_{j}^{(2)} + R^{33}_{5}\,
    p_{j}^{(5)} \\ 
    & =  i\left(p_{j}^{(1)} + \mathrm{e}^{\frac{4i\pi}{5}} p_{j}^{(2)} +
      \mathrm{e}^{-\frac{4i\pi}{5}} p_{j}^{(5)}\right)\,,
  \end{aligned}
\end{equation}
where the $R^{ab}_{c}$ belong to one of the sets of $R$-moves (\ref{Rmoves}).
These operators satisfy the relations:
\begin{equation*}
\begin{aligned}
  G_{j}G_{j+1}G_{j} & = G_{j+1}G_{j}G_{j+1} \\
  E_{j}E_{j\pm1}E_{j} & =  E_{j} \\
  G_{j}G_{j\pm1}E_{j} & = E_{j}G_{j\pm1}G_{j} \, = \, E_{j\pm1}E_{j} \\
  G_{j\pm1}E_{j}G_{j\pm1} & = G_{j}^{-1}E_{j\pm1}G_{j}^{-1} \\
  E_{j}G_{j\pm1}E_{j} & =  -iE_{j} \\
  G_{j}E_{j} & = E_{j}G_{j} \, = \, i E_{j} \\
  G_{j} - G_{j}^{-1} & = -i\phi(\mathbf{1} - E_{j}) \\
  E_{j}^{2} & = \sqrt{5} E_{j}
\end{aligned}
\end{equation*}
and
\begin{equation*}
\begin{aligned}
  G_{j}G_{j'} & =  G_{j'}G_{j} \\
  E_{j}E_{j'} & =  E_{j'}E_{j} 
\end{aligned}
\end{equation*}
for $|j-j'|>1$.  As mentioned above there exist three other sets of $R$-moves
consistent with our choice of the modular $\mathcal{S}$-matrix (\ref{Smat}).  Using one
of these in (\ref{repBMW}) would be equivalent to replacing $G_{j}$ with
either $G_{j}^{-1}$, $U^{\dagger}G_{j}U$ or $U^{\dagger}G_{j}^{-1}U$ with the
unitary transformation (\ref{Unonloc}) and modifying the above relations
suitably.


The BMW algebra contains a copy of the Temperley-Lieb algebra \cite{TeLi71} as
a subalgebra, from which we can construct the solution to the YBE
\begin{equation*}
\begin{aligned}
  R_{j}(u) 
  & = \sinh(\gamma + u)\,\mathbf{1} - \sinh(u)\,E_{j}, \\
  & = \sinh(\gamma-u)\,p^{(1)} + \sinh(\gamma+u)\left[ p^{(2)}_j + p^{(5)}_j \right]
\end{aligned}
\end{equation*}
with $\cosh\gamma = \frac{\sqrt{5}}{2}$.  Expanding of the resulting transfer
matrix around $u_0=0$ one obtains the anyon model (\ref{eqnHamTheta}) with
$(\alpha',\alpha'')=\pm (\frac{1}{2\sqrt{2}}, \frac{1}{2\sqrt{10}})$ for
$\theta=0$ and $\pi$.  As for other Temperley-Lieb models it can be written as
\begin{equation*}
  \mathcal{H}_{0,\pi} = \frac{\mp1}{\sqrt{10}} \sum_{j=1}^\L
  E_{j}+\mathrm{const.}\,.
\end{equation*}

Using the \emph{full} BMW algebra one can find two additional solutions to the
YBE \cite{ChGx91}.  One of them is \cite{CGLX92,Grim94}
\begin{equation}
\label{RmatBMW}
\begin{aligned}
  R_{j}(u) =&
  \sinh\left(\frac{7i\pi}{10}\right)\sinh\left(\frac{i\pi}{10}\right)\mathbf{1}
  + 
  \frac{1}{2}\sinh(u)\left(\mathrm{e}^{\frac{i\pi}{10}-u} G_{j} -
    \mathrm{e}^{u-\frac{i\pi}{10}}G_{j}^{-1}\right) \\ 
  =&
  \sinh\left(u+\frac{i\pi}{10}\right)
  \sinh\left(u+\frac{3i\pi}{10}\right)p^{(1)}_j +
  \sinh\left(u-\frac{i\pi}{10}\right)
  \sinh\left(u-\frac{3i\pi}{10}\right)p^{(2)}_j 
  \\ 
  & +
  \sinh\left(u+\frac{9i\pi}{10}\right)
  \sinh\left(u+\frac{3i\pi}{10}\right)p^{(5)}_j   \,.
\end{aligned}
\end{equation}
This solution leads to integrable points at $\theta=\eta,\eta+\pi$ where $\eta
= \mathrm{atan}\left(\frac{1+\sqrt{5}}{4}\right)-\frac{\pi}{4} $.  The other
$R$-matrix obtained by Baxterization of the BMW representation is related to
(\ref{RmatBMW}) through the transformation (\ref{Unonloc}) and gives rise to
integrable points at $\theta=-\eta,-\eta+\pi$.  However, these integrable
points are equivalent to the ones they are mapped from and are subsequently
omitted from further analysis.

As stated earlier, the weights given by (\ref{rmat}) will lead to a solution
of the YBE and a commuting transfer matrix (\ref{eqTrans}).  It turns out that
this transfer matrix belongs to a family of commuting transfer matrices.  To
define these transfer matrices we present the YBE for face weights
\begin{equation}
  \label{eqnWBYE}
  \begin{aligned}
    & \sum_{g} \WopAlt[a][g][b][c][u][\alpha][\beta]
    \WopAlt[g][f][c][d][u+v][\alpha][\gamma]
    \WopAlt[a][e][g][f][v][\beta][\gamma] \\ 
    & \quad = \quad \sum_{g} \WopAlt[b][g][c][d][v][\alpha][\beta]
    \WopAlt[a][e][b][g][u+v][\alpha][\gamma]
    \WopAlt[e][f][g][d][u][\alpha][\beta].  
\end{aligned}
\end{equation}
The general form we use for these face-weights is
\begin{equation}
  \label{eqnWeightGen}
  \WopAlt[a][b][c][d][u][\alpha][\beta] = \sum_{l=1}^{6}
  w^{\alpha,\beta}_{l}(u)
  \left[\left(F^{a\beta\alpha}_{d}\right)^{b}_{l}\right]^{*}
  \left(F^{a\alpha\beta}_{d}\right)^{c}_{l}. 
\end{equation}
In the case $\alpha=\beta=\gamma=3$ and $w^{3,3}_{l}(u)= w_{l}(u)$, equations
(\ref{eqnYBE}) and (\ref{eqnWBYE}) are equivalent.  We are able to determine
the required set of weights (see Appendix~\ref{app_BMW_tm}), such that we have
a family of commuting transfer matrices
\begin{equation}
  \label{eqnCommTranFam}
  \begin{aligned}
    \bra[b_{1}...b_{\L}]t^{(\ell)}(u)\ket[a_{1}...a_{\L}] & = \prod_{i=1}^{\L}
    \WopAlt[b_{i}][b_{i+1}][a_{i}][a_{i+1}][u][\ell][3]\,,\\ 
    \left[t^{(\ell)}(u) , t^{(\ell')}(v)\right] & = 0\,,
  \end{aligned}
\end{equation}
for $\ell,\ell'=1,\ldots,6$ and $u,v\in\C$.  An additional important feature
of these transfer matrices is that the limit of each is also a $Y$-operator
(\ref{Ytopo}), i.e.
\begin{eqnarray*}
  \lim_{u\rightarrow \pm \infty} t^{(\ell)}(u) & = & const \times Y_{\ell}\,.
\end{eqnarray*}
The appearance of the topological operators as a limit of these transfer
matrices is not unexpected as the braiding governed by these weights and the
underlying category coincide when $u\rightarrow \pm \infty$.

\section{Bethe Ansatz and low energy spectrum} 
\subsection{The Temperley-Lieb point}
The Temperley-Lieb algebra underlying the model at $\theta=0,\pi$ implies that
the spectrum of the anyon chain can be related to that of the XXZ spin-1/2
quantum chain with anisotropy $|\Delta|=\cosh\gamma=\frac{\sqrt{5}}{2}$
\cite{TeLi71,OwBa87}.  For the periodic boundary conditions considered here
the Temperley-Lieb equivalence implies that the spectrum of the transfer
matrix or the derived Hamiltonian coincides (up to degeneracies) with that of
the corresponding operator of the XXZ chain subject to suitably twisted
boundary conditions.  The eigenvalues of the latter are parametrized by
solutions $\{ u_j\}_{j=1}^n$ to the Bethe equations
\begin{equation}
\label{TL_BAE}
  \left[\frac{\sinh(u_{j}-\frac{\gamma}{2})}{\sinh(u_{j}+\frac{\gamma}{2})}
  \right]^{\L}
  = - \zeta^{-2} \prod_{k=1}^{n}
  \frac{\sinh(u_{j}-u_{k}-\gamma)}{\sinh(u_{j}-u_{k}+\gamma)}\,, 
\end{equation}
where the number of roots is related to the total magnetization of the spin
chain through $S^z=\frac{\L}{2}-n$.  In general the twist $\zeta$ can take
different values for different states depending on on the symmetry sector of
the Temperley-Lieb equivalent model \cite{AuKl10}.
To determine the allowed values of the twist the Bethe equations
(\ref{TL_BAE}) should be obtained directly from the spectral problem of the
model considered, e.g.\ by deriving functional equations for the eigenvalues
of the transfer matrices from the set of fusion relations between the finite
set (\ref{eqnCommTranFam}) \cite{KuRS81,BaRe89}.  In the case of
Temperley-Lieb type models the fusion hierarchy does not close in general.
Motivated by the functional equations for the XXZ chain we therefore introduce
an infinite set of functions $\Lambda_d(u)$
\begin{equation}
\label{TL_fusion_0}
\begin{aligned}
  \Lambda_{d+1}(u) & = \Lambda\left(u+(d-1)\gamma\right)\,
  \Lambda_{d}\left(u\right) +
  \Delta(u+(d-1)\gamma)\,\Lambda_{d-1}\left(u\right)\,
  \quad \mathrm{for~}d\ge2\,,\\
  \Delta(u) & = \left[4\sinh(\gamma-u) \sinh(\gamma+u)\right]^{\L}\,,
\end{aligned}
\end{equation}
starting from a given eigenvalue $\Lambda(u)\equiv\Lambda_2(u)$ of the
transfer matrix $t^{(3)}(u)$ of the anyon chain (we set
$\Lambda_1(u)\equiv1$).\footnote{%
  Note that $\Delta(u)$ is determined by the normalization of the $R$-matrix
  via the unitarity condition
  $R(u)R(-u)=4\sinh(\gamma-u)\sinh(\gamma+u)\,\mathbf{1}$.}
Computing this hierarchy for anyon chains of length up to $\L=4$ we find that
the functions $\Lambda_d(u)$ are of the form
\begin{equation}
  \Lambda_{d}(u+(1-d)\gamma) =  \prod_{k=1}^{d-2}
  \left[\sinh(u-k\gamma)\right]^{\L} \, c_d f_{d}(u)\,.
\end{equation}
In this expression, the $c_d$ are complex constants while the $f_d(u)$ with
the following properties: (i) they are analytic functions, (ii) their finite
zeroes converge to a set of $n=O(\L)$ complex numbers for sufficiently large
$d$ and (iii) they cycle, as a function of $d$, with period $p=1,2$ or $3$
depending on the eigenvalue $\Lambda_2(u)$ used as a seed in the recursion
(\ref{TL_fusion_0}), i.e.\ $f_{d+p}(u)=f_d(u)$.  This generalizes the
observation for spin-1/2 chains where the $d\to\infty$ limit of the functions
$\Lambda_d(u)$ has been found to exist (corresponding to a cycle with $p=1$)
and is related to the eigenvalues of the so-called $Q$-operator
\cite{Pron00,YaNZ06}.  The periodicity in $d$ allows to rewrite the recursion
relation (\ref{TL_fusion_0}) in terms of a linear combination $q_d(u)$ of the
functions $f_{d}(u)$ to $f_{d+p-1}(u)$ such that $q_d(u) \simeq \zeta^d q(u)
\equiv \zeta^d \prod_{\ell=1}^n\sinh(u-u_\ell+\frac{\gamma}{2})$ for
$d\to\infty$.  As a result the spectral problem can be formulated as a
so-called $TQ$-equation \cite{Baxter:book}
\begin{equation}
\label{TL_fusion}
\begin{aligned}
  \Lambda(u)\,q(u) & = \zeta^{-1}[\sinh(u+\gamma)]^\L\, q(u-\gamma) +
  \zeta\,[\sinh(u)]^\L\, q(u+\gamma)\,.
\end{aligned}
\end{equation}
As a consequence of the analyticity of the transfer matrix eigenvalues
$\Lambda(u)$ the zeroes $\{u_\ell\}_{\ell=1}^n$ of the $q$-functions are
solutions to the Bethe equations (\ref{TL_BAE}).  In terms of these Bethe
roots energy and momentum of the corresponding state is given by
\begin{equation}
\label{TL_spec}
\begin{aligned}
  E &= \pm \frac{1}{2\sqrt{2}}\left\{2\L + \frac{1}{2\sqrt{5}} \sum_{k=1}^{n}
    \frac{1}{\sinh(u_{k}-\frac{\gamma}{2})\sinh(u_{k}+\frac{\gamma}{2})}
  \right\} \\
  P &= -i\log\frac{\Lambda(0)}{\sinh^\L\gamma} = i \sum_{k=1}^{n} \log\left(
    \frac{\sinh(u_{k}-\frac{\gamma}{2})}{\sinh(u_{k}+\frac{\gamma}{2})}
  \right) + i\log\zeta\,. 
\end{aligned}
\end{equation}
 
With the expression (\ref{TL_fusion}) for the spectrum of the transfer matrix,
the value of the twist corresponding to a given eigenvalue can be obtained
from its asymptotics as $u\to\pm\infty$: computing the eigenvalues
$\lambda_\pm$ of the operators
\begin{equation*}
  t_{\pm} = \lim_{u\rightarrow\pm\infty} \left[
    \left(\frac{2}{\sqrt{\phi^{\pm1}}}\,\mathrm{e}^{\mp u}\right)^\L
    t(u) \right] 
\end{equation*}
for chains with up to $\L=10$ sites we find that they take values
\begin{equation}
\label{TL_lasy}
  \lambda_{\pm} \in \left\{0,\, \pm 1,\, \pm \sqrt{5}\right\} \cup
   \left\{ 2\cosh\left(k\gamma \pm l\frac{i\pi}{k}\right) \left| 0\le l \le
       k=1,\ldots,\frac{\L}{2} \right.\right\}.
\end{equation}
We find that $t_{+}$ and $t_{-}$ are real matrices and
$t_+ = [t_-]^\top$ (more generally, $[t(u)]^\top = t(-u-\gamma)$).
Consequently, if $\lambda_{\pm}$ are eigenvalues sharing the same eigenvector 
then $\lambda_+=\lambda_-^*$.
Using this in the $TQ$-equation (\ref{TL_fusion}) we can relate these
eigenvalues to the twist
\begin{equation*}
  \lambda_{\pm} = \left( \mathrm{e}^{\pm(\frac{\mathcal{L}}{2}-n)\gamma}\zeta^{-1}  +
    \mathrm{e}^{\mp(\frac{\mathcal{L}}{2}-n)\gamma} \zeta \right) 
\end{equation*}
or
\begin{equation}
\label{TL_zetas}
  \lambda_{+}+\lambda_{-} = 
  2 \cosh\left(\left(\frac{\mathcal{L}}{2}-n\right)\gamma\right)
    \left( \zeta + \zeta^{-1} \right) \,.
\end{equation}
%
Note that changing the twist $\zeta\to\zeta^{-1}$ does not affect the spectrum
of the anyon chain: a solution $\{u_{k}\}_{k=1}^{n}$ of the Bethe equations
(\ref{TL_BAE}) at twist $\zeta$ with energy $E$ and momentum $P$ can be mapped
to the solution $\{-u_{k}\}_{k=1}^{n}$ for twist $\zeta^{-1}$ having the same
energy but momentum $(-P)$, see (\ref{TL_spec}).

Additional constraints on the allowed twists follow from the translational
invariance of the system: taking the product over all Bethe equations
(\ref{TL_BAE}) we obtain
\begin{equation*}
  \prod_{j=1}^{n}
  \left[\frac{\sinh(u_{j}-\frac{\gamma}{2})}{
      \sinh(u_{j}+\frac{\gamma}{2})}\right]^{\L}
  = \zeta^{-2n}.
\end{equation*}
Taking the logarithm of this equation the Bethe roots $\{u_j\}$ can be
eliminated from the expression for the momentum eigenvalue (\ref{TL_spec})
giving $ \L P = {2\pi k} + i(\L-2n)\log\zeta$ for some integer $k$.
As a consequence of periodic boundary conditions the momentum eigenvalues are
quantized, $\L P \in 2\pi\,\Z$.  For states corresponding to root
configurations $\{u_j\}_{j=1}^{\L/2}$ this is always true.  From our numerical
data we find, however, that the corresponding eigenvalues of the transfer
matrix take asymptotic values $\lambda_\pm\in\{0,\pm1,\pm\sqrt{5}\}$ only.
Configurations with $n\neq\frac{\L}{2}$ roots, on the other hand, are only
possible for twists $\zeta^{\L-2n}=1$ all of which are all allowed by
(\ref{TL_zetas}) with (\ref{TL_lasy}).

In summary we have used the Temperley-Lieb equivalence to relate the complete
spectrum of the anyon chain to those found in the $S^z=\frac{\L}{2}-n$
eigenspaces of XXZ spin-1/2 chains with anisotropy $|\Delta|=\sqrt{5}/2$ and
periodic boundaries twisted by
\begin{equation}
\label{TL_zetasbyn}
\begin{aligned}
 n=\frac{\L}{2}&:\qquad \zeta,\zeta^{-1}\in\left\{
    i,\pm\mathrm{e}^{\frac{\pi}{3}i},\pm\phi \right\}\,,\\
 n\neq\frac{\L}{2}&:\qquad \zeta,\zeta^{-1} \in \left\{ \mathrm{e}^{\frac{2\pi
         i}{\L-2n}\ell}: \ell = 0\ldots \L-2n\right\}\,.
\end{aligned}
\end{equation}
Since the spectral properties of the model should not depend on the boundary
conditions in the thermodynamic limit this relation to the XXZ spin chain
implies that the anyon chain has gapped excitations over a possibly degenerate
ground state for $\theta=0$ and $\pi$.

\subsubsection{Ground state and low lying excitations for $\theta=0$}
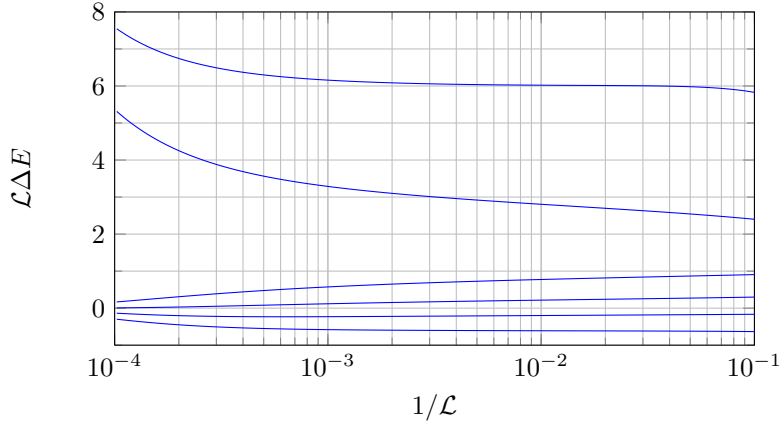
\begin{figure}[t]
\begin{center}
\begin{tikzpicture}
	\begin{axis}[width=10.0cm,height=6.0cm, grid=both, xmin=0.0001,xmax=0.1,xmode=log, ymin=-1.0,ymax=8.0, minor tick num=1, xlabel=$1/\L$,ylabel=$\L\Delta E$]
		\addplot[smooth,blue,mark=.] coordinates {(0.16667,-0.65487) (0.125,-0.64065) (0.1,-0.63355) (0.083333,-0.62934) (0.071429,-0.62655) (0.0625,-0.62454) (0.055556,-0.62303) (0.05,-0.62183) (0.041667,-0.62003) (0.035714,-0.61872) (0.03125,-0.61769) (0.027778,-0.61685) (0.025,-0.61614) (0.021739,-0.61523) (0.019231,-0.61447) (0.017241,-0.61381) (0.015625,-0.61322) (0.013889,-0.61252) (0.0125,-0.61189) (0.011111,-0.61117) (0.01,-0.61053) (0.0089286,-0.60981) (0.0080645,-0.60915) (0.0072464,-0.60844) (0.0065789,-0.60776) (0.0059524,-0.60704) (0.0053763,-0.60627) (0.0048544,-0.60545) (0.004386,-0.6046) (0.0039683,-0.6037) (0.0035971,-0.60277) (0.003268,-0.60179) (0.0029586,-0.60071) (0.0026882,-0.59959) (0.002439,-0.59837) (0.0022124,-0.59705) (0.002008,-0.59562) (0.0018248,-0.59409) (0.0016556,-0.59239) (0.0015015,-0.59053) (0.0013624,-0.58849) (0.0012376,-0.58628) (0.0011236,-0.58384) (0.0010204,-0.58115) (0.00092593,-0.57816) (0.00084034,-0.57484) (0.00076336,-0.5712) (0.000693
 48,-0.56717) (0.00062972,-0.56266) (0.00057208,-0.55768) (0.00051975,-0.55214) (0.00047214,-0.54596) (0.00042918,-0.53914) (0.00039002,-0.53153) (0.00035436,-0.52305) (0.00032196,-0.51363) (0.00029257,-0.50319) (0.00026596,-0.49167) (0.00024166,-0.47889) (0.00021968,-0.46485) (0.00019968,-0.44941) (0.00018149,-0.4325) (0.00016496,-0.41407) (0.00014993,-0.39406) (0.00013628,-0.37252) (0.00012389,-0.34949) (0.00011261,-0.32503) (0.00010235,-0.29928)};
		\addplot[smooth,blue,mark=.] coordinates {(0.16667,-0.15914) (0.125,-0.16283) (0.1,-0.16602) (0.083333,-0.16873) (0.071429,-0.17103) (0.0625,-0.17304) (0.055556,-0.1748) (0.05,-0.17637) (0.041667,-0.17907) (0.035714,-0.18133) (0.03125,-0.18328) (0.027778,-0.18498) (0.025,-0.1865) (0.021739,-0.1885) (0.019231,-0.19024) (0.017241,-0.19179) (0.015625,-0.19319) (0.013889,-0.19486) (0.0125,-0.19635) (0.011111,-0.19802) (0.01,-0.19951) (0.0089286,-0.20111) (0.0080645,-0.20256) (0.0072464,-0.20408) (0.0065789,-0.20546) (0.0059524,-0.20689) (0.0053763,-0.20835) (0.0048544,-0.20981) (0.004386,-0.21127) (0.0039683,-0.21271) (0.0035971,-0.21412) (0.003268,-0.2155) (0.0029586,-0.21692) (0.0026882,-0.21828) (0.002439,-0.21965) (0.0022124,-0.221) (0.002008,-0.22233) (0.0018248,-0.22362) (0.0016556,-0.22489) (0.0015015,-0.22612) (0.0013624,-0.2273) (0.0012376,-0.22839) (0.0011236,-0.22942) (0.0010204,-0.23034) (0.00092593,-0.23117) (0.00084034,-0.23185) (0.00076336,-0.23238) (0.00069348,-
 0.23272) (0.00062972,-0.23286) (0.00057208,-0.23275) (0.00051975,-0.23236) (0.00047214,-0.23165) (0.00042918,-0.23058) (0.00039002,-0.22909) (0.00035436,-0.22714) (0.00032196,-0.22467) (0.00029257,-0.22162) (0.00026596,-0.21795) (0.00024166,-0.21358) (0.00021968,-0.20847) (0.00019968,-0.20257) (0.00018149,-0.19581) (0.00016496,-0.18819) (0.00014993,-0.17966) (0.00013628,-0.17026) (0.00012389,-0.16002) (0.00011261,-0.14898) (0.00010235,-0.13723)};
		\addplot[smooth,blue,mark=.] coordinates {(0.16667,0.32378) (0.125,0.30872) (0.1,0.29854) (0.083333,0.29088) (0.071429,0.28474) (0.0625,0.27963) (0.055556,0.27524) (0.05,0.2714) (0.041667,0.26489) (0.035714,0.25951) (0.03125,0.25491) (0.027778,0.25088) (0.025,0.2473) (0.021739,0.24256) (0.019231,0.23841) (0.017241,0.23471) (0.015625,0.23137) (0.013889,0.22736) (0.0125,0.22374) (0.011111,0.21967) (0.01,0.216) (0.0089286,0.212) (0.0080645,0.20838) (0.0072464,0.20452) (0.0065789,0.20099) (0.0059524,0.19728) (0.0053763,0.19345) (0.0048544,0.18955) (0.004386,0.18561) (0.0039683,0.18165) (0.0035971,0.17769) (0.003268,0.17375) (0.0029586,0.16959) (0.0026882,0.16551) (0.002439,0.16128) (0.0022124,0.15694) (0.002008,0.15255) (0.0018248,0.14813) (0.0016556,0.14353) (0.0015015,0.13882) (0.0013624,0.13403) (0.0012376,0.12921) (0.0011236,0.12426) (0.0010204,0.11924) (0.00092593,0.11408) (0.00084034,0.10884) (0.00076336,0.10358) (0.00069348,0.098243) (0.00062972,0.092825) (0.00057208,0.0
 87381) (0.00051975,0.081906) (0.00047214,0.076405) (0.00042918,0.070942) (0.00039002,0.065483) (0.00035436,0.06006) (0.00032196,0.05471) (0.00029257,0.049472) (0.00026596,0.044386) (0.00024166,0.039442) (0.00021968,0.034721) (0.00019968,0.030225) (0.00018149,0.025991) (0.00016496,0.022054) (0.00014993,0.018434) (0.00013628,0.015162) (0.00012389,0.01225) (0.00011261,0.0097003) (0.00010235,0.0075112) };
		\addplot[smooth,blue,mark=.] coordinates {(0.16667,0.93778) (0.125,0.91938) (0.1,0.90617) (0.083333,0.89576) (0.071429,0.88712) (0.0625,0.87971) (0.055556,0.8732) (0.05,0.86739) (0.041667,0.85733) (0.035714,0.8488) (0.03125,0.84135) (0.027778,0.83474) (0.025,0.82878) (0.021739,0.82079) (0.019231,0.81368) (0.017241,0.80728) (0.015625,0.80143) (0.013889,0.79433) (0.0125,0.78788) (0.011111,0.78055) (0.01,0.77386) (0.0089286,0.76653) (0.0080645,0.75981) (0.0072464,0.7526) (0.0065789,0.74594) (0.0059524,0.7389) (0.0053763,0.73157) (0.0048544,0.72403) (0.004386,0.71634) (0.0039683,0.70856) (0.0035971,0.70072) (0.003268,0.69285) (0.0029586,0.68445) (0.0026882,0.67613) (0.002439,0.66743) (0.0022124,0.65843) (0.002008,0.64921) (0.0018248,0.63981) (0.0016556,0.62994) (0.0015015,0.6197) (0.0013624,0.60915) (0.0012376,0.59837) (0.0011236,0.58715) (0.0010204,0.57556) (0.00092593,0.56346) (0.00084034,0.55095) (0.00076336,0.5381) (0.00069348,0.52481) (0.00062972,0.51098) (0.00057208,0.496
 73) (0.00051975,0.48198) (0.00047214,0.46671) (0.00042918,0.45103) (0.00039002,0.43479) (0.00035436,0.41801) (0.00032196,0.40071) (0.00029257,0.38295) (0.00026596,0.36477) (0.00024166,0.34607) (0.00021968,0.32704) (0.00019968,0.30762) (0.00018149,0.28787) (0.00016496,0.2679) (0.00014993,0.24775) (0.00013628,0.22757) (0.00012389,0.20747) (0.00011261,0.18754) (0.00010235,0.16792) };
		\addplot[smooth,blue,mark=.] coordinates {(0.16667,2.2658) (0.125,2.3455) (0.1,2.4005) (0.083333,2.4417) (0.071429,2.4743) (0.0625,2.5011) (0.055556,2.5239) (0.05,2.5435) (0.041667,2.5762) (0.035714,2.6028) (0.03125,2.6252) (0.027778,2.6445) (0.025,2.6615) (0.021739,2.6837) (0.019231,2.7031) (0.017241,2.7201) (0.015625,2.7355) (0.013889,2.7538) (0.0125,2.7702) (0.011111,2.7886) (0.01,2.8052) (0.0089286,2.8231) (0.0080645,2.8394) (0.0072464,2.8568) (0.0065789,2.8727) (0.0059524,2.8894) (0.0053763,2.9067) (0.0048544,2.9244) (0.004386,2.9424) (0.0039683,2.9606) (0.0035971,2.9788) (0.003268,2.9972) (0.0029586,3.0168) (0.0026882,3.0362) (0.002439,3.0566) (0.0022124,3.0778) (0.002008,3.0996) (0.0018248,3.122) (0.0016556,3.1457) (0.0015015,3.1706) (0.0013624,3.1964) (0.0012376,3.2232) (0.0011236,3.2515) (0.0010204,3.2813) (0.00092593,3.3129) (0.00084034,3.3463) (0.00076336,3.3814) (0.00069348,3.4186) (0.00062972,3.4584) (0.00057208,3.5008) (0.00051975,3.546) (0.00047214,3.5947) (0
 .00042918,3.6466) (0.00039002,3.7027) (0.00035436,3.7634) (0.00032196,3.8292) (0.00029257,3.9005) (0.00026596,3.9777) (0.00024166,4.0623) (0.00021968,4.1542) (0.00019968,4.2551) (0.00018149,4.3657) (0.00016496,4.4873) (0.00014993,4.6213) (0.00013628,4.769) (0.00012389,4.932) (0.00011261,5.1125) (0.00010235,5.3127)};
		\addplot[smooth,blue,mark=.] coordinates {(0.16667,5.4276) (0.125,5.7028) (0.1,5.8276) (0.083333,5.8931) (0.071429,5.931) (0.0625,5.9546) (0.055556,5.97) (0.05,5.9806) (0.041667,5.9935) (0.035714,6.0006) (0.03125,6.005) (0.027778,6.0078) (0.025,6.0097) (0.021739,6.0118) (0.019231,6.0133) (0.017241,6.0145) (0.015625,6.0155) (0.013889,6.0168) (0.0125,6.0181) (0.011111,6.0196) (0.01,6.0212) (0.0089286,6.0231) (0.0080645,6.025) (0.0072464,6.0272) (0.0065789,6.0295) (0.0059524,6.0322) (0.0053763,6.0351) (0.0048544,6.0384) (0.004386,6.042) (0.0039683,6.046) (0.0035971,6.0502) (0.003268,6.0547) (0.0029586,6.0598) (0.0026882,6.0652) (0.002439,6.0711) (0.0022124,6.0776) (0.002008,6.0847) (0.0018248,6.0923) (0.0016556,6.1007) (0.0015015,6.1099) (0.0013624,6.1198) (0.0012376,6.1306) (0.0011236,6.1425) (0.0010204,6.1554) (0.00092593,6.1696) (0.00084034,6.1852) (0.00076336,6.2021) (0.00069348,6.2207) (0.00062972,6.2413) (0.00057208,6.2638) (0.00051975,6.2886) (0.00047214,6.316) (0.00042
 918,6.3461) (0.00039002,6.3795) (0.00035436,6.4166) (0.00032196,6.4578) (0.00029257,6.5036) (0.00026596,6.5543) (0.00024166,6.6111) (0.00021968,6.6741) (0.00019968,6.7447) (0.00018149,6.8236) (0.00016496,6.912) (0.00014993,7.0112) (0.00013628,7.1223) (0.00012389,7.2469) (0.00011261,7.3871) (0.00010235,7.5449) };
	\end{axis} 
\end{tikzpicture}
\caption{\label{FigTLEnScaling} $\L\Delta E$ as a function of $1/\L$ for the
  ground state and 1st, 2nd, 3rd, 4th and 6th lowest excitations. We have
  omitted data for the 5th lowest excitation due to insufficient solutions of
  the Bethe equations. The ground state has multiplicity two with states in
  symmetry sectors $\psi_{1}$ and $\psi_{6}$. The next three energies have
  multiplicities two, four and two, respectively, and are in sectors
  $\psi_{3}\oplus\psi_{4}$, $\psi_{2}\oplus\psi_{5}$ and
  $\psi_{3}\oplus\psi_{4}$, respectively.}
\end{center}
\end{figure}

For $\theta=0$ the ground state of the related anti-ferromagnetic XXZ chain is
a completely filled band of $n=\L/2$ imaginary Bethe roots $-\pi/2<i
u_j\le\pi/2$.  We have solved the Bethe equations (\ref{TL_BAE}) subject to
the twists appearing in the anyon model according to (\ref{TL_zetasbyn}) for
such configurations numerically.  The corresponding energies and momenta
(\ref{TL_spec}) of these Bethe configurations have been identified with levels
in the spectrum of the model for $\theta=0$ as found by exact diagonalization
of the anyon Hamiltonian for $\L\le12$, the ground state being the one with
twist $\zeta=\pm\phi$.  For larger $\L$ the differences between their energies
become exponentially small indicating the emergence of a manifold of ten
degenerate ground states spanning all topological sectors in the thermodynamic
limit, see Figure \ref{FigTLEnScaling}.
The ground state energy density coincides with that of the anti-ferromagnetic
XXZ chain \cite{Walk59}
\begin{equation}
  \frac{E^{(0)}_{\theta=0}}{\L}  = \frac{1}{\sqrt{2}} - \frac{1}{2\sqrt{10}}
  \sum_{k\in\Z} \frac{\mathrm{e}^{-\gamma|k|}}{\cosh(\gamma k)}
  \approx 0.238822178\,.
\end{equation}
The other excitations are separated from the ground states by a gap.  From the
solution of the XXZ model this gap is \cite{JoKM73} 
%
\begin{equation}
  \Delta E = \frac{2}{\pi\,\sqrt{10}}\,K(k)\,\sqrt{1-k^2}
  \simeq 2.9\,10^{-4}\,.
\end{equation}
Here $K(k)$ is the complete elliptic integral with modulus defined by
\begin{equation*}
  \pi\frac{K'(k)}{K(k)} = \gamma\,.
\end{equation*}

\subsubsection{Ground state and low lying excitations for $\theta=\pi$}
For coupling $\theta=\pi$ the Temperley-Lieb equivalent spin chain is the
ferromagnetic XXZ model.  The ground states appear in the sector with $n=0$
Bethe roots.  They have energy $E^{(0)}_{\theta=\pi}=-\L/\sqrt{2}$ and all
possible momenta.
A family of excitations can be found in the sector with precisely one Bethe
root:
solving the Bethe equation (\ref{TL_BAE}) for given twist $\zeta$ we obtain
\begin{equation}
  u = \frac{1}{2}\ln\left[
    \frac{\sinh\left(-\frac{i\pi}{\L}k+\frac{\gamma}{2}+\frac{1}{\L}\ln(\zeta)
      \right)}{\sinh\left(-\frac{i\pi}{\L}k-\frac{\gamma}{2} +
        \frac{1}{\L}\ln(\zeta)\right)} \right] \,,
  \quad 0\le k\le \L-1
\end{equation}
where, as a consequence of (\ref{TL_zetasbyn}), $\zeta$ can take values
$\zeta=\exp(\frac{2\pi i}{\L-2} \ell)$ for integer $\ell$.  From
(\ref{TL_spec}) energy and momentum of these solutions are obtained to be
\begin{equation}
\begin{aligned}
    E&= E^{(0)}_{\theta=\pi} + \frac{1}{\sqrt{2}} -
    \frac{2}{\sqrt{10}}\cos\left( \frac{2\pi}{\L}k 
    - \frac{4\pi}{\L(\L-2)}\ell\right)\,,\\
  P &= \frac{2\pi}{L}(k+\ell)\,.
\end{aligned}
\end{equation}
The lowest of these excitations is the one with quantum numbers $k=0=\ell$,
implying a gap $\Delta E=E-E_{\theta=\pi}^{(0)}=1/\sqrt{2} - 2/\sqrt{10}$.  In our
numerical analysis of the spectrum we find for $\theta=\pi$ that the ground
state and excitations with finite energy show degeneracies growing with the
system size $\L$.

\subsection{The BMW points}
The existence of a family of commuting transfer matrices for the integrable
points $\theta=\eta,\eta+\pi$, allows for a straightforward construction of
Bethe equations which are necessarily complete. Specifically, we can find
functional relations (\ref{BMW_fusion}) for the transfer matrices resembling
the $SO(5)_{2}$ fusion rules
\begin{equation}
\label{eqnFusRel}
\begin{aligned}
  t^{(6)}(u) t^{(6)}(u) & =  \mathbf{1}, \\
  t^{(2)}(u) t^{(3)}(u) & =
  \left[\sinh\left(u-\frac{i\pi}{10}\right)\right]^{\L}
  t^{(3)}(u+\frac{2i\pi}{5}) + \left[i\sinh\left(u\right)\right]^{\L}
  t^{(6)}(u) t^{(3)}(u-\frac{2i\pi}{5})\,.
\end{aligned}
\end{equation}
By its construction (\ref{eqTrans}) the transfer matrix $t(u)\equiv
t^{(3)}(u)$ is a Laurent polynomial in $\mathrm{e}^u$.  Therefore, the
eigenvalues $\Lambda(u) = c \prod_{k=1}^{n}\sinh(u-u_{k}-\frac{i\pi}{20})$ can
be parametrized in terms of their zeroes $\{u_k\}_{k=1}^n$ and an amplitude
$c$.  Analyticity of these expressions implies that the parameters $\{u_k\}$
are given by the Bethe equations%
\footnote{This is different to the usual parametrization of the transfer
  matrix eigenvalues in terms of the zeroes of the $q$-functions, see e.g.\ in
  the TL-models (\ref{TL_fusion}).  A similar observation has been made in the
  $D(D_3)$ vertex and fusion path models \cite{FiFr13}.}
\begin{equation}
\label{BMW_BAE}
\left(i\,\frac{\sinh\left(u_{j}+\frac{i\pi}{20}\right)}{
    \sinh\left(u_{j}-\frac{i\pi}{20}\right)} 
\right)^{\L} = - y_{6} \prod_{k=1}^{n} \left(
  \frac{\sinh(u_{j}-u_{k}+\frac{2i\pi}{5})}{
    \sinh(u_{j}-u_{k}-\frac{2i\pi}{5})}\right)  
\end{equation}
where $y_{6}=\pm1$ is the eigenvalue of $t^{(6)}(u)=Y_{6}$. By inspection we
see that the Bethe equations are equivalent to the ones appearing in the
$Z_{5}$ Fateev-Zamolodchikov (FZ) model, sometimes referred to as the
self-dual chiral Potts model, up to a twist \cite{FaZa82,Albe94}. This
connection to the $Z_{5}$ FZ model is not unexpected as the family of
commuting transfer matrices implies that this $SO(5)_{2}$ model is also a
descendant of the zero-field six-vertex model, like the FZ model
\cite{BaSt90}.

Given a set of solutions to the Bethe equations, $\{u_{j}\}$, we are able to
determine energy and momentum of the corresponding state from (\ref{eqnPandH})
\begin{equation}
\label{BMW_spec}
\begin{aligned}
  E & = \pm \frac{1}{2} \cos(\eta + \frac{\pi}{4}) \left\{ \L -
    \frac{1}{4i\cos(\frac{\pi}{10})}\sum_{k=1}^{n}
    \frac{\cosh(u_{k}+\frac{i\pi}{20})}{\sinh(u_{k}+\frac{i\pi}{20})}
  \right\}, \\ 
  P 
  & = -i\left[(2\L-n)\ln(2) + \ln(c) +
    \sum_{k=1}^{n}\ln\left(\mathrm{e}^{-(u_{k}+\frac{i\pi}{20})} -
      \mathrm{e}^{(u_{k}+\frac{i\pi}{20})}\right) \right]. 
\end{aligned}
\end{equation}
By construction, every eigenstate of the model corresponds to a particular
solution of the Bethe equations. Among all the possible solutions of the
latter, however, the physical ones for the anyon chain have to obey certain
constraints.  For example, the Hermitecity of the Hamiltonian requires that
only root configurations giving a real energy should be considered.
Similarly, the momentum eigenvalues are real which gives a constraint on the
constant $c$ in the above expression.  Using the asymptotics of the transfer
matrix, i.e.\
\begin{equation*}
 (Y_{3})^{2} = \lim_{u\rightarrow+\infty} \left[ \mathrm{e}^{-4u\mathcal{L}} \times
   t^{(3)}(-u) t^{(3)}(u) \right] \,,
\end{equation*}
we find that states in the sectors $\psi_{1}$ or $\psi_{6}$ are parametrized
by $n=2\L$ Bethe roots and have a constant $c^{2}=5$ appearing in the
momentum.  Similarly, for states in the $\psi_{3}$ or $\psi_{4}$ sector we
find $n=2\L$ and $c^{2}=1$ while for the states in the $\psi_{2}$ or
$\psi_{5}$ sector we find $n<2\L$ and $c\in \C$ (in our numerical studies we
find $n=2\L-4$ or $n=2\L-2$ for the latter).
We can gain additional information from the asymptotics of the transfer
matrices (which become the $Y$-operators) appearing in the fusion relations
(\ref{eqnFusRel}), determining the relation
\begin{equation*}
  y_{2} =  \mathrm{e}^{-\frac{2i\pi}{5}(2\L-n)} + 
           y_{6}\,\mathrm{e}^{\frac{2i\pi}{5}(2\L-n)} \,.
\end{equation*}
Thus simply knowing $y_{6}$ and how many finite Bethe roots there are
determines the eigenvalue of $Y_{2}$. This information is summarized in
table \ref{tabBMW_topo}.
\begin{table}[t]
\begin{tabularx}{\textwidth}{CCCCC} \hline \hline
  Sector(s) & Bethe Roots  & $y_{6}$ & $y_2$ & $c$ 
  \tabularnewline \hline \hline
  $\psi_{1}$ or $\psi_{6}$ & $2\L$ & $+1$ & $2$ & $\pm \sqrt{5}$ \tabularnewline
  $\psi_{3}$ or $\psi_{4}$ & $2\L$ & $-1$ & $0$ & $\pm 1$ \tabularnewline 
  $\psi_{2}$ & $2\L-2$ & $+1$ & $-\phi$ & $\C$ \tabularnewline
  $\psi_{5}$ & $2\L-4$ & $+1$ & $\phi^{-1}$ & $\C$ \tabularnewline \hline \hline
\end{tabularx}
\caption{\label{tabBMW_topo}Characterization of Bethe states according to
  topological sectors.}
\end{table}

\noindent
As this integrable point is related to the $Z_{5}$ FZ model and the Bethe
equations are equivalent up to an allowed phase factor, we can use the Bethe
root classification of Albertini \cite{Albe94}. Thus one expects that the Bethe
roots come in the following varieties
\begin{quote}
\begin{description}
\item[$\pm$-strings] $u_{k}+\frac{i\pi}{4}(1-p)$ with $u_{k}\in \R$ and $p=\pm1$,
\item[2-strings] $\{u_{k}+\frac{i\pi}{4}(1-p)-\frac{3i\pi}{5}+\frac{2i\pi}{5}l \}_{l=1}^{2}$ with $u_{k}\in \R$ and $p=\pm1$,
\item[4-strings] $\{u_{k}+\frac{i\pi}{4}(1-p)+\frac{2i\pi}{5}l \}_{l=1}^{4}$ with $u_{k}\in \R$ and $p=\pm1$,
\item[pairs] $\{u_{k} \pm \frac{i\pi}{4}\}$ with $u_{k}\in \R$,
\item[sextets] $\{u_{k} \pm (\frac{i\pi}{4}+\frac{i\pi}{10}l)\}_{l=-1}^{1}$ with $u_{k}\in \R$.
\end{description}
\end{quote}
We have confirmed this by exact diagonalization of the transfer matrix for
systems with small chain lengths, $\L \leq 12$.


\subsubsection{Thermodynamic limit of the BMW model at $\theta=\eta$}
\begin{figure}
\includegraphics[width=0.7\textwidth]{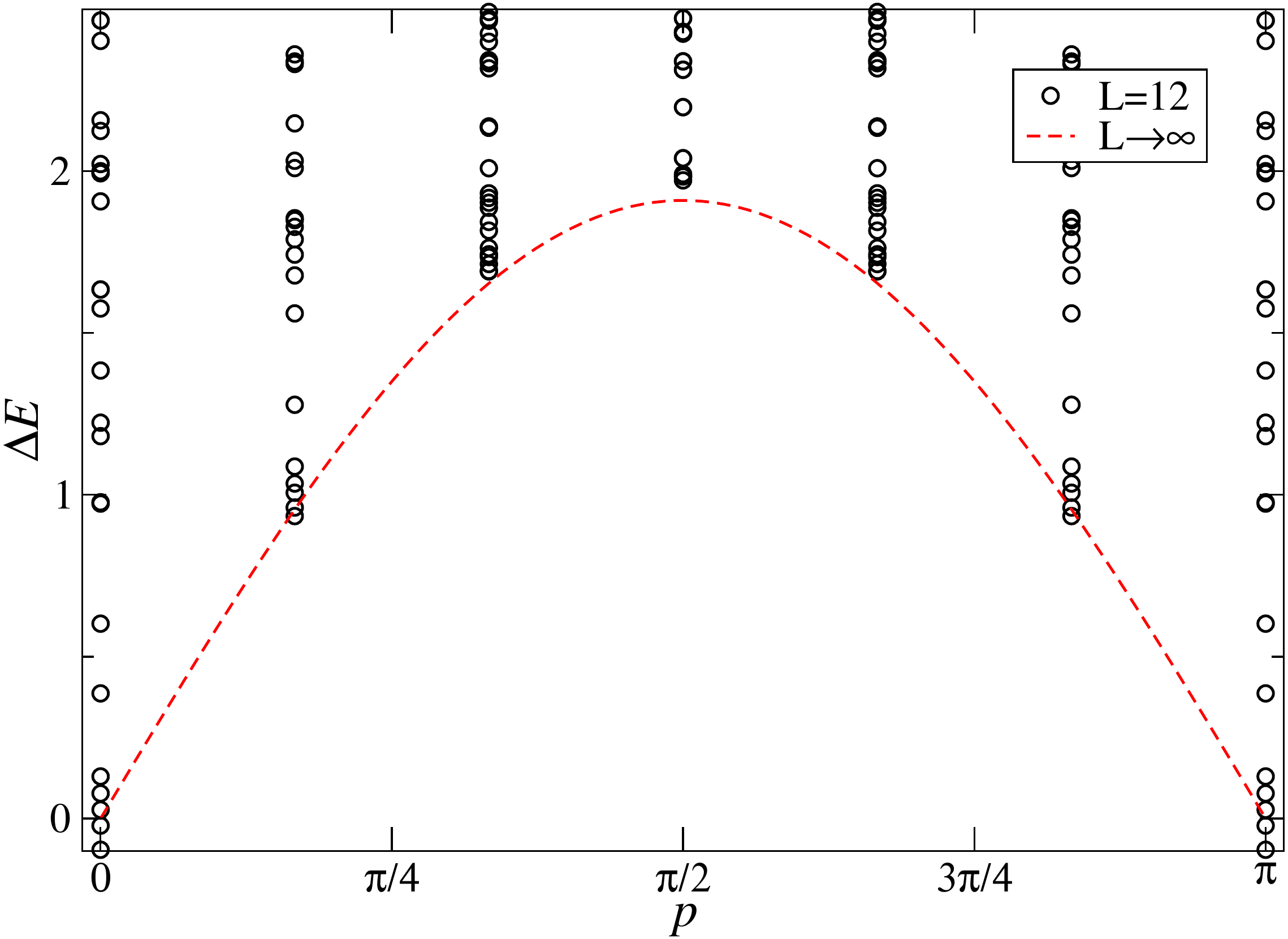}
\caption{\label{fig:specBMW1}Finite size spectrum of the BMW model at
  $\theta=\eta$ for $\L=12$: energies $\Delta E=E(\L)-E_{\theta=\eta}^{(0)}$
  are shown in units of $2\pi v_F/\L$.  The dashed line indicates the
  dispersion (\ref{eqndressedmom}) of low energy excitations in the
  thermodynamic limit.}
\end{figure}

In the thermodynamic limit the ground state for the $\theta=\eta$ point
consists of $\frac{\L}{2}$ 4-strings with only even chain lengths admissible.
The corresponding energy density can be computed using the root density
formalism \cite{YaYa69} where the configuration of Bethe roots is described by
the density of 4-strings.  The latter is given as the solution of the linear
integral equation
\begin{equation*}
\begin{aligned}
  \rho(u) & =  A'(u;\frac{19}{20}) + A'(u;\frac{7}{20}) + A'(u;\frac{15}{20})
  + A'(u;\frac{3}{20}) - \int_{-\infty}^{\infty} \mathrm{d}v\,
  \rho(v)A'(u-v;\frac{2}{5})\,, \\ 
  &\quad A'(u;t)  = -\frac{1}{2\pi}
  \frac{\sin(2t\pi)}{\sinh(u-it\pi)\sinh(u+it\pi)}\,.
\end{aligned}
\end{equation*}
Solving this equation by Fourier transform the ground energy density can be
computed to give
\begin{equation*}
\begin{aligned}
  \frac{E^{(0)}_{\theta=\eta}}{\L}
  &= \frac{\sqrt{22+2\sqrt{5}}}{1450\pi} \,\left(40\pi
    +28\pi\sqrt{5} +20\sqrt{5}\sqrt{10+2\sqrt{5}}-75\sqrt{10+2\sqrt{5}}
  \right) \,\\
  &\approx \,\, 0.2339952817 \,. 
\end{aligned}
\end{equation*}
The low energy excitations are gapless with linear dispersion near the Fermi
points at $k_F=0,\pi$, see Fig.~\ref{fig:specBMW1}.
To calculate the corresponding Fermi velocity, we observe that the
contribution of each individual string to the energy and momentum can be
expressed in terms of their root density in the thermodynamic limit
\begin{equation}
\label{eqndressedmom}
\begin{aligned}
  \epsilon(u) &= \frac{\pi\cos(\frac{\pi}{4}+\eta)}{8\cos(\frac{\pi}{10})}
  \rho(u)\,,\\
  p(u) &= \pi \int^u \mathrm{d}v\, \rho(v)\,.
\end{aligned}
\end{equation}
Eliminating the rapidity $u$ we obtain the dispersion relation $\epsilon(p)$
of the elementary excitations in the system and the Fermi velocity is obtained
to be
\begin{equation*}
  v_{F} = \frac{5\cos(\frac{\pi}{4}+\eta)}{4\cos(\frac{\pi}{10})} \,\, \approx
  \,\, 1.021807736 \,.
\end{equation*}

\subsubsection{Thermodynamic limit of the BMW model at  $\theta=\eta+\pi$}
\begin{figure}
\includegraphics[width=0.7\textwidth]{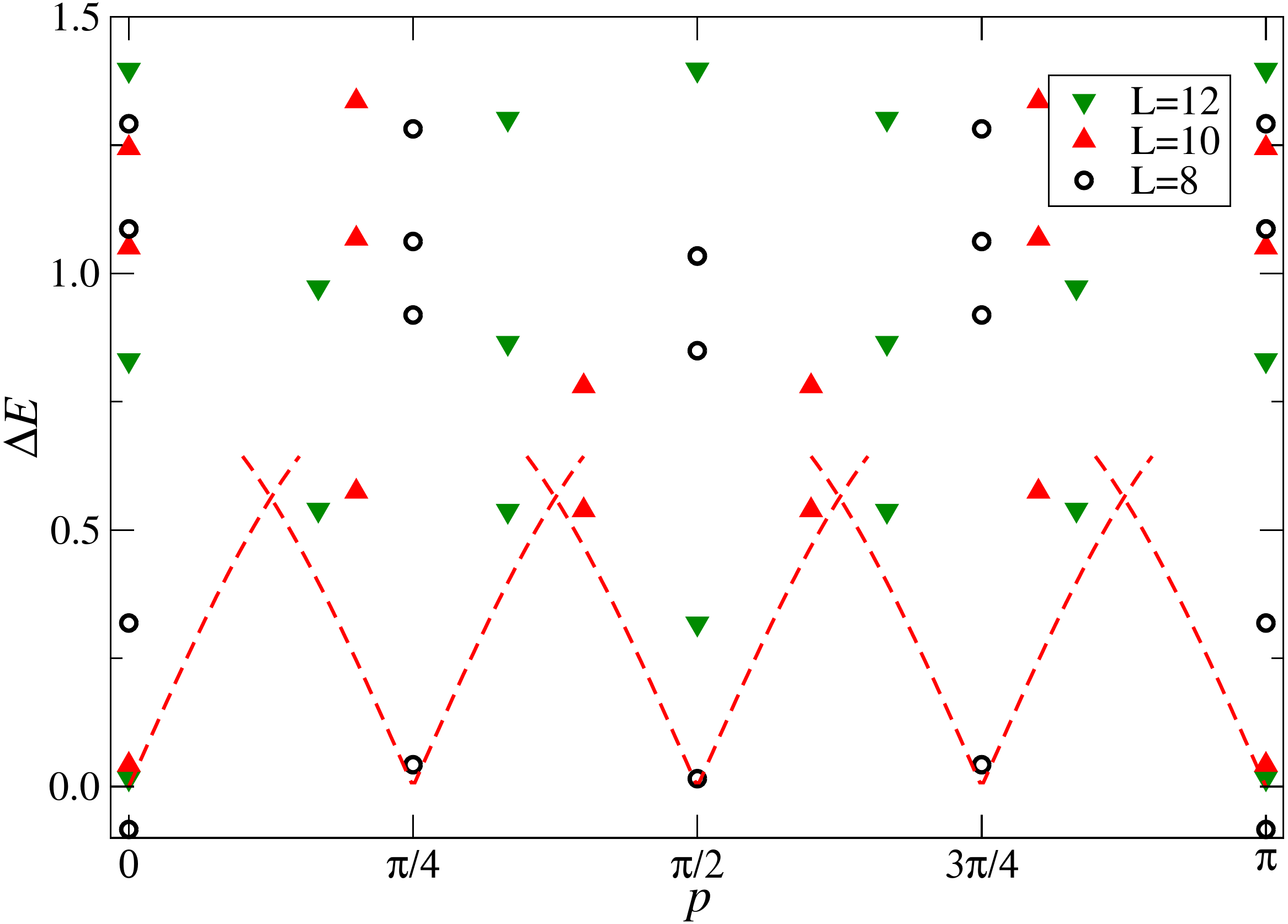}
\caption{\label{fig:specBMW2}Finite size spectrum of the BMW model at
  $\theta=\eta+\pi$: energies $\Delta E=E(\L)-E_{\theta=\eta+\pi}^{(0)}$ are
  shown in units of $2\pi v_F/\L$ for $\L=8$, $10$, $12$.  The dashed line
  indicates the dispersion of low energy excitations in the thermodynamic
  limit (scaled for $\L=10$),  see also Ref.~\onlinecite{Finch.etal14}.}
\end{figure}

For $\theta=\eta+\pi$ the ground state in the thermodynamic limit consists
of only $+$- and $-$-strings appearing in a ratio of five to three. For finite
size systems we find that the ground state is only realized when $\L$ is a
multiple of 8.  As we have two different types of string configurations
appearing we have two root densities defined by coupled integral equations
\begin{equation}
\nonumber
\begin{aligned}
  \rho_{+}(u) & = - A'(u;\frac{1}{20}) + \int_{-\infty}^{\infty}\mathrm{d}v\,
  \rho_{+}(v) A'(u-v;\frac{2}{5}) + \int_{-\infty}^{\infty}\mathrm{d}v\,
  \rho_{-}(v)
  A'(u-v;\frac{9}{10}) \,,\\
  \rho_{-}(u) & = - A'(u;\frac{11}{20}) + \int_{-\infty}^{\infty}\mathrm{d}v\,
  \rho_{+}(v) A'(u-v;\frac{9}{10}) +\int_{-\infty}^{\infty}
  \mathrm{d}v\,\rho_{-}(v) A'(u-v;\frac{2}{5})\,.
\end{aligned}
\end{equation}
Again, these integral equations are straightforward to solve by Fourier
transform giving
\begin{equation*}
  \rho_{\pm}(u)  = \frac{5}{2\pi} \left( \frac{\pm1 +
      \sqrt{2}\cos(\frac{3\pi}{8})\cosh(\frac{5}{2}u) +
      \sqrt{2}\cos(\frac{\pi}{8})\cosh(\frac{15}{2}u)}{\cosh(10u)} \right) \,.
\end{equation*}
As was the case with the previous critical integrable point,
the dressed energies of $\pm$-strings are scalar multiples of the
corresponding root densities,
\begin{equation*}
  \epsilon_{\pm}(u)  =
  \frac{\pi\cos(\frac{\pi}{4}+\eta)}{8\cos(\frac{\pi}{10})} \rho_{\pm}(u) \,.
\end{equation*}
Again the system has massless excitations with linear dispersion near its
Fermi points, which in this case are at multiples of $\frac{\pi}{4}$, see
Fig.~\ref{fig:specBMW2}. 
The ground state energy density and Fermi velocity are
\begin{equation*}
\begin{aligned}
  \frac{E^{(0)}_{\theta=\eta+\pi}}{\L} & =  \frac{1}{2} \cos\eta\left\{1 +
    \frac{1}{4i\cos(\frac{\pi}{10})} \int_{-\infty}^{\infty}\mathrm{d}u\,
    \left[\rho_{+}(u)
      \frac{\cosh(u+\frac{i\pi}{20})}{\sinh(u+\frac{i\pi}{20})} +
      \rho_{-}(u)\frac{\cosh(u+\frac{11i\pi}{20})}{
                       \sinh(u+\frac{11i\pi}{20})}\right]
  \right\} 
  \,\\
  &\approx \,-0.7578846235\,, \\
  v_{F} & =  \frac{5}{128\sqrt{\frac{11}{8}+\frac{\sqrt{5}}{8}}
    \left(\cos(\frac{\pi}{40})\right)^{4}
    -128\sqrt{\frac{11}{8}+\frac{\sqrt{5}}{8}} 
    \left(\cos(\frac{\pi}{40})\right)^{2}
    +16\sqrt{\frac{11}{8}+\frac{\sqrt{5}}{8}} 
  } \,\\ 
  &\approx \, 0.2554519349 \,.
\end{aligned}
\end{equation*}

\section{Finite size spectra and conformal field theory}
The models at $\theta=\eta,\pi+\eta$ derived from the BMW transfer matrices
constructed above are critical and therefore expected to be described by
conformal field theories (CFTs).  As a consequence of conformal invariance in
the continuum limit the scaling behavior of the ground state energy is
predicted to be \cite{BlCN86,Affl86}
\begin{equation}
  \label{cft0}
  E^{(0)}(\L) = \epsilon_{\infty}\L - \frac{\pi v_F}{6\L}\,c +
  o(\L^{-1}), 
\end{equation}
where $c$ is the central charge of the underlying Virasoro algebra and $v_{F}$
is the Fermi velocity.  The primary fields present in the critical model
determine the finite size energies and momenta of the excited states
($P^{(0)}$ being the momentum at one of the Fermi points)
\begin{equation}
\label{cft}
  E(\L) - E^{(0)}(\L) = \frac{2\pi v_F}{\L} \left(X+n+\bar{n}\right)\,,\quad
  P(\L) - P^{(0)} = \frac{2\pi}{\L} \left(s+n-\bar{n}\right) + \mathrm{const.}\,.
\end{equation}
This allows us to determine the scaling dimensions $X=h+\bar{h}$ and conformal
spins $s = h-\bar{h}$ of the primary fields present in the CFT from the
spectrum of the lattice model ($n$, $\bar{n}$ are non-negative integers).
Solving the Bethe equations numerically for root configurations corresponding
to a particular excitation for various lattice sizes we obtain a sequence of
finite size estimations for the scaling dimensions
\begin{equation}
  \label{cftnum}
  X^{\mathrm{num}}_\theta (\mathcal{L})=\frac{\mathcal{L}}{2\pi v_F}
  \left(E(\mathcal{L }) - E^{(0)}(\mathcal{L })\right)\,.
\end{equation}
This sequence is then extrapolated to obtain a numerical approximation
$X_\theta^{\mathrm{ext}}$ to the scaling dimension, which is subsequently
identified with a pair $(h,\bar{h})$ of conformal weights.  Similarly, the
central charge can be computed by finite size scaling analysis of the ground
state energy based on (\ref{cft0}).

\subsection{The BMW model at $\theta=\eta$}
Solving the Bethe equations for the ground state configuration of $4$-strings
in systems with even $\L$ we find  the scaling of the ground state energy
\begin{equation*}
  E^{(0)}(\L) = E^{(0)}_{\theta=\eta} - 0.6114471233\,\L^{-1} + o(\L^{-1})\,.
\end{equation*}
This implies that the central charge is $c=1.14285(1)$, predicted to be
actually $\frac{8}{7}$ in agreement with the ferromagnetic $Z_{5}$
Fateev-Zamolodchikov model.  
This central charge appears in several rational CFTs from the minimal series
of Casimir-type $\mathcal{W}$-algebras, see Appendix~\ref{appCFT}.
Restricting ourselves to unitary ones these are
\begin{align*}
&	A_4: & &\mathcal{W}(2,3,4,5) & & c_{A_4}(6,7)\,,\\
&	B_2: & &\mathcal{W}(2,4)     & & c_{B_2}(5,7)\,,\\
&	C_2: & &\mathcal{W}(2,4)     & & c_{C_2}(5,7)\,.
\end{align*}
Note that $A_4={SL}(5)$ with real form ${SU}(5)$ contains
$B_2={SO}(5)$ and $C_2={Sp}(4)$ as a subalgebra.  
Since $B_2\cong C_2$ we only consider the former.\footnote{%
  We also note that the microscopic description of the model in terms of
  ${SO}(5)$ anyons does not go together with the symplectic structure of
  $C_2$.}

To identify the low energy effective theory of this anyon model we extend our
analysis to the scaling dimensions obtained from the finite size spectrum of
excited states, see Table~\ref{tabXeta}.
\begin{table}[t]
\begin{tabularx}{\textwidth}{CCCCCCC} \hline \hline
	Sector(s) & $X^{\mathrm{ext}}$  & $X^{\mathrm{CFT}}$ & $s$ & $(h,\bar{h})$ & $N_{4s}$ & $N_{0}$ \tabularnewline \hline \hline
	$\psi_{1}\oplus\psi_{6}$ & 0.00000(1) & 0 & 0.00 & $(0,0)$ & $\frac{\L}{2}$ & 0 \tabularnewline
	$\psi_{3}\oplus\psi_{4}$ & 0.07144(1) & $\frac{1}{14}$ & 0.00 & $(\frac{1}{28},\frac{1}{28})$ & $\frac{\L}{2}$ & 0 \tabularnewline
	$\psi_{2}$ & 0.117(1) & $\frac{4}{35}$ & 0.00 & $(\frac{2}{35},\frac{2}{35})$ & $\frac{\L}{2}-1$ & 2 \tabularnewline
	$\psi_{5}$ & 0.17146(1) & $\frac{6}{35}$ & 0.00 & $(\frac{3}{35},\frac{3}{35})$ & $\frac{\L}{2}-1$ & 0 \tabularnewline
	$\psi_{3}\oplus\psi_{4}$ & 0.2147(2) & $\frac{3}{14}$ & 0.00  & $(\frac{3}{28},\frac{3}{28})$ & $\frac{\L}{2}-1$ & 4 \tabularnewline
	$\psi_{3}\oplus\psi_{4}$ & 0.50001(1)  & $\frac{1}{2}$& 0.00 & $(\frac{1}{4},\frac{1}{4})$ & $\frac{\L}{2}-1$ & 4 \tabularnewline
	$\psi_{1}\oplus\psi_{6}$ & 0.579(2)  & $\frac{4}{7}$ & 0.00 & $(\frac{2}{7},\frac{2}{7})$ & $\frac{\L}{2}-2$ & 8 \tabularnewline
	$\psi_{3}\oplus\psi_{4}$ & 1.068(2)  & $1\frac{1}{14}$ & 1.00 & - & $\frac{\L}{2}-1$ & 4 \tabularnewline
	$\psi_{3}\oplus\psi_{4}$ & 1.07148(1) & $1\frac{1}{14}$ & 1.00  & - & $\frac{\L}{2}-1$ & 4 \tabularnewline
	$\psi_{3}\oplus\psi_{4}$ & 1.07211(1) & $\frac{15}{14}$ & 0.00  & $(\frac{15}{28},\frac{15}{28})$ & $\frac{\L}{2}-2$ & 8 \tabularnewline
	\hline \hline
\end{tabularx}
\caption{Scaling dimensions obtained from the finite size spectrum of the
  BMW model $\mathcal{H}_{\theta=\eta}$ with even $\L$ corresponding 
  to a $c=\frac{8}{7}$ model. The scaling dimensions, $X^{\mathrm{ext}}$, and
  the corresponding error estimates are obtained by extrapolation of the
  numerical finite size energies (\ref{cftnum}) for $\L$ up to 
  300.  $X^{\mathrm{CFT}}$ is the scaling dimension from the conjectured
  CFT, $s$ is the conformal spin of the level as obtained from the finite size
  data.  For primary fields we also display the corresponding pair of conformal
  weights ($h,\bar{h})$.  $N_{4s}$ and $N_{0}$ denote the number of 4-strings
  and other Bethe roots, respectively, appearing in the Bethe root
  configuration. 
  \label{tabXeta}}    
\end{table}
%
Clearly our data rule out the $Z_5$ parafermions since the conformal weights
observed in the sector $\psi_3\oplus\psi_4$, e.g.\ $\frac{1}{28}$,
$\frac{3}{28}$ and $\frac{15}{28}$, are not present in the spectrum
(\ref{eq:h-paraferm}) for $k=5$.
On the other hand the finite size data are compatible with the
$\mathcal{W}B_2(5,7)$ minimal model (\ref{cft_specWB2}): the conformal weights
up to $h=2/7$ are complete, the first one missing is $h=17/35$.  This is not
surprising as the numerical analysis gets more demanding for higher values of
the conformal weight.  Therefore, we conclude that if the continuum limit of
this BMW model is described by a rational conformal field theory, it
presumably is the $(5,7)$ minimal model of the $\mathcal{W}{B}_2$ series.

\subsection{The BMW model at $\theta=\pi+\eta$}
Using the finite size data for the ground state we find the scaling behavior,
\begin{equation*}
  E^{(0)}(\L) = E^{(0)}_{\theta=\eta+\pi} - 0.1337534147\, \L^{-1} + o(\L^{-1})
\end{equation*}
from which we can calculate the central charge $c=0.99999(1)$ in agreement
with $c=1$ for the anti-ferromagnetic $Z_{5}$ FZ model.  
Again there are several rational CFTs in the minimal series of Casimir-type
$\mathcal{W}$-algebras with the same central charge:
\begin{align*}
&	A_3: & &\mathcal{W}(2,3,4) & & 
	c_{A_3}(5,6)\,,\\
&	B_n: & &\mathcal{W}(2,4,\ldots,2n) & & 
	c_{B_n}(n,n+1)\ \ \textrm{and}\ \ 
        c_{B_n}(2n-1,2n)\,,\\
& 	     & &\mathcal{W}(2,4,\ldots,2n,{\textstyle\frac12}(2n+1)) & &
	c_{\mathcal{B}_{0,n}}(2n,2n+1)\,,\\
&	C_n: & &\mathcal{W}(2,4,\ldots,2n) & & 
	c_{C_n}(n,2n-1)\ \ \textrm{and}\ \ 
        c_{C_n}(n+1,2n)\,,\\
&	D_n: & &\mathcal{W}(2,4,\ldots,2n-2,n) & & 
	c_{D_n}(2n-1,2n)\,.
\end{align*}
The unitary models ones among these are $\mathcal{W}A_3(5,6)$, the $C_3$-model
$\mathcal{W}C_3(3,5)$, and the generic series of minimal models for
$\mathcal{B}_{0,\ell}=OSp(1,2\ell)$ and 
$D_\ell={SO}({2\ell})$, 
i.e.\ $\mathcal{WB}_{0,\ell}(2\ell,2\ell+1)$ and
$\mathcal{W}D_\ell(2\ell-1,2\ell)$. 
Keeping in mind the 
${SO}(5)$ 
anyon structure of the underlying
lattice model, the $C_3$-model with its symplectic 
${Sp}(6)$ symmetry appears to be an unlikely candidate for the low energy
effective theory.

To identify which $c=1$ CFT is realized we can look at the finite size
properties of low-lying excitations presented in 
Tables~\ref{tabScalingDimAF0}--\ref{tabScalingDimAF6}.
\begin{table}[t] 
\begin{tabularx}{\textwidth}{CCCCCCCCC} \hline \hline
 Sector(s) & $P^{(0)}$ & $X^{\mathrm{ext}}$ & $X^{\mathrm{CFT}}$ & 
 $s$ & $(h,\bar{h})$ & $N_{+}$ & $N_{-}$ & $N_{0}$ \tabularnewline \hline \hline
 $\psi_{1}\oplus\psi_{6}$ & $0$ & 0.00000(1) & 0 & 
 0.00 & $(0,0)$ & $\frac{5\L}{4}$ & $\frac{3\L}{4}$ & 0 
 \tabularnewline
 $\psi_{2}$ & $\frac{\pi}{2}$ & 0.10000(1) & $\frac{1}{10}$ & 
 0.00 & $(\frac{1}{20},\frac{1}{20})$ & $\frac{5\L}{4}-1$ & $\frac{3\L}{4}-1$ & 0 
 \tabularnewline
 $\psi_{3}\oplus\psi_{4}$ & $\frac{\pi}{4}$,$\frac{3\pi}{4}$ & 0.12500(1) & $\frac{1}{8}$ & 
 0.00 & $(\frac{1}{16},\frac{1}{16})$ & $\frac{5\L}{4}$ & $\frac{3\L}{4}$ & 0 
 \tabularnewline
 $\psi_{5}$ & $0$ & 0.40000(1) & $\frac{2}{5}$ & 
 0.00 & $(\frac{1}{5},\frac{1}{5})$ & $\frac{5\L}{4}-2$ & $\frac{3\L}{4}-2$ & 0 
 \tabularnewline
 $\psi_{5}$ & $\frac{\pi}{2}$ & 0.90000(1) & $\frac{9}{10}$ & 
 0.00 & $(\frac{9}{20},\frac{9}{20})$ & $\frac{5\L}{4}-3$ & $\frac{3\L}{4}-1$ & 0
 \tabularnewline
 $\psi_{2}$ & $\frac{\pi}{2}$ & 1.10000(1) & $1\frac{1}{10}$ & 
 1.00 & - & $\frac{5\L}{4}-2$ & $\frac{3\L}{4}-2$ & 2
 \tabularnewline
 $\psi_{2}$ & $\frac{\pi}{2}$ & 1.10000(1) & $1\frac{1}{10}$ & 
 $-1.00$ & - & 
 \tabularnewline
 $\psi_{3}\oplus\psi_{4}$ & ${\frac{\pi}{4}}$ & 1.12471(8) & $1\frac{1}{8}$ & 
 1.00 &  - & $\frac{5\L}{4}-2$ & $\frac{3\L}{4}-2$ & 4
 \tabularnewline
 $\psi_{3}\oplus\psi_{4}$ & ${\frac{3\pi}{4}}$ & 1.12471(8) & $1\frac{1}{8}$ & 
 $-1.00$ &  - 
 \tabularnewline
	\hline \hline
\end{tabularx}
\caption{Scaling dimensions appearing in the spectrum of
  $\mathcal{H}_{\theta=\eta+\pi}$ for $\L \mod 8=0$. $P^{(0)}$ lists the Fermi
  points near which the state is observed in the finite size
  spectrum. $X^{\mathrm{ext}}$ is prediction for the scaling dimension
  obtained by extrapolation of the finite size data.  $X^{\mathrm{CFT}}$ are
  the conjectured values for the scaling dimension and $s$ the conformal
  spins, as obtained from the finite size data.  For primary fields we also
  display the corresponding pair of conformal weights $(h,\bar{h})$.  The
  quantities $N_{+}$, $N_{-}$ and $N_{0}$ correspond respectively to the
  number of $+$-strings, $-$-strings and other Bethe roots in the Bethe root
  configuration.  Note that among the states with scaling dimension
  $X=1\frac18$ we have only identified the $(1,0)$ and $(0,1)$ descendents of
  the primaries $(h,\bar{h})=(\frac1{16},\frac1{16})$ at the Fermi points
  $\frac{\pi}{4}$ and $\frac{3\pi}{4}$, respectively.  For the other two
  descendents as well as the primary fields with $(\frac9{16},\frac9{16})$
  (with momenta $P^{(0)}=\frac{\pi}4$, $\frac{3\pi}{4}$) we have not
  identified the corresponding root configuration.
  \label{tabScalingDimAF0}} 
\end{table}
%
\begin{table}[t] 
\begin{tabularx}{\textwidth}{CCCCCCCCC} \hline \hline
  Sector(s) & $P^{(0)}$ & $X^{\mathrm{ext}}$ & $X^{\mathrm{CFT}}$ & $s$ &
  $(h,\bar{h})$ & $N_{+}$ & $N_{-}$ & $N_{0}$ 
\tabularnewline \hline \hline
  $\psi_{3}\oplus\psi_{4}$ & 0 & 0.12500(1) & $\frac{1}{8}$ & 0.00 &
  $(\frac{1}{16},\frac{1}{16})$ & $\frac{5(\L-2)}{4}+2$ & $\frac{3(\L-2)}{4}2$
  & 0 
\tabularnewline 
  $\psi_{3}\oplus\psi_{4}$ & $\frac{\pi}2$ & 0.62500(1) & $\frac{5}{8}$ &
  $0.50$ & $(\frac{9}{16},\frac{1}{16})$ & $\frac{5(\L-2)}{4}+2$ &
  $\frac{3(\L-2)}{4}+0$ & 2 
\tabularnewline
  $\psi_{3}\oplus\psi_{4}$ & $\frac{\pi}2$ & 0.62500(1) & $\frac{5}{8}$ &
  $-0.50$ & $(\frac{1}{16},\frac{9}{16})$ 
\tabularnewline 
  $\psi_{5}$ & $\frac{3\pi}4$&  0.65000(1) & $\frac{13}{20}$ & $0.25$ &
  $(\frac{9}{20},\frac{1}{5})$ & $\frac{5(\L-2)}{4}$ & $\frac{3(\L-2)}{4}$ & 0
\tabularnewline
  $\psi_{5}$ & $\frac{\pi}4$ &  0.65000(1) & $\frac{13}{20}$ & $-0.25$ &
  $(\frac{1}{5},\frac{9}{20})$ 
\tabularnewline
  $\psi_{2}$ & $\frac{\pi}4$ & 0.85000(1) & $\frac{17}{20}$ & $0.75$ &
  $(\frac{4}{5},\frac{1}{20})$ & $\frac{5(\L-2)}{4}$ & $\frac{3(\L-2)}{4}$ & 2
  \tabularnewline
  $\psi_{2}$ &$\frac{3\pi}4$ & 0.85000(1) & $\frac{17}{20}$ & $-0.75$ &
  $(\frac{1}{20},\frac{4}{5})$ 
\tabularnewline 
  $\psi_{3}\oplus\psi_{4}$ & 0 & 1.12500(1) & $\frac{9}{8}$ & 0.00 &
  $(\frac{9}{16},\frac{9}{16})$ & $\frac{5(\L-2)}{4}$ & $\frac{3(\L-2)}{4}$ &
  4 
\tabularnewline 
  $\psi_{3}\oplus\psi_{4}$ & 0 & 1.12505(6) & $1\frac{1}{8}$ & 1.00 & - &
  $\frac{5(\L-2)}{4}$ & $\frac{3(\L-2)}{4}$ & 4 
\tabularnewline 
  $\psi_{3}\oplus\psi_{4}$ & 0 & 1.12505(6) & $1\frac{1}{8}$ & $-1.00$ & - &
\tabularnewline 
	\hline \hline
\end{tabularx}
\caption{Same as Table~\ref{tabScalingDimAF0} but for  $\L \mod
  8=2$.\label{tabScalingDimAF2}}  
\end{table}
%
\begin{table}[t] 
\begin{tabularx}{\textwidth}{CCCCCCCCC} \hline \hline
  Sector(s) & $P^{(0)}$ & $X^{\mathrm{ext}}$ & $X^{\mathrm{CFT}}$ & $s$ &
  $(h,\bar{h})$ & $N_{+}$ & $N_{-}$ & $N_{0}$ 
\tabularnewline \hline \hline
  $\psi_{2}$ & 0 & 0.10000(1) & $\frac{1}{10}$ & 
 	0.00 & $(\frac{1}{20},\frac{1}{20})$ & $\frac{5\L}{4}-1$ & $\frac{3\L}{4}-1$ & 0 
  \tabularnewline
  $\psi_{5}$ & $\frac{\pi}{2}$ & 0.40000(1) & $\frac{2}{5}$ & 
  0.00 & $(\frac{1}{5},\frac{1}{5})$ & $\frac{5\L}{4}-2$ & $\frac{3\L}{4}-2$ & 0 
  \tabularnewline
  $\psi_{3}\oplus\psi_{4}$ & $\frac{3\pi}{4}$ & 0.62500(1) & $\frac{5}{8}$ & 
  $0.50$ & $(\frac{9}{16},\frac{1}{16})$ & $\frac{5L}{4}-1$ & $\frac{3\L}{4}-1$ & 2 
  \tabularnewline
  $\psi_{3}\oplus\psi_{4}$ & $\frac{\pi}{4}$ & 0.62500(1) & $\frac{5}{8}$ & 
  $-0.50$ & $(\frac{1}{16},\frac{9}{16})$ 
  \tabularnewline
  $\psi_{3}\oplus\psi_{4}$ & $\frac{\pi}{4}$ & 0.62500(1) & $\frac{5}{8}$ & 
  $0.50$ & $(\frac{9}{16},\frac{1}{16})$ & $\frac{5L}{4}-1$ & $\frac{3\L}{4}-1$ & 2 
  \tabularnewline
  $\psi_{3}\oplus\psi_{4}$ & $\frac{3\pi}{4}$ & 0.62500(1) & $\frac{5}{8}$ & 
  $-0.50$ & $(\frac{1}{16},\frac{9}{16})$ 
  \tabularnewline
  $\psi_{5}$ & 0 & 0.90000(1) & $\frac{9}{10}$ & 
  0.00 & $(\frac{9}{20},\frac{9}{20})$ & $\frac{5\L}{4}-3$ & $\frac{3\L}{4}-1$ & 0
  \tabularnewline
  $\psi_{1}\oplus\psi_{6}$ & $\frac{\pi}{2}$ & 1.00000(1) & 1 & 
  $1.00$ & $(1,0)$ & $\frac{5\L}{4}-1$ & $\frac{3\L}{4}-1$ & 2
  \tabularnewline
  $\psi_{1}\oplus\psi_{6}$ & $\frac{\pi}{2}$ & 1.00000(1) & 1 &
  $-1.00$ & $(0,1)$ 
  \tabularnewline
  $\psi_{2}$ & 0 & 1.10000(1) & $1\frac{1}{10}$ & 
  $1.00$ & - & $\frac{5\L}{4}-2$ & $\frac{3\L}{4}-2$ & 2
  \tabularnewline
  $\psi_{2}$ & 0 & 1.10000(1) & $1\frac{1}{10}$ & 
  $-1.00$ & - 
  \tabularnewline
	\hline \hline
\end{tabularx}
\caption{Same as Table~\ref{tabScalingDimAF0} but for  $\L \mod
  8=4$.\label{tabScalingDimAF4}}  
\end{table}
%
\begin{table}[t] 
\begin{tabularx}{\textwidth}{CCCCCCCCC} \hline \hline
  Sector(s) & $P^{(0)}$ & $X^{\mathrm{ext}}$ & $X^{\mathrm{CFT}}$ & $s$ &
  $(h,\bar{h})$ & $N_{+}$ & $N_{-}$ & $N_{0}$ 
\tabularnewline \hline \hline
  $\psi_{3}\oplus\psi_{4}$ & 0 & 0.12500(1) & $\frac{1}{8}$ & 
  0.00 & $(\frac{1}{16},\frac{1}{16})$ & $\frac{5(\L-2)}{4}+3$ & $\frac{3(\L-2)}{4}+1$ & 0
  \tabularnewline
  $\psi_{3}\oplus\psi_{4}$ & $\frac{\pi}{2}$ & 0.62500(1) & $\frac{5}{8}$ & 
  $0.50$ & $(\frac{9}{16},\frac{1}{16})$ & $\frac{5(\L-2)}{4}+1$ & $\frac{3(\L-2)}{4}+1$ & 2 
  \tabularnewline
  $\psi_{3}\oplus\psi_{4}$ & $\frac{\pi}{2}$ & 0.62500(1) & $\frac{5}{8}$ & 
  $-0.50$ & $(\frac{1}{16},\frac{9}{16})$ 
  \tabularnewline
  $\psi_{5}$ & $\frac{\pi}{4}$ & 0.65000(1) & $\frac{13}{20}$ & 
  $0.25$ & $(\frac{9}{20},\frac{1}{5})$ & $\frac{5(\L-2)}{4}$ & $\frac{3(\L-2)}{4}$ & 0 
  \tabularnewline
  $\psi_{5}$ & $\frac{3\pi}{4}$ & 0.65000(1) & $\frac{13}{20}$ & 
  $-0.25$ & $(\frac{1}{5},\frac{9}{20})$ 
  \tabularnewline
  $\psi_{2}$ & $\frac{3\pi}{4}$ & 0.85000(1) & $\frac{17}{20}$ & 
  $0.75$ & $(\frac{4}{5},\frac{1}{20})$ & $\frac{5(\L-2)}{4}$ & $\frac{3(\L-2)}{4}$ & 2 
  \tabularnewline
  $\psi_{2}$ & $\frac{\pi}{4}$ & 0.85000(1) & $\frac{17}{20}$ & 
  $-0.75$ & $(\frac{1}{20},\frac{4}{5})$ 
  \tabularnewline
  $\psi_{3}\oplus\psi_{4}$ & 0 & 1.1248(1) & $1\frac{1}{8}$ & 
  1.00 & - & $\frac{5(\L-2)}{4}+1$ & $\frac{3(\L-2)}{4}-1$ & 4 
  \tabularnewline
  $\psi_{3}\oplus\psi_{4}$ & 0 & 1.1248(1) & $1\frac{1}{8}$ & 
  $-1.00$ & - 
  \tabularnewline
	\hline \hline
\end{tabularx}
\caption{Same as Table~\ref{tabScalingDimAF0} but for  $\L \mod
  8=6$.\label{tabScalingDimAF6}} 
\end{table}
%
Clearly, the observed spectrum contains conformal weights not present in the
${Z}_4$ parafermion theory, thereby excluding the $A_3$-model.
Furthermore, we can rule out the $\mathcal{W}{C}_3(4,6)$ model: even without
explicit computation of the spectrum we just know from the general formula
(\ref{cft_WGspec}) for the conformal weights, that their denominators are of
the form $2pq$.  Therefore, to yield the denominators $5$ and $20$ observed
numerically in the anyon model $2pq$ has to be divisible by $5$.

Among the remaining rational CFTs the smallest ones respecting the five-fold
discrete symmetry of the BMW model are the $\mathcal{WB}_{0,2}(4,5)$ model and
the $\mathcal{W}{D}_5(9,10)$ model.  Comparing their spectra
(\ref{cft_specWBf2}) and (\ref{cft_specWD5}) with the numerical data we find
that the former does not accommodate all observed weights while all weights of
the $(9,10)$ model of the $\mathcal{W}{D}_5$ minimal series
\begin{equation}
\label{WD5_weights}
  \left\{0,1,\frac54,\frac{1}{20},\frac45,\frac{9}{20},\frac15,\frac{1}{16},
  \frac{9}{16}\right\}\,
\end{equation}
with $h\le1$ have been identified.

A peculiar feature of the low energy spectrum observed numerically is the
identification of levels corresponding to operators with non-integer spin $s$,
see Tables~\ref{tabScalingDimAF2}--\ref{tabScalingDimAF6}.  
For the sectors $\psi_2$ and $\psi_5$, where the total momentum is not
determined uniquely in terms of the Bethe roots, see Eq.~(\ref{BMW_spec}) one
can use this fact to determine the Fermi point $P^{(0)}$ at which these levels
occur: for example, the levels $(h,\bar{h}) \in \{(\frac{1}{5},\frac{9}{20}),
(\frac{4}{5},\frac{1}{20})\}$ observed for $\L\mod4=2$ can only appear at
$P^{(0)}=\frac{\pi}{4}$.  Changing the length of the system to $\L\mod4=6$
they are observed at $P^{(0)}=\frac{3\pi}{4}$, indicating that they have to
appear with a multiplicity of two in the full partition function.

To provide additional evidence for our proposal that the anyon model is a
lattice regularization of a $\mathcal{W}{D}_5(9,10)$ rational CFT the modular
invariant partition function of the model has to be expressed in terms of the
Virasoro characters of the relevant representations of the
$\mathcal{W}$-algebra.
All representations corresponding to the conformal weights (\ref{WD5_weights})
have multiplicity one, except $h=5/4$, $1/16$ and $9/16$, which have trivial
multiplicities two.  This means that our theory should possess nine
\emph{independent} characters.

From the known embedding structure of Virasoro modules, we know that for a CFT
with central charge $c=1$ there is just one null state at level one in the
Virasoro vacuum representation, $h=0$.  Furthermore, the symmetry generators
in the chiral symmetry algebra $\mathcal{W}{D}_5$ have dimension $d>3$.
Therefore, additional states beyond pure Virasoro descendents can only appear
at level four or above.  This also implies that any additional null states
cannot occur below level eight.  The reason is that the Virasoro module
generated by $W^{(4)}_{-4}|0\rangle$ has just one Virasoro null state, namely
at level five, i.e.\ at level nine with respect to $|0\rangle$. Furthermore,
the lowest possible $\mathcal{W}$-algebra null state could be a linear
combination containing $W^{(4)}_{-4}W^{(4)}_{-4}|0\rangle$ at level eight with
respect to $|0\rangle$.  Analogous statements hold for the Virasoro submodule
generated by $W^{(5)}_{-5}|0\rangle$.  Therefore, we obtain
\begin{equation*}
  \chi_{h=0}(q) = \frac{1}{\eta(q)}(1-q+q^4+q^5\ldots)\,,
\end{equation*}
where $\eta(q)=q^{1/24}\prod_{n=1}^\infty (1-q^n)$ is the Dedekind
eta-function.

This insight, together with similar considerations for the other
representations, is sufficient to set up a differential equation from which
the characters of the rational CFT can be computed to arbitrary powers without
additional prior knowledge of the detailed structure of the representations:
the modular differential equation encodes the fact that the characters of the
nine inequivalent irreps of the CFT form a nine-dimensional representation of
the modular group in terms of modular functions with asymptotic behavior
fixed by the numbers $h_i-\frac{c}{24}$ \cite{MaMS88}.  Assuming that all
characters can be written as modular forms of weight $1/2$ divided by the
Dedekind eta-function we find
\begin{equation}
\label{WD5_modforms}
\begin{aligned}
\chi_0(q)    &=
	\frac{1}{2\eta(q)}\big(\Theta_{0,5}(q)+\Theta_{0,4}(q)-\Theta_{4,4}(q)\big)
	\,,\\
\chi_1(q)    &=
	\frac{1}{2\eta(q)}\big(\Theta_{0,5}(q)-\Theta_{0,4}(q)+\Theta_{4,4}(q)\big)
	\,,\\
\chi_{5/4}(q)   &=
	\frac{1}{\eta(q)}\,\Theta_{5,5}(q)\,,\\
\chi_{1/20}(q) &=
	\frac{1}{\eta(q)}\,\Theta_{1,5}(q)\,,\qquad
\chi_{4/5}(q)   =
	\frac{1}{\eta(q)}\,\Theta_{4,5}(q)\,,\\
\chi_{9/20}(q)  &=
	\frac{1}{\eta(q)}\,\Theta_{3,5}(q)\,,\qquad
\chi_{1/5}(q)   =
	\frac{1}{\eta(q)}\,\Theta_{2,5}(q)\,,\\
\chi_{1/16}(q)  &=
	\frac{2}{\eta(q)}\,\Theta_{1,4}(q)\,,\qquad
\chi_{9/16}(q)  =
	\frac{2}{\eta(q)}\,\Theta_{3,4}(q)\,.
\end{aligned}
\end{equation}
Here we have used the Jacobi-Riemann Theta-functions
\begin{equation*}
  \Theta_{\lambda,k}(q) = \sum_{n\in\mathbb{Z}}
  q^{\frac{(2kn+\lambda)^2}{4k}}\,.
\end{equation*}
Note that we have introduced an explicit factor of two in the definition of
$\chi_{1/16}(q)$ and $\chi_{9/16}(q)$.  In fact, these characters as well as
$\chi_{5/4}(q)=2\,q^{-1/24+5/4} \left(1+q+2q^2+3q^3+5q^4+\ldots\right)$
should be read as $\chi_h(q)=\chi_h^+(q)+\chi_h^-(q)$ being the sum of two
identical characters for the two equivalent representations appearing in the
Kac-table for $h\in\{\frac54,\frac{1}{16}, \frac{9}{16}\}$, see
(\ref{cft_specWD5}).  For the present discussion of the conformal spectrum
this is sufficient.  In Appendix~\ref{app_WD5fus}, where we compute the fusion
rules for the $\mathcal{W}D_5((9,10)$ rational CFT, these representations have
to be disentangled in a consistent way.

Given (\ref{WD5_modforms}) the modular $\mathcal{S}$-matrix is easily
computed, for the result see Appendix~\ref{app_WD5fus}.  Modular invariance of
the partition function of the model, expressed as $Z=\sum_{ij} \chi_i
M_{ij}\bar{\chi}_j$ in terms of the characters (\ref{WD5_modforms}),
translates into the condition $M = \mathcal{S}^\dagger M \mathcal{S}$ for the
matrix $M$ of multiplicities.\footnote{%
  Strictly speaking, we have invariance under a subgroup of the modular group
  only: while the partition function is invariant under the action of
  $\mathcal{S}:\tau\to-1/\tau$, the presence of fields with quarter spin
  implies that this is true only for quadruples of the modular transformation
  $\mathcal{T}:\tau\to\tau+1$.}
With the $\mathcal{S}$-matrix (\ref{WD5_smat}) at hand this condition can be
solved and we find
\begin{equation}
\label{WD5_Mmat}
\begin{aligned}
  &M_{ij}(n,m,p,q) =\\ 
  &\left(\begin{array}{ccccccccc}
 n  & n-q &  \frac12 p  &        &        &      &      &       &       \\   
n-q &  n  &  \frac12 p  &        &        &      &      &       &       \\   
\frac12 p & \frac12 p &n-p-\frac12 q&        &        &      &      &       &       \\   
    &     &       &2(n-p)-q&    p   &     &   &       &       \\   
    &     &       &    p   &  2n-q  &     &   &       &       \\   
    &     &       &        &        & 2(n-p)-q&  p         &       \\   
    &     &       &        &   &    p   & 2n-q  &       &       \\   
    &     &       &        &        &      &      &  \frac12 m  &\frac12 (q-m)\\   
    &     &       &        &        &      &      &\frac12 (q-m)&  \frac12 m    
  \end{array}\right)\,.
\end{aligned}
\end{equation}
Rows and columns refer to representations in the order listed in Eq.~(\ref{WD5_weights}).
The four blocks in $M$ match with the decomposition of the spectrum of the
lattice model according to topological sectors $\psi_1\oplus\psi_6$, $\psi_2$,
$\psi_5$, and $\psi_3\oplus\psi_4$ of the lattice model, see
Tables~\ref{tabBMW_topo} and \ref{tabScalingDimAF0}--\ref{tabScalingDimAF6}.
In Eq.~(\ref{WD5_Mmat}) $n,m,p,q$ are non-negative integers such that $M$ has
no negative entries and full rank.
Half-odd integer entries in (\ref{WD5_Mmat}) are a consequence of our choice
for the characters of representations with multiplicity two: for example, from
the diagonal partition function,
$M_{ij}(1,1,0,1)=\mathrm{diag}(1,1,\frac12,1,1,1,1,\frac12,\frac12)$, one can
infer that only either the combination $\chi_{5/4}^+\bar\chi_{5/4}^+ +
\chi_{5/4}^-\bar\chi_{5/4}^-$ or the combination $\chi_{5/4}^+\bar\chi_{5/4}^-
+ \chi_{5/4}^-\bar\chi_{5/4}^+$ enters.  The same holds for the two
representations with $h=1/16$ and $h=9/16$.
Of course, linear combinations of arbitrary solutions $M_{ij}(n,m,p,q)$
with non-negative integer coefficients are also possible, but yield nothing 
new.  

From the finite size spectrum of the anyon model we know that
$M_{h\bar{h}}=\frac12$ for $h,\bar{h}\in\{\frac{1}{16},\frac{9}{16}\}$ which
implies $m=1$, $q=2$.  Similarly, as discussed above, the quarter spin fields
in the symmetry sectors $\psi_2$ and $\psi_5$ appear twice (with different
momentum) in the low energy spectrum, hence $p=2$.  As a result the smallest
off-diagonal partition function which incorporates all non-diagonal
combinations found in our model is $M_{ij}(4,1,2,2)$.  This leads to a
fourfold degenerate ground state in the model while it is unique in the
lattice model with periodic boundary conditions studied here.  We expect that
the other ground states (corresponding to the degenerate minima of the
dispersion at momenta $P^{(0)}=k\frac{\pi}{4}$, $k=1,2,3$) are realized when
considering more general (twisted) boundary conditions for the anyon chain.

Finally, we note that in the CFT describing the collective behavior of the
BMW model at $\theta=\pi+\eta$ the symmetry of the underlying anyon model
appears to be modified.  If we understand the
${SO}(5)$ symmetry of the anyons as a \emph{discrete} one (similar as the
cyclic subgroup ${Z}_5$ of
${SU}(2)$), however, ${SO}(10)$ could correspond to the dihedral subgroup
${D}_5$ of
${SU}(2)$.
This is appealing for two reasons: firstly, it would fit with the well known
$A$-$D$-$E$ classification of rational conformal field theories with $c=1$ as
conformal field theories with extended symmetries given by modding out
discrete symmetries of 
$\widehat{{SU}}(2)$ \cite{Ginsparg87}.  Secondly, a similar construction with
$D(D_3)$ anyons also has a $c=1$ point, where it coincides with the ${Z}_4$
parafermions \cite{FiFr12,FiFr13}. These, however, are
$\widehat{{SU}}(4)\cong\widehat{{SO}}(6)$, i.e.\ $\hat A_3\cong \hat D_3$.
\section{Discussion}
In this paper we have constructed several integrable one-dimensional models of
interacting anyons satisfying the fusion rules for $SO(5)_2$.  For particles
carrying topological charge $\psi_3$ we have constructed representations of
the Birman-Murakami-Wenzl (BMW) algebra (or its Temperley-Lieb (TL)
subalgebra) in terms of the local projection operators appearing in the anyon
model.  Based on these representations we found $R$-matrices solving the
Yang-Baxter equation from which commuting transfer matrices have been
obtained.
The spectrum of these models is parametrized in terms of solutions from Bethe
equations which have been derived using the Temperley-Lieb equivalence to the
six-vertex model and the fusion hierarchy of transfer matrices, respectively.
The topological charges characterizing the spectrum of the anyon chains have
been identified with the transfer matrices in the limit of infinite spectral
parameter.  This allows to classify the spectrum.

By solving the Bethe equations we have identified the ground states and low
energy excitations of these models.  The TL-models are equivalent to the
six-vertex model in its (anti-)ferromagnetic massive phases.  The other two
integrable models are critical at zero temperature.  From the finite size
spectrum we have been able to identify the conformal field theories describing
their scaling limit.  
Based on the Bethe ansatz solution we find that these systems are closely
related to the $Z_{5}$ Fateev-Zamolodchikov (FZ) or self-dual chiral Potts
model.  In particular, they have the same Virasoro central charges, i.e.\
$c=\frac{8}{7}$ and $c=1$, as the ferro- and anti-ferromagnetic $Z_5$ FZ
model.
Careful analysis of the excitation spectrum reveals, however, that the
boundary conditions imposed by the conserved topological charges of the anyon
model lead to rational conformal field theories with the same central charges
but invariant under extensions of the Virasoro algebra, i.e.\ $\mathcal{W}B_2$
and $\mathcal{W}D_5$, respectively.  The difference between, e.g., the
$\mathcal{W}B_2(5,7)$ model and $Z_5$ parafermions proposed previously for the
FZ model, shows up in the presence of additional conformal weights
$h\in\{\frac{1}{28},\frac{3}{28},\frac{1}{4},\ldots\}$ in the finite size
spectrum which do not appear in the spectrum of the $Z_5$ parafermionic
theory.  For the $\mathcal{W}D_5(9,10)$ rational CFT we have computed the
characters of the Virasoro representations, $\mathcal{S}$-matrix and fusion
rules.  Based on the data obtained from the finite size spectrum, an
off-diagonal modular invariant partition function has been proposed.  This is
a first step towards the construction of lattice operators corresponding to
the fields appearing in the continuum theory (see e.g.\ \cite{MCAL14}).

These findings lead us to conjecture that the sequence of $c=1$ theories
describing the critical properties of the anti-ferromagnetic $Z_n$ FZ models
for $n$ odd \cite{Albe94} are in fact rational CFTs with $\mathcal{W}D_n$
symmetry.  For the case $n=3$, corresponding to the 3-state Potts chain, the
modular invariant partition function has been computed \cite{KeMc93} leading
to the identification of a low energy effective theory with $Z_4$ parafermions
(see also Ref.~\cite{FiFr13} for another related anyon chain).  This CFT
happens to have the same spectrum as the $\mathcal{W}D_3$ model.  Further
support for this conjecture is obtained from the numerical solution of the
Bethe equations of the FZ models for $n=7,9$ indicating the presence of
$\mathcal{W}D_n$ primary fields with conformal weight $h \in
\{\frac{k^2}{4n}\,|\, k=0\ldots n\}\cup \{\frac{1}{16}, \frac{9}{16},1\}$
(note that the exponents $\frac{1}{16}$, $\frac{9}{16}$ appear in the
spectrum of the FZ model with $\pi$-twisted boundary conditions).

\begin{figure}[t]
\begin{center}
\begin{tikzpicture}[scale=1.0]  
	\tikzstyle{every node}=[circle,draw,thin,fill=blue!40,minimum size=14pt,inner sep=0pt]
	\tikzstyle{every loop}=[]
	\node (n0) at (0.0,0.0) {$\psi_{1}$};
	\node (n1) at (1.5,0.0) {$\psi_{2}$};
	\node (n4) at (1.5,1.0) {$\psi_{5}$};
	\node (n5) at (3.0,0.0) {$\psi_{6}$};
	\node (n2) at (4.5,0.0) {$\psi_{3}$};
	\node (n3) at (6.0,0.0) {$\psi_{4}$};
	\foreach \from/\to in {n0/n1,n1/n4,n1/n5,n2/n2,n2/n3,n3/n3,n4/n4} \draw (\from) -- (\to);
	\foreach \Lnode in {n2,n3,n4} \path (\Lnode) edge[loop] (\Lnode);
\end{tikzpicture}
\end{center}
\caption{The same as Figure \ref{FigPsi3Neig} except for a chain of $\psi_{2}$
  anyons. \label{FigPsi2Neig}} 
\end{figure}
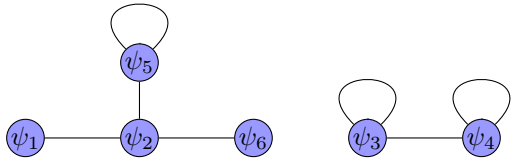

Finally, let us remark that different anyon chains can be constructed from the
$SO(5)_2$ fusion rules, Table~\ref{tabso52fus}: considering fusion paths
(\ref{fuspath}) for $\psi_2$-anyons the neighboring labels are restricted to
be adjacent nodes on the graph shown in Figure \ref{FigPsi2Neig}.
In this case the fusion path basis can be decomposed into two decoupled
subspaces: the Hilbert space spanned by states with particles $\{\psi_3,
\psi_4\}$ is isomorphic to that of a spin-$\frac{1}{2}$ chain.  Hence it can
be written as a tensor product of local spaces $\mathbb{C}^2$ .
The complementary set of particles $\{\psi_{1},\psi_{2},\psi_{5},\psi_{6} \}$
forms a fusion category equivalent to categories of irreducible representation
of the dihedral group of order 10.  Using this insight we can expect that the
model is connected to the one-dimensional spin-$\frac{1}{2}$ XXZ model
\cite{Finch13}.  Indeed we find that the $R$-matrix of the six-vertex model
with anisotropy parameter $\gamma$ can be expressed in terms of the local
projection operators appearing in this $\psi_2$-anyon chain:
\begin{equation*}
  R(u)  = \left(\sinh(\gamma)+\sinh(2u)\right)p^{(1)} +
  \left(\sinh(\gamma+2u)\right)p^{(5)} +
  \left(\sinh(\gamma)-\sinh(2u)\right)p^{(6)} \,.
\end{equation*}
The resulting model shares bulk properties with the spin-$\frac{1}{2}$ XXZ
chain.  As in the case of the $\psi_3$-model this does not, however, mean that
the excitation spectrum or the operator content will be the same.  This is
again a consequence of the presence of commuting topological charges modifying
the boundary conditions in the anyon model.

Even more models for anyons satisfying a given set of fusion rules can be
obtained by using other sets of $F$-moves consistent with the pentagon
equation.  This may lead to different models similar as in the case of
$SU(2)_3$ fusion rules leading to both the Fibonacci and the (non-unitary)
Yang-Lee anyon chains \cite{FTLT07,AGLT11}.  This question and the possibility
of additional integrable models for interacting $SO(5)_2$ anyons will be
studied in future work.

\begin{acknowledgments}
This work has been supported by the Deutsche Forschungsgemeinschaft under
grant no.\ Fr~737/7.
\end{acknowledgments}

\newpage
\appendix
\section{Transfer matrices for the BMW integrable point}
\label{app_BMW_tm}
To construct the family (\ref{eqnCommTranFam}) of transfer matrices
$t^{(\ell)}(u)$ we define the functions appearing in the generalized Boltzmann
weights (\ref{eqnWeightGen}) to be
\begin{align*}
    w^{1,3}_{3}(u) & = 1\,,\\
    w^{2,3}_{3}(u) & = i\cosh(u)\,, \qquad 
    w^{2,3}_{4}(u)   = \sinh(u)\,, \\
    w^{3,3}_{1}(u) & = \sinh(u+\frac{i\pi}{10})\sinh(u+\frac{3i\pi}{10})\,, \\
    w^{3,3}_{2}(u) & = \sinh(u-\frac{i\pi}{10})\sinh(u-\frac{3i\pi}{10})\,, \\
    w^{3,3}_{5}(u) & = \sinh(u+\frac{9i\pi}{10})\sinh(u+\frac{3i\pi}{10})\,, \\
    w^{4,3}_{2}(u) & = -\sinh(u+\frac{2i\pi}{5})\sinh(u+\frac{i\pi}{5})\,, \\
    w^{4,3}_{3}(u) & = -\sinh(u+\frac{2i\pi}{5})\sinh(u-\frac{i\pi}{5})\,, \\
    w^{4,3}_{6}(u) & = \sinh(u-\frac{2i\pi}{5})\sinh(u-\frac{i\pi}{5})\,, \\
    w^{5,3}_{3}(u) & = \mathrm{e}^{\frac{8i\pi}{10}}\,, \qquad 
    w^{5,3}_{4}(u)   = \mathrm{e}^{\frac{3i\pi}{10}}\,, \\
    w^{6,3}_{4}(u) & = -i\,.
\end{align*}
The additional weights are constructed from (\ref{RmatBMW}) as descendents
\cite{BaSt90}. 

Among the resulting transfer matrices $t^{(\ell)}(u)$, the
ones for $\ell=1,5,6$ are found to be independent of the spectral parameter
$u$.  For $t^{(1)}(u)$ and $t^{(6)}(u)$ this follows from $\psi_{1} \tp
\psi_{3}$ and $\psi_{6} \tp \psi_{3}$ both being isomorphic to simple
objects. The absence of a parameter in $t^{(5)}(u)$ is connected to our choice
of representation of the BMW algebra. The different representations and their
corresponding $R$-matrices lead to different sets of transfer matrices and we
find that it is always the case that either $t^{(2)}(u)$ or $t^{(5)}(u)$ is
parameter independent.
 
From the analysis of small systems, $\L \leq 10$, we find that the topological
$Y$-operators (\ref{Ytopo}) are obtained as limits of the transfer matrices,
i.e.\
\begin{eqnarray*}
  t^{(1)}(u) & = & Y_{1}\,,\\
  \lim_{u\rightarrow\pm\infty} \left[ 2^{\L}
    \mathrm{e}^{\mp\left(u+\frac{7i\pi}{10}\right)\mathcal{L}} \times t^{(2)}(u)
  \right] & = & Y_{2}\,, \\
  \lim_{u\rightarrow\pm\infty} \left[ 4^{\L}
    \mathrm{e}^{\mp\left(2u+\frac{9i\pi}{10}\right)\mathcal{L}} \times
    t^{(3)}(u) 
  \right] & = & Y_{3}\,, \\
  \lim_{u\rightarrow\pm\infty} \left[ 4^{\L}
    \mathrm{e}^{\mp\left(2u+\frac{2i\pi}{5}\right)\mathcal{L}} \times t^{(4)}(u)
  \right] & = & Y_{4}\,, \\
  t^{(5)}(u) & = &  Y_{5}\,,\\
  t^{(6)}(u) & = &  Y_{6}\,.
\end{eqnarray*}
Given the interpretation of both the $t^{(\ell)}(u)$ and the $Y_\ell$ as
descriptions of certain braiding processes we expect this relation to hold for
arbitrary system sizes.
Using the fusion procedure \cite{BaRe89} for face models we find closed
relations for transfer matrices
\begin{equation}
\label{BMW_fusion}
\begin{aligned}
  t^{(6)}(u)\,t^{(6)}(u) & =  \mathbf{1}\,, \\
  t^{(6)}(u)\,t^{(3)}(u) & =  i^{\L}\,t^{(4)}(u+\frac{i\pi}{2})\,, \\
  t^{(2)}(u)\,t^{(3)}(u) & =
  \left[\sinh\left(u-\frac{i\pi}{10}\right)\right]^{\L}\,
  t^{(3)}(u+\frac{2i\pi}{5}) +
  \left[\sinh\left(u\right)\right]^{\L}\,t^{(4)}(u+\frac{i\pi}{10}) \\ 
  & =  \left[\sinh\left(u-\frac{i\pi}{10}\right)\right]^{\L}\,
  t^{(3)}(u+\frac{2i\pi}{5}) + \left[i\sinh\left(u\right)\right]^{\L}\,
  t^{(6)}(u)\,t^{(3)}(u-\frac{2i\pi}{5}) \,.
\end{aligned}
\end{equation}
Examining Table \ref{tabso52fus} we see that the relations between transfer
matrices mimic the fusions rules of the underlying category.

\section{Rational CFTs with extended symmetries}
\label{appCFT}
As discussed in the main text the finite size spectrum of a critical
one-dimensional lattice model is completely determined by the central charge
$c$ of the underlying Virasoro algebra and the spectrum of conformal weights
$\{h\}$, i.e.\ the eigenvalues of the Virasoro zero mode $L_0$ on the primary
(or Virasoro highest weight) states.
Often, however, the Virasoro algebra alone is not sufficient to decompose the
state space of the system into a finite set of irreducible highest weight
representations. We collect here some general facts about rational CFTs with
extended chiral symmetry algebras, mainly taken from
\cite{Figu90,LuFa90,BoSc93,FrKW92,BEHH95} and the reprint volume
\cite{BoSc95}.

\subsection{Chiral symmetry algebras}
This is true in particular for rational conformal field theories with
$c\geq1$: these theories having a \emph{finite} number of admissible highest
weight representations with respect to its maximally extended chiral symmetry
algebra $\mathcal{W}(d_1=2,d_2,\ldots,d_\ell)$.  Here $d_k$ denote the
conformal scaling dimensions of chiral primary fields $W^{(d_k)}$ which,
together with the energy-momentum tensor $T\equiv W^{(d_1)}$, generate the
chiral symmetry algebra.  A highest weight state $\ket[h\equiv
  w^{(1)},w^{(2)},w^{(3)}, \ldots w^{(\ell)}]$ has then $\ell$ quantum numbers
$w^{(k)}$ with respect to the zero modes $W^{(d_k)}_0$,
i.e.\ $W^{(d_k)}_0\ket[h,w^{(2)},w^{(3)}, \ldots w^{(\ell)}] =
w^{(k)}\ket[h,w^{(2)},w^{(3)}, \ldots w^{(\ell)}]$.  The highest weight
representations are closed under fusion and hence give rise to a
\emph{finite}-dimensional representation of the modular group
$\mathrm{PSL}(2,\mathbb{Z})$.
As all fields in the symmetry algebra must be mutually local, the
scaling dimensions are restricted to be integers or half-integers,
$2\,d_k\in\mathbb{N}$.  Therefore, chiral symmetry algebras constitute
meromorphic conformal field theories.

A class of chiral symmetry algebras whose representation theory is
particularly well understood are $\mathcal{W}$-algebras associated with
Lie-algebras $\mathfrak{g}$, where the scaling dimensions $d_k$ are related to
the exponents of the Lie-algebra $\mathfrak{g}$, i.e.\ where one has one
$W^{(d_k)}$ field associated to each independent Casimir operator.  For a
Lie-algebra of rank $\ell$, the $\mathcal{W}$-algebra is generated by $\ell$
such fields, one of those always being the Virasoro field associated to the
quadratic Casimir operator.  For this class of $\mathcal{W}$-algebras, free
field realizations are known and hence the algebras can, in principle, be
explicitly constructed \cite{FrKW92}. For all semi-simple Lie-algebras the
corresponding $\mathcal{W}$-algebras are known together with their so-called
minimal series of rational theories they admit.

In the following let $\mathfrak{g}$ denote a simple Lie-algebra of rank
$\ell$, and $\mathfrak{h}$ a Cartan subalgebra of it. We denote by
$d_{\mathfrak{g}}$ the dimension of the Lie-algebra and by $g^*$ its dual
Coxeter number. Let $\Phi_+$ denote the set of positive roots $\alpha$, and
$\mathrm{ht}\,\alpha$ the height of the root $\alpha$.  The height of a
positive root is determined from its unique decomposition
$\alpha=\sum_{i=1}^rn_i\alpha_i$ into a linear combination of the simple roots
$\alpha_i$ with non-negative integers $n_i$:
$\mathrm{ht}\,\alpha=\sum_{i=1}^rn_i$.  Finally, the maximal root may be
denoted by $\rho$.  Under affinization of the Lie-algebra to
$\hat{\mathfrak{g}}$, we introduce the level $x = 2k/|\rho|^2$, where $k$ is
the central extension of $\hat{\mathfrak{g}}$. Choosing the Killing form as
our invariant symmetric bilinear form on $\mathfrak{g}$, one finds that
$|\rho|^2=1/g^*$ and the central charge for $\mathcal{W}_{\hat{\mathfrak{g}}}$
at level $x$ is given by \cite{Figu90}
\begin{align}
\label{cft_WGcc}
	c_{\mathfrak{g}}(x) &= \ell - \frac{d_{\mathfrak{g}}g^*}{x+g^*}
	-\frac{12}{g^*}(x+g^*)\sum_{\alpha\in\Phi_+}(\mathrm{ht}\,\alpha)^2
	+ 12\sum_{\alpha\in\Phi_+}\mathrm{ht}\,\alpha\,.
\end{align}
We see that the central charge is entirely expressed in terms of basic
properties of the Lie-algebra and the sums of the (squares of the) heights of
the positive roots. The former are standard and can be found in every text
book on Lie-algebras, the latter can easily be computed from the explicit
choice of set of simple roots $\Delta$.  In fact, the values of these sums do
not depend on the choice of $\Delta$ after all.  The data needed to
explicitely compute the central charge for a simple Lie algebra are given in
Table~\ref{tab:cftLie}.
\begin{table}[t]
\begin{tabularx}{\textwidth}{
	X >{$}c<{$} X >{$}c<{$} X >{$}c<{$} X >{$}c<{$} X >{$}c<{$} X
}
\hline\hline
& \mathfrak{g} & & d_{\mathfrak{g}} & & g^* & & 
\sum_{\alpha\in\Phi_+} \mathrm{ht}\,\alpha & & 
\sum_{\alpha\in\Phi_+}(\mathrm{ht}\,\alpha)^2 & \\[0.2cm]
\hline\\[-0.35cm]
& A_\ell & & \ell(\ell+2) & & \ell+1 & & \frac16\ell(\ell+1)(\ell+2) & & 
\frac{1}{12}\ell(\ell+1)^2(\ell+2) & \\[0.07cm]
& B_\ell & & \ell(2\ell+1) & & 2\ell-1 & & \frac16\ell(\ell+1)(4\ell-1) & &
\frac16\ell(\ell+1)(2\ell-1)(2\ell+1) & \\[0.07cm]
& C_\ell & & \ell(2\ell+1) & & \ell+1 & & \frac16\ell(\ell+1)(4\ell-1) & &
\frac16\ell(\ell+1)(2\ell-1)(2\ell+1) & \\[0.07cm]
& D_\ell & & \ell(2\ell-1) & & 2(\ell-1) & & \frac13\ell(\ell-1)(2\ell-1) & &
\frac13\ell(\ell-1)^2(2\ell-1) & \\[0.07cm]
& E_6 & &  78 & & 12 & &  156 & &   936 & \\[0.07cm]
& E_7 & & 133 & & 18 & &  399 & &  3591 & \\[0.07cm]
& E_8 & & 248 & & 30 & & 1240 & & 18600 & \\[0.07cm]
& F_4 & &  52 & &  9 & &  110 & &   702 & \\[0.07cm]
& G_2 & &  14 & &  4 & &   16 & &    56 & \\[0.07cm]
&\mathcal{B}_{0,\ell} & & \ell(2\ell+1) & & 2\ell-1 & &   
\frac16\ell(2\ell-1)(2\ell+1) & & \frac{1}{12}\ell(2\ell-1)^2(2\ell+1) & 
\\[0.15cm]
\hline\hline
\end{tabularx}
\caption{\label{tab:cftLie}Lie algebra data.}
\end{table}

\subsection{Minimal series of Casimir-type $\mathcal{W}$-algebras}
Setting $x+g^*=p/q\in\mathbb{Q}$ with $p,q$ coprime positive integers in
(\ref{cft_WGcc}) one obtains the minimal series of a chiral symmetry algebra,
i.e. a series of values of the central charge $c_{\mathfrak{g}}(p,q)$, where
due to an exceptional high number of null states per Verma module the
irreducible representations are as small as possible.  For these values of $c$
only finitely many irreducible highest weight representations are needed to
build a complete, rational conformal field theory. The corresponding $L_0$
eigenvalues of the highest weight states are all known.

In Table~\ref{tab:cftW} we give a list of all the Casimir-type
$\mathcal{W}_{\hat{\mathfrak{g}}}$-algebras, explicitly denoting the
dimensions $d_k$ of their generators.  
\begin{table}[t]
\begin{tabularx}{\textwidth}{
	X >{$}c<{$} X >{$}l<{$} >{$}l<{$} 
}
\hline\hline\\[-0.35cm]
& \mathfrak{g} & &
\mathcal{W}_{\hat{\mathfrak{g}}} & 
c_{\mathfrak{g}}(p,q)
\\[0.2cm]
\hline\\[-0.35cm]
& A_\ell & & \mathcal{W}A_\ell=\mathcal{W}(2,3,\ldots,d,\ldots,\ell+1) & 
	\ell\left(1 - (\ell+1)(\ell+2)\frac{(p-q)^2}{pq}\right)
\\[0.18cm]
& B_\ell & & \mathcal{W}B_\ell=\mathcal{W}(2,4,\ldots,2d,\ldots,2\ell) & 
	\ell\left(1 + 2(\ell+1)(4\ell-1) - (2\ell+1)\Big(
	\frac{(2\ell-1)q}{p} + \frac{2(\ell+1)p}{q}\Big)\right)
\\[0.18cm]
& C_\ell & & \mathcal{W}C_\ell=\mathcal{W}(2,4,\ldots,2d,\ldots,2\ell) & 
	\ell\left(1 + 2(\ell+1)(4\ell-1) - (2\ell+1)\Big(
	\frac{2(2\ell-1)q}{p} + \frac{(\ell+1)p}{q}\Big)\right)
\\[0.18cm]
& D_\ell & & \mathcal{W}D_\ell=\mathcal{W}(2,4,\ldots,2d,\ldots,2(\ell-1),\ell) & 
	\ell\left(1 - (2\ell-1)(2\ell-2)\frac{(p-q)^2}{pq}\right)
\\[0.18cm]
& E_6 & & \mathcal{W}E_6=\mathcal{W}(2,5,6,8,9,12) & 
	6\left(1 - 156\frac{(p-q)^2}{pq}\right)
\\[0.18cm]
& E_7 & & \mathcal{W}E_7=\mathcal{W}(2,6,8,10,12,14,18) & 
	7\left(1 - 342\frac{(p-q)^2}{pq}\right)
\\[0.18cm]
& E_8 & & \mathcal{W}E_8=\mathcal{W}(2,8,12,14,18,20,24,30) & 
	8\left(1 - 930\frac{(p-q)^2}{pq}\right)
\\[0.18cm]
& F_4 & & \mathcal{W}F_4=\mathcal{W}(2,6,8,12) & 
	4\left(1 + 330 - 117\Big(\frac{2p}{q} + \frac{q}{p}\Big)\right)
\\[0.18cm]
& G_2 & & \mathcal{W}G_2=\mathcal{W}(2,6) & 
	2\left(1 + 96 - 28\Big(\frac{3p}{q}+\frac{q}{p}\Big)\right)
\\[0.18cm]
& \!\!\mathcal{B}_{0,\ell} & & \mathcal{WB}_{0,\ell}=\mathcal{W}(2,4,\ldots,2\ell,
	\frac12(2\ell+1)) & 
	(\ell+\frac12)\left(1-2\ell(2\ell-1)\frac{(p-q)^2}{pq}\right)
\\[0.3cm]
\hline\hline
\end{tabularx}
\caption{\label{tab:cftW}Casimir-type $\mathcal{W}$-algebras and their
  corresponding minimal series.}
\end{table}
By definition, the first generator, always of scaling dimension two, is the
energy-momentum tensor, all other generators are Virasoro primary fields.
Casimir-type $\mathcal{W}$-algebras are purely bosonic algebras, i.e.\ all
generators have integer scaling dimensions.  For each of the algebras, we also
list the central charges of their corresponding series of minimal models for
level $x+g^*=p/q\in\mathbb{Q}$.

For non-simply-laced Lie-algebras alternative constructions of extended chiral
symmetries are possible if one allows for generators with half-integer
spin. In particular \cite{LuFa90}, one can construct an alternative
$\mathcal{W}$-algebra for the $B_\ell$ series, i.e.\ for
${SO}(2\ell+1)$, 
which contains
precisely one fermionic generator. 
In the literature, these algebras are often
denoted $\mathcal{W}_{\hat{B}f}$-algebras, and are given as
$\mathcal{W}(2,4,\ldots,2d,\ldots,2\ell,\frac12(2\ell+1))$.   We will denote
them as $\mathcal{WB}_{0}$-algebras in the following.  These
algebras also admit a minimal series whose members have the central charges
\begin{align}
\label{cft_WBfcc}
	c_{\mathcal{B}_{0,\ell}}(p,q) &= \Big(\ell+\frac12\Big)\left(
	1 - 2\ell(2\ell-1)\frac{(p-q)^2}{pq}\right)\,.
\end{align}
We note that the first member, $\mathcal{W}_{\hat{\mathcal{B}}_{0,1}}=
\mathcal{W}(2,\frac32)$, yields the minimal series of the $N=1$ supersymmetric
extension of the Virasoro algebra, when $p$ and $q$ are replaced by
half-integers $p/2$ and $q/2$.
In general, the $\mathcal{WB}_{0}$-algebras can be realized from the
Lie-superalgebras 
$\mathcal{B}_{0,\ell} ={OSp}(1|2\ell)$, 
which
explains our notation.  
This is the only example of a Lie-superalgebra leading to a true Casimir-type
$\mathcal{W}$-algebra.
Note that the central charge (\ref{cft_WBfcc}) can formally be obtained from
Eq.~(\ref{cft_WGcc}) for $D_\ell$ by replacing $\ell\mapsto\ell+\frac12$.

\subsection{Spectra of Casimir-type $\mathcal{W}$-algebras}
In a similar way as the central charges (\ref{cft_WGcc}) the conformal weights
appearing in minimal models of Casimir-type $\mathcal{W}$-algebras are
essentially determined by data from the underlying Lie-algebra. Let again
$\ell$ denote the rank of a simple Lie-algebra $\mathfrak{g}$. 

Highest-weight representations of $\mathfrak{g}$ to weights
$\lambda=\sum_{i=1}^\ell r_i\lambda^i$ with $\lambda^i$ denoting the
fundamental weights, and $r_i\in\mathbb{N}$ can be labeled by the positive
integers $\mathbf{r}=(r_1,r_2,\ldots,r_\ell)$. The weight lattice has an
associated dual lattice, spanned by the fundamental co-roots $\lambda^*_i$.  A
co-weight is then given by $\lambda^*=\sum_{i=1}^\ell s^i\lambda^*_i$ with
$s^i\in \mathbb{N}$ and labeled by the positive integers
$\mathbf{s}=(s^1,s^2,\ldots,s^\ell)$.
The conformal weights of a minimal model of a Casimir-type
$\mathcal{W}$-algebra $\mathcal{W}_{\hat{\mathfrak{g}}}$ with central charge
$c_{\mathfrak{g}}(p,q)$ are given by \cite{LuFa90,BEHH95}
\begin{align}
\label{cft_WGspec}
	h_{\mathbf{r},\mathbf{s}}(p,q) &= 
	\frac{q^2\,\mathbf{r}\cdot(C^{-1}D)\cdot\mathbf{r}
	-2pq\,\mathbf{r}\cdot(C^{-1})\cdot\mathbf{s}
	+p^2\,\mathbf{s}\cdot(D^{-1}C^{-1})\cdot\mathbf{s}
	}{2pq} - \frac{\ell - c_{\mathfrak{g}}(p,q)}{24}\,,
\end{align}
where $C$ denotes the Cartan matrix of the simple Lie-algebra $\mathfrak{g}$,
and where the weights and co-weights must fulfill the conditions
$\sum_{i=1}^\ell r_im^i\leq p-1$ and $\sum_{i=1}^\ell s^im^*_i\leq q-1$. Here, 
the $m^i$ are the normalized
components of the highest root $\psi$ in the directions of the simple roots
$\alpha_i$, i.e.\ $\frac{\psi}{\psi^2}=\sum_{i=1}^\ell m^i
\frac{\alpha_i}{\alpha_i^2}$ and $m^*_i = \frac{2}{\alpha_i^2}m^i$. 
Thus, all we need to know are the integers $m^i$ and $m^*_i$ and the matrices
$C$ and $D$, which give the scalar products $\lambda^i\lambda^j=(C^{-1}D)^{ij}$,
$\lambda^*_i\lambda^*_j=(D^{-1}C^{-1})_{ij}$ and finally
$\lambda^i\lambda^*_j = (C^{-1})^i_{\ j}$. 
To denote them in the following table as concise as possible, we denote by
$E_{ij}$ the $\ell\times\ell$ matrix with entries
$(E_{ij})_{kl}=\delta_{ik}\delta_{jl}$.  Next, we denote by $\mathbf{1}$ the
$\ell\times\ell$ identity matrix.  We denote the Cartan-matrix of $A_\ell$ as
$A=2\,\mathbf{1}-\sum_{i=1}^{\ell-1}(E_{i,i+1}+E_{i+1,i})$. As all
Cartan-matrices of simple Lie-algebras are deviations from the $A_\ell$-case,
we give all other Cartan-matrices in terms of $A$ up to corrections in terms
of some $E_{ij}$. The matrix $D$ is always diagonal and just the identity in
the simply-laced case. As a consequence, Eq.~(\ref{cft_WGspec}) can be
factorized in the simply laced case to give
\begin{align*}
	h_{\mathbf{r},\mathbf{s}}(p,q) &= 
	\frac{(q\,\mathbf{r}- p\,\mathbf{s})\cdot(C^{-1})\cdot
	(q\,\mathbf{r}- p\,\mathbf{s})
	}{2pq} - \frac{\ell - c_{\mathfrak{g}}(p,q)}{24}\,.
\end{align*}
A similar formula yields the conformal weights for the series of
$\mathcal{WB}_{0}$-algebra minimal models, where \cite{LuFa90}
\begin{align}
\label{cft_WBfspec}
	h_{\mathbf{r},\mathbf{s}}(p,q) &= 
	\frac{(q\,\mathbf{r}- p\,\mathbf{s})\cdot(C^{-1})\cdot
	(q\,\mathbf{r}- p\,\mathbf{s})
	}{2pq} - \frac{(\ell+\frac12) - 
          c_{\mathcal{B}_{0,\ell}}(p,q)}{24} + 
	\frac{\epsilon_{\mathbf{r},\mathbf{s}}}{16}\,.
\end{align}
Here, $\epsilon_{\mathbf{r},\mathbf{s}}=(r_\ell-s_\ell\mod 2)$
distinguishes between the Neveu-Schwarz sector ($r_\ell-s_\ell\equiv 0\
\mod 2$) and the Ramond sector ($r_\ell-s_\ell\equiv 1\mod2$). 
These sectors correspond to periodic or anti-periodic boundary conditions,
respectively.

Table~\ref{tab:cftLie2} lists all Lie-algebra data needed for explicit
computations of conformal weights for the Casimir-type
$\mathcal{W}_{\hat{\mathfrak{g}}}$-algebras.
\begin{table}[t]
\noindent\begin{tabularx}{\textwidth}{
	X >{$}c<{$} X >{$}c<{$} >{$}c<{$} >{$}c<{$} X
}
\hline\hline
& \mathfrak{g} & & (m^i) & (m^*_i) & C\hfill D &\\
\hline
& A_\ell & & (1,\ldots,1) & (1,\ldots,1)& 
A \hfill
\mathbf{1} & \\
& B_\ell & & (1,2,\ldots,2,1) & (1,2,\ldots,2) & 
A\!-\!E_{\ell-1,\ell} \hfill
\mathbf{1}\!-\!\frac12E_{\ell\ell} & \\
& C_\ell & & (1,\ldots,1) & (2,\ldots,2,1) & 
A\!-\!E_{\ell,\ell-1} \hfill
\frac12\mathbf{1}\!+\!\frac12E_{\ell\ell} & \\
& D_\ell & & (1,2\ldots,2,1,1) & (1,2,\ldots,2,1,1) & 
A\!+\!E_{\ell,\ell-1}\!+\!E_{\ell-1,\ell}\!-\!E_{\ell,\ell-2}\!-\!E_{\ell-2,
	\ell} \hfill
\mathbf{1} & \\
& E_6 & & (1,2,2,3,2,1) & (1,2,2,3,2,1) & 
A\!+\!E_{12}\!+\!E_{21}\!-\!E_{13}\!-\!E_{31}\!-\!E_{24}\!-\!E_{42} \hfill
\ \ \ \mathbf{1} & \\
& E_7 & & (2,2,3,4,3,2,1) & (2,2,3,4,3,2,1) & 
A\!+\!E_{12}\!+\!E_{21}\!-\!E_{13}\!-\!E_{31}\!-\!E_{24}\!-\!E_{42} \hfill
\ \ \ \mathbf{1} & \\
& E_8 & & \!(2,3,4,6,5,4,3,2)\! & \!(2,3,4,6,5,4,3,2)\! & 
A\!+\!E_{12}\!+\!E_{21}\!-\!E_{13}\!-\!E_{31}\!-\!E_{24}\!-\!E_{42} \hfill
\ \ \ \phantom{mlml}\mathbf{1}& \\
& F_4 & & (1,2,3,2) & (2,4,3,2) & 
A\!-\!E_{32} \hfill
\frac12\mathbf{1}\!+\!\frac12E_{33}\!+\!\frac12E_{44} & \\
& G_2 & & (2,1) & (2,3) & 
A\!-\!2E_{12} \hfill
E_{11}\!+\!\frac13E_{22} \\
& \!\mathcal{B}_{0,\ell}\! & & (1,2,\ldots,2,1) & (1,2,\ldots,2,1) & 
A\!-\!E_{\ell,\ell-1}\!-\!E_{\ell-1,\ell}\!+\!2E_{\ell,\ell} \hfill
\mathbf{1} & \\[0.15cm]
\hline\hline
\end{tabularx}
\caption{\label{tab:cftLie2}Further Lie algebra data.}
\end{table}

\subsection{Some examples}
In the following we present the spectra of some rational CFTs with central
charge $c=1$ and $c=\frac{8}{7}$, respectively.
\subsubsection{$Z_k$ parafermions}
The $Z_k$ parafermion conformal field theory has central charge
$c_k=2\frac{k-1}{k+2}$.  Obviously, we get $c=1$ for $k=4$ and $c=8/7$ for
$k=5$.  The conformal spectrum is known to be the set \cite{ZaFa85,GeQi87}
\begin{align*}
  h_{\ell,m} &= \frac{1}{2}\frac{\ell(k-\ell)}{k(k+2)} +
  \frac{(\ell+m)(\ell-m)}{4k}\,, 
	\ \ \ \
	1\leq \ell\leq k\,,
	\ \ 
	-\ell\leq m\leq \ell\,,
	\ \ 
	\textrm{and}
	\ \
	\ell+m\equiv 0\ \mathrm{mod}\ 2\,,
\end{align*}
of conformal weights.  The fields with conformal weights $h_{\ell,\ell}$ and
$h_{\ell,-\ell}$ constitute the order and disorder fields, respectively.

As rational conformal field theories, ${Z}_k$ parafermions possess an
extension of the Virasoro algebra to a $\mathcal{W}A_k$-algebra.  In the
minimal series they appear for $(p,q)=(k+1,k+2)$.  It is
straightforward to check that Eqs.~(\ref{cft_WGcc}) and (\ref{cft_WGspec})
reproduce the central charge and conformal weights of the parafermions, e.g.\
\begin{equation}
\label{eq:h-paraferm}\begin{aligned}
  k=4: \quad & c = 1\,,\quad
        h \in \left\{0,\frac{1}{16},\frac1{12},\frac13,
           \frac9{16},\frac34,1\right\}\,,\\
  k=5: \quad & c=\frac87\,,\quad
        h \in \left\{0,\frac{2}{35},\frac{3}{35},\frac27,
           \frac{17}{35},\frac{23}{35},\frac45,\frac67,\frac65 \right\}\,. 
\end{aligned}
\end{equation}

\subsubsection{Casimir-type $\mathcal{W}$-algebras related to $B$,
  $\mathcal{B}_0$ and $D$}
Based on the discrete symmetries of the BMW anyon model, the
$\mathcal{W}B_2(5,7)$ CFT has been identified as the most likely candidate
for the $c=1$ case.  The integers $r_i$ ($s^i$) parameterizing the highest
weights (co-weights) are restricted by $r_1+r_2\le4$ and $s^1+2s^2\le6$.  From
(\ref{cft_WGspec}) we obtain the conformal weights which can be given in the
following compact way
\begin{align}
\label{cft_specWB2}
	h_{\mathbf{r},\mathbf{s}}[\mathcal{W}B_{2}(5,7)] &=
\left[ \begin{array}{ccccccc} 
	{{\bf r}}\backslash{{\bf s}} & 
	\left[\!\begin{array}{c} 
		1\\[-0.1cm] 1
	\end{array}\!\right] & 
	\left[\!\begin{array}{c} 
 		2\\[-0.1cm] 1
	\end{array}\!\right] &
  	\left[\!\begin{array}{c} 
		1\\[-0.1cm] 2
	\end{array}\!\right] & 
	\left[\!\begin{array}{c} 
		3\\[-0.1cm] 1
	\end{array}\!\right] &
   	\left[\!\begin{array}{c} 
		2\\[-0.1cm] 2
	\end{array}\!\right] & 
	\left[\!\begin{array}{c} 
		4\\[-0.1cm] 1
	\end{array}\!\right] \\ 
	\noalign{\medskip}[1,1]&
	0&
	\frac{2}{7}&
	\frac{6}{7}&
	\frac{9}{7}&
	\frac{13}{7}&
	3\\ 
	\noalign{\medskip}[1,2]&
	\frac{1}{4}&
	\frac{1}{28}&
	\frac{3}{28}&
	\frac{15}{28}&
	\frac{17}{28}&
	\frac{7}{4}\\ 
	\noalign{\medskip}[2,1]&
	\frac{4}{5}&
	\frac{3}{35}&
	\frac{23}{35}&
	\frac{3}{35}&
	\frac{23}{35}&
	\frac{4}{5}\\
	\noalign{\medskip}[1,3]&
	\frac{6}{5}&
	\frac{17}{35}&
	\frac{2}{35}&
	\frac{17}{35}&
	\frac{2}{35}&
	\frac{6}{5}\\
	\noalign{\medskip}[2,2]&
	\frac{7}{4}&
	\frac{15}{28}&
	\frac{17}{28}&
	\frac{1}{28}&
	\frac{3}{28}&
	\frac{1}{4}\\ 
	\noalign{\medskip}[3,1]&
	3&
	\frac{9}{7}&
	\frac{13}{7}&
	\frac{2}{7}&
	\frac{6}{7}&
	0
\end{array} \right]\,.
\end{align}
Eqs.~(\ref{cft_WGspec}) and (\ref{cft_WBfspec}) yield the complete spectrum of
a minimal model of a Casimir-type $\mathcal{W}$-algebra including all
non-trivial multiplicities.  To determine the true multiplicity, however, ones
has to take into account symmetries relating different labels
$(\mathbf{r},\mathbf{s})$ within the weight lattice.  To obtain the true
multiplicities it is often sufficient to divide the multiplicities read off
from the conformal grid by the number of times the vacuum representation with
$h=0$ appears.  In the example for $\mathcal{W}B_2(5,7)$ above the true
multiplicities of all weights are one.

For the anyon model with central charge $c=\frac{8}{7}$ we have identified two
series of minimal models for $\mathcal{B}_{0,\ell}$ and
$D_{\ell}={SO}(2\ell)$.
  The smallest ones respecting the five-fold
discrete symmetry of the anyon model are $\mathcal{WB}_{0,2}(4,5)$ and
$\mathcal{W}D_{5}(9,10)$, respectively.  Again the spectra can be given by
conformal grids: for the $\mathcal{WB}_{0,2}(4,5)$ model we find
\begin{align}
\label{cft_specWBf2}
	h_{\mathbf{r},\mathbf{s}}[\mathcal{WB}_{0,2}(4,5)] &=
\left[ \begin{array}{ccccccc} 
	{{\bf r}}\backslash{{\bf s}} & 
	\left[\!\begin{array}{c} 
		1\\[-0.1cm] 1
	\end{array}\!\right] & 
	\left[\!\begin{array}{c} 
 		1\\[-0.1cm] 2
	\end{array}\!\right] &
  	\left[\!\begin{array}{c} 
		2\\[-0.1cm] 1
	\end{array}\!\right] & 
	\left[\!\begin{array}{c} 
		1\\[-0.1cm] 3
	\end{array}\!\right] &
   	\left[\!\begin{array}{c} 
		2\\[-0.1cm] 2
	\end{array}\!\right] & 
	\left[\!\begin{array}{c} 
		3\\[-0.1cm] 1
	\end{array}\!\right] \\ 
	\noalign{\medskip}[1,1]&
	0&
	\frac{1}{16}&
	\frac{1}{10}&
	\frac{2}{5}&
	\frac{9}{16}&
	1\\ 
	\noalign{\medskip}[1,2]&
	\frac{5}{8}&
	\frac{1}{16}&
	\frac{9}{40}&
	\frac{1}{40}&
	\frac{1}{16}&
	\frac{5}{8}\\ 
	\noalign{\medskip}[2,1]&
	1&
	\frac{9}{16}&
	\frac{1}{10}&
	\frac{2}{5}&
	\frac{1}{16}&
	0
\end{array} \right]\,.
\end{align}
The corresponding table for the model with $\mathcal{W}D_5$ symmetry is
already quite large
\begin{align}
  \label{cft_specWD5}
  & h_{\mathbf{r},\mathbf{s}}[\mathcal{W}{D}_{5}(9,10)] =
  \left[ \begin{array}{cc@{}c@{}c@{}c@{}c@{}c@{}c@{}c@{}c@{}c@{}c@{}c} 
	{{\bf r}}\backslash{{\bf s}} & 
	\left[\!\begin{array}{c} 
		1\\[-0.1cm] 1\\[-0.1cm] 1\\[-0.1cm] 1\\[-0.1cm] 1
	\end{array}\!\right] & 
	\left[\!\begin{array}{c} 
		1\\[-0.1cm] 1\\[-0.1cm] 1\\[-0.1cm] 1\\[-0.1cm] 2
	\end{array}\!\right] & 
	\left[\!\begin{array}{c} 
		1\\[-0.1cm] 1\\[-0.1cm] 1\\[-0.1cm] 2\\[-0.1cm] 1
	\end{array}\!\right] & 
	\left[\!\begin{array}{c} 
		2\\[-0.1cm] 1\\[-0.1cm] 1\\[-0.1cm] 1\\[-0.1cm] 1
	\end{array}\!\right] & 
	\left[\!\begin{array}{c} 
		1\\[-0.1cm] 1\\[-0.1cm] 1\\[-0.1cm] 1\\[-0.1cm] 3
	\end{array}\!\right] & 
	\left[\!\begin{array}{c} 
		1\\[-0.1cm] 1\\[-0.1cm] 1\\[-0.1cm] 2\\[-0.1cm] 2
	\end{array}\!\right] & 
	\left[\!\begin{array}{c} 
		1\\[-0.1cm] 1\\[-0.1cm] 1\\[-0.1cm] 3\\[-0.1cm] 1
	\end{array}\!\right] & 
	\left[\!\begin{array}{c} 
		1\\[-0.1cm] 1\\[-0.1cm] 2\\[-0.1cm] 1\\[-0.1cm] 1
	\end{array}\!\right] &
  	\left[\!\begin{array}{c} 
		1\\[-0.1cm] 2\\[-0.1cm] 1\\[-0.1cm] 1\\[-0.1cm] 1
	\end{array}\!\right] & 
	\left[\!\begin{array}{c} 
		2\\[-0.1cm] 1\\[-0.1cm] 1\\[-0.1cm] 1\\[-0.1cm] 2
	\end{array}\!\right] &
   	\left[\!\begin{array}{c} 
		2\\[-0.1cm] 1\\[-0.1cm] 1\\[-0.1cm] 2\\[-0.1cm] 1
	\end{array}\!\right] & 
	\left[\!\begin{array}{c} 
		3\\[-0.1cm] 1\\[-0.1cm] 1\\[-0.1cm] 1\\[-0.1cm] 1
	\end{array}\!\right] \\ 
	\noalign{\medskip}[1,1,1,1,1]&
	0&
	\frac{1}{16}&
	\frac{1}{16}&
	\frac{1}{20}&
	\frac{5}{4}&
	\frac{4}{5}&
	\frac{5}{4}&
	\frac{9}{20}&
	\frac{1}{5}&
	\frac{9}{16}&
	\frac{9}{16}&
	1\\ 
	\noalign{\medskip}[1,1,1,1,2]&
	\frac{5}{4}&
	\frac{1}{16}&
	\frac{9}{16}&
	\frac{4}{5}&
	0&
	\frac{1}{20}&
	1&
	\frac{1}{5}&
	\frac{9}{20}&
	\frac{1}{16}&
	\frac{9}{16}&
	\frac{5}{4}\\
	\noalign{\medskip}[1,1,1,2,1]&
	\frac{5}{4}&
	\frac{9}{16}&
	\frac{1}{16}&
	\frac{4}{5}&
	1&
	\frac{1}{20}&
	0&
	\frac{1}{5}&
	\frac{9}{20}&
	\frac{9}{16}&
	\frac{1}{16}&
	\frac{5}{4}\\
	\noalign{\medskip}[2,1,1,1,1]&
	1&
	\frac{9}{16}&
	\frac{9}{16}&
	\frac{1}{20}&
	\frac{5}{4}&
	\frac{4}{5}&
	\frac{5}{4}&
	\frac{9}{20}&
	\frac{1}{5}&
	\frac{1}{16}&
	\frac{1}{16}&
	0
\end{array} \right]\,.
\end{align}
Note, however, that the $D_\ell$ models with odd $\ell$ have a ${Z}_4$
symmetry in the conformal spectrum.  Therefore the true multiplicities of the
conformal weights are one except for
$h\in\{\frac{1}{16},\frac{9}{16},\frac{5}{4}\}$, which appear with
multiplicity two.

\section{$\mathcal{S}$-matrix and fusion rules for $\mathcal{W}D_5(9,10)$}
\label{app_WD5fus}
With the characters (\ref{WD5_modforms}) the $\mathcal{S}$-matrix for the
modular transformation $\chi_i(\mathrm{e}^{-2\pi i/\tau}) =
\sum_{ij}\mathcal{S}_{ij}\, \chi_j(\mathrm{e}^{2\pi i\tau})$ is easily found
to be ($\phi = \frac{1+\sqrt{5}}{2}$, as in Eq.~(\ref{Smat}))
\begin{equation}
  \label{WD5_smat}
  \mathcal{S}_{\mathcal{W}{D}_5(9,10)} = \frac{1}{2\sqrt{10}}
  \left(\begin{array}{ccccccccc}
      1 &  1  &  1 &  2  &  2  &  2  &  2  &  \sqrt{5}  &  \sqrt{5}  \\
      1 &  1  &  1 &  2  &  2  &  2  &  2  & -\sqrt{5}  & -\sqrt{5}  \\
      2 &  2  & -2 & -4  &  4  &  -4  & 4  &  0    &  0    \\
 2 &  2  & -2 &  2\phi  & -2\phi  & -2\phi^{-1}  & +2\phi^{-1}  & 0 & 0 \\
 2 &  2  &  2 & -2\phi  & -2\phi  &  2\phi^{-1}  & 2\phi^{-1}  & 0 & 0 \\
 2 &  2  & -2 & -2\phi^{-1} &  2\phi^{-1} &  2\phi  & -2\phi   & 0 & 0 \\
 2 &  2  &  2 &  2\phi^{-1} &  2\phi^{-1} & -2\phi  & -2\phi   & 0 & 0 \\
 2\sqrt{5}&-2\sqrt{5}&  0   &  0    &  0    &  0    &  0    & \sqrt{10}&-\sqrt{10}\\
 2\sqrt{5}&-2\sqrt{5}&  0   &  0    &  0    &  0    &  0    &-\sqrt{10}& \sqrt{10}
	\end{array}\right)\,,
\end{equation}
where rows and columns refer to the representations in the order listed in
(\ref{WD5_weights}).  Note that while this $\mathcal{S}$-matrix fulfills
$\mathcal{S}^2=(\mathcal{ST})^3= \mathbf{1}$ it is not symmetric.  This is to
be expected, as we have characters with non-trivial multiplicities.

As a final test for a theory to be a \emph{bona fide} rational CFT the fusion
rules, as computed from the characters and their $\mathcal{S}$-matrix via the
Verlinde formula\footnote{The index $\mathrm{vac}$ labels the vacuum
  representation.}
\begin{equation}
  N_{ij}^{k} = 
	\sum_r\frac{\mathcal{S}_{i,r} \mathcal{S}_{j,r}
            \mathcal{S}^\dagger_{k,r}}{\mathcal{S}_{\mathrm{vac},r}}\,,
\end{equation}
have to be admissible, i.e.\ all fusion coefficients must be non-negative
integers.  

To obtain the fusion rules from the $\mathcal{S}$-matrix (\ref{WD5_smat}) of
the $\mathcal{W}D_5(9,10)$ rational CFT we have to keep in mind that the
representations with $h\in\{\frac54,\frac1{16},\frac9{16}\}$ appear with
multiplicity two.  As a consequence, the corresponding representations should
be replaced by sums over their multiplicities, e.g.\ $\Phi_{5/4} =
\Phi_{5/4}^+ \oplus \Phi_{5/4}^-$, and the resulting identities need to be
disentangled in a consistent way.  In practice this amounts to enlarging the
$\mathcal{S}$-matrix by adding rows and columns for each representation
according to their true multiplicities.  Consistency requires that the
enlarged $\mathcal{S}$-Matrix, $\tilde{\mathcal{S}}$, is unitary and satisfies
$\tilde{\mathcal{S}}^2=\mathcal{C}$ with the conjugation matrix $\mathcal{C}$
such that $\mathcal{C}(\chi_h)=\chi_h$ for representations with multiplicity
one, while $\mathcal{C}(\chi_h^\pm) = \chi_h^\mp$ for
$h\in\{\frac54,\frac1{16},\frac9{16}\}$.  The resulting
$\tilde{\mathcal{S}}$-matrix gives rise to the following fusion rules:
\begin{align*}
  &\Phi_0 \otimes \Phi_h = \Phi_h\,, \\[0.5em]
  & \Phi_1 \otimes \Phi_1 = \Phi_0 \,,
  && \Phi_1 \otimes \Phi_{5/4}^\pm = \Phi_{5/4}^\mp \,,\\
  & \Phi_1 \otimes \Phi_{1/20} = \Phi_{1/20}\,,
  && \Phi_1 \otimes \Phi_{4/5} = \Phi_{4/5}\,,\\
  & \Phi_1 \otimes \Phi_{9/20} = \Phi_{9/20}\,, 
  && \Phi_1 \otimes \Phi_{1/5} = \Phi_{1/5}\,, \\
  & \Phi_1 \otimes \Phi_{1/16}^\pm = \Phi_{9/16}^\pm\,,
  && \Phi_1 \otimes \Phi_{9/16}^\pm = \Phi_{1/16}^\pm\,, \\[0.5em]
  & \Phi_{5/4}^\pm \otimes \Phi_{5/4}^\pm = \Phi_1 \,,
  && \Phi_{5/4}^\pm \otimes \Phi_{5/4}^\mp = \Phi_0 \,,\\
  & \Phi_{5/4}^\pm \otimes \Phi_{1/20} = \Phi_{4/5}\,,
  && \Phi_{5/4}^\pm \otimes \Phi_{4/5} = \Phi_{1/20}\,,\\
  & \Phi_{5/4}^\pm \otimes \Phi_{9/20} = \Phi_{1/5}\,,
  && \Phi_{5/4}^\pm \otimes \Phi_{1/5} = \Phi_{9/20}\,,\\
  & \Phi_{5/4}^\pm \otimes \Phi_{1/16}^\pm = \Phi_{9/16}^\mp\,,
  && \Phi_{5/4}^\pm \otimes \Phi_{1/16}^\mp = \Phi_{1/16}^\pm\,,\\
  & \Phi_{5/4}^\pm \otimes \Phi_{9/16}^\pm = \Phi_{1/16}^\mp\,,
  && \Phi_{5/4}^\pm \otimes \Phi_{9/16}^\mp = \Phi_{9/16}^\pm\,,\\[0.5em]
  & \Phi_{1/20} \otimes \Phi_{1/20} = \Phi_0 \oplus \Phi_1 \oplus \Phi_{1/5}\,,
  && \Phi_{1/20} \otimes \Phi_{4/5} = \Phi_{5/4}^+\oplus\Phi_{5/4}^-
                                  \oplus \Phi_{9/20} \,,\\
  & \Phi_{1/20} \otimes \Phi_{9/20} = \Phi_{1/5} \oplus \Phi_{4/5}  \,,
  && \Phi_{1/20} \otimes \Phi_{1/5} = \Phi_{1/20} \oplus \Phi_{9/20}\,, \\
  & \Phi_{1/20} \otimes \Phi_{1/16}^\pm = \Phi_{1/16}^\mp \oplus \Phi_{9/16}^\mp\,,  
  && \Phi_{1/20} \otimes \Phi_{9/16}^\pm = \Phi_{1/16}^\mp \oplus \Phi_{9/16}^\mp\,,
  \\[0.5em]
\end{align*}
\begin{align*}
  & \Phi_{4/5}  \otimes \Phi_{4/5} = \Phi_0 \oplus \Phi_1 \oplus \Phi_{1/5}\,,
  && \Phi_{4/5} \otimes \Phi_{9/20} = \Phi_{1/20} \oplus \Phi_{9/20}\,,\\
  & \Phi_{4/5}  \otimes \Phi_{1/5} = \Phi_{1/5} \oplus \Phi_{4/5}\,,
  && \Phi_{4/5}  \otimes \Phi_{9/16}^\pm  = \Phi_{1/16}^\pm \oplus \Phi_{9/16}^\pm \,,  \\
  & \Phi_{4/5}  \otimes \Phi_{1/16}^\pm  = \Phi_{1/16}^\pm \oplus \Phi_{9/16}^\pm\,, 
  \\[0.5em]
  & \Phi_{9/20} \otimes \Phi_{9/20}  = \Phi_0 \oplus \Phi_1 \oplus \Phi_{4/5} \,,
  && \Phi_{9/20} \otimes \Phi_{1/5}   = \Phi_{5/4}^+ \oplus\Phi_{5/4}^- 
                                     \oplus \Phi_{1/20} \,,\\
  & \Phi_{9/20} \otimes \Phi_{1/16}^\pm = \Phi_{1/16}^\mp \oplus \Phi_{9/16}^\mp\,, 
  && \Phi_{9/20} \otimes \Phi_{9/16}^\pm = \Phi_{1/16}^\mp \oplus \Phi_{9/16}^\mp\,,
  \\[0.5em]
  & \Phi_{1/5}  \otimes \Phi_{1/5} = \Phi_0 \oplus \Phi_1 \oplus \Phi_{4/5}\,,
  && \Phi_{1/5}  \otimes \Phi_{9/16}^\pm  = \Phi_{1/16}^\pm \oplus \Phi_{9/16}^\pm\,,\\
  & \Phi_{1/5}  \otimes \Phi_{1/16}^\pm  = \Phi_{1/16}^\pm \oplus \Phi_{9/16}^\pm\,,
  \\[0.5em]
  & \Phi_{1/16}^\pm \otimes \Phi_{1/16}^\pm  = \Phi_{5/4}^\pm
                           \oplus \Phi_{1/20} \oplus \Phi_{9/20}\,
  && \Phi_{1/16}^\pm \otimes \Phi_{1/16}^\mp  = \Phi_0 \oplus
                              \Phi_{1/5} \oplus \Phi_{4/5}\,,\\ 
  & \Phi_{1/16}^\pm \otimes \Phi_{9/16}^\pm  = \Phi_{5/4}^\mp \oplus 
                              \Phi_{1/20} \oplus \Phi_{9/20} \,,
  && \Phi_{1/16}^\pm \otimes \Phi_{9/16}^\mp  = \Phi_1 \oplus
                              \Phi_{1/5} \oplus \Phi_{4/5}\,,\\[0.5em]
  & \Phi_{9/16}^\pm \otimes \Phi_{9/16}^\pm  = \Phi_{5/4}^\pm \oplus 
                             \Phi_{1/20} \oplus \Phi_{9/20} \,,
  && \Phi_{9/16}^\pm \otimes \Phi_{9/16}^\mp = \Phi_0 \oplus
                             \Phi_{1/5} \oplus \Phi_{4/5}\,.
\end{align*}


\begin{thebibliography}{68}%
\makeatletter
\providecommand \@ifxundefined [1]{%
 \@ifx{#1\undefined}
}%
\providecommand \@ifnum [1]{%
 \ifnum #1\expandafter \@firstoftwo
 \else \expandafter \@secondoftwo
 \fi
}%
\providecommand \@ifx [1]{%
 \ifx #1\expandafter \@firstoftwo
 \else \expandafter \@secondoftwo
 \fi
}%
\providecommand \natexlab [1]{#1}%
\providecommand \enquote  [1]{``#1''}%
\providecommand \bibnamefont  [1]{#1}%
\providecommand \bibfnamefont [1]{#1}%
\providecommand \citenamefont [1]{#1}%
\providecommand \href@noop [0]{\@secondoftwo}%
\providecommand \href [0]{\begingroup \@sanitize@url \@href}%
\providecommand \@href[1]{\@@startlink{#1}\@@href}%
\providecommand \@@href[1]{\endgroup#1\@@endlink}%
\providecommand \@sanitize@url [0]{\catcode `\\12\catcode `\$12\catcode
  `\&12\catcode `\#12\catcode `\^12\catcode `\_12\catcode `\%12\relax}%
\providecommand \@@startlink[1]{}%
\providecommand \@@endlink[0]{}%
\providecommand \url  [0]{\begingroup\@sanitize@url \@url }%
\providecommand \@url [1]{\endgroup\@href {#1}{\urlprefix }}%
\providecommand \urlprefix  [0]{URL }%
\providecommand \Eprint [0]{\href }%
\providecommand \doibase [0]{http://dx.doi.org/}%
\providecommand \selectlanguage [0]{\@gobble}%
\providecommand \bibinfo  [0]{\@secondoftwo}%
\providecommand \bibfield  [0]{\@secondoftwo}%
\providecommand \translation [1]{[#1]}%
\providecommand \BibitemOpen [0]{}%
\providecommand \bibitemStop [0]{}%
\providecommand \bibitemNoStop [0]{.\EOS\space}%
\providecommand \EOS [0]{\spacefactor3000\relax}%
\providecommand \BibitemShut  [1]{\csname bibitem#1\endcsname}%
\let\auto@bib@innerbib\@empty
\bibitem [{\citenamefont {Bethe}(1931)}]{Bethe31}%
  \BibitemOpen
  \bibfield  {author} {\bibinfo {author} {\bibfnamefont {H.}~\bibnamefont
  {Bethe}},\ }\href@noop {} {\bibfield  {journal} {\bibinfo  {journal} {Z.
  Phys.}\ }\textbf {\bibinfo {volume} {71}},\ \bibinfo {pages} {205} (\bibinfo
  {year} {1931})}\BibitemShut {NoStop}%
\bibitem [{\citenamefont {Baxter}(1982)}]{Baxter:book}%
  \BibitemOpen
  \bibfield  {author} {\bibinfo {author} {\bibfnamefont {R.~J.}\ \bibnamefont
  {Baxter}},\ }\href@noop {} {\emph {\bibinfo {title} {{E}xactly {S}olved
  {M}odels in {S}tatistical {M}echanics}}}\ (\bibinfo  {publisher} {Academic
  Press},\ \bibinfo {address} {London},\ \bibinfo {year} {1982})\BibitemShut
  {NoStop}%
\bibitem [{\citenamefont {Korepin}\ \emph {et~al.}(1993)\citenamefont
  {Korepin}, \citenamefont {Bogoliubov},\ and\ \citenamefont
  {Izergin}}]{VladB}%
  \BibitemOpen
  \bibfield  {author} {\bibinfo {author} {\bibfnamefont {V.~E.}\ \bibnamefont
  {Korepin}}, \bibinfo {author} {\bibfnamefont {N.~M.}\ \bibnamefont
  {Bogoliubov}}, \ and\ \bibinfo {author} {\bibfnamefont {A.~G.}\ \bibnamefont
  {Izergin}},\ }\href@noop {} {\emph {\bibinfo {title} {{Quantum Inverse
  Scattering Method and Correlation Functions}}}}\ (\bibinfo  {publisher}
  {Cambridge University Press},\ \bibinfo {address} {Cambridge},\ \bibinfo
  {year} {1993})\BibitemShut {NoStop}%
\bibitem [{\citenamefont {Essler}\ \emph {et~al.}(2005)\citenamefont {Essler},
  \citenamefont {Frahm}, \citenamefont {G{\"o}hmann}, \citenamefont
  {Kl{\"u}mper},\ and\ \citenamefont {Korepin}}]{HUBBARD}%
  \BibitemOpen
  \bibfield  {author} {\bibinfo {author} {\bibfnamefont {F.~H.~L.}\
  \bibnamefont {Essler}}, \bibinfo {author} {\bibfnamefont {H.}~\bibnamefont
  {Frahm}}, \bibinfo {author} {\bibfnamefont {F.}~\bibnamefont {G{\"o}hmann}},
  \bibinfo {author} {\bibfnamefont {A.}~\bibnamefont {Kl{\"u}mper}}, \ and\
  \bibinfo {author} {\bibfnamefont {V.~E.}\ \bibnamefont {Korepin}},\ }\href
  {\doibase 10.1017/CBO9780511534843} {\emph {\bibinfo {title} {{T}he
  {O}ne-{D}imensional {Hubbard} {M}odel}}}\ (\bibinfo  {publisher} {Cambridge
  University Press},\ \bibinfo {address} {Cambridge (UK)},\ \bibinfo {year}
  {2005})\BibitemShut {NoStop}%
\bibitem [{\citenamefont {Laughlin}(1983)}]{Laug83}%
  \BibitemOpen
  \bibfield  {author} {\bibinfo {author} {\bibfnamefont {R.~B.}\ \bibnamefont
  {Laughlin}},\ }\href@noop {} {\bibfield  {journal} {\bibinfo  {journal}
  {Phys. Rev. Lett.}\ }\textbf {\bibinfo {volume} {50}},\ \bibinfo {pages}
  {1395} (\bibinfo {year} {1983})}\BibitemShut {NoStop}%
\bibitem [{\citenamefont {Moessner}\ and\ \citenamefont
  {Sondhi}(2001)}]{MoSo01}%
  \BibitemOpen
  \bibfield  {author} {\bibinfo {author} {\bibfnamefont {R.}~\bibnamefont
  {Moessner}}\ and\ \bibinfo {author} {\bibfnamefont {S.~L.}\ \bibnamefont
  {Sondhi}},\ }\href@noop {} {\bibfield  {journal} {\bibinfo  {journal} {Phys.
  Rev. Lett.}\ }\textbf {\bibinfo {volume} {86}},\ \bibinfo {pages} {1881}
  (\bibinfo {year} {2001})},\ \Eprint {http://arxiv.org/abs/cond-mat/0007378}
  {cond-mat/0007378} \BibitemShut {NoStop}%
\bibitem [{\citenamefont {Balents}\ \emph {et~al.}(2002)\citenamefont
  {Balents}, \citenamefont {Fisher},\ and\ \citenamefont {Girvin}}]{BaFG02}%
  \BibitemOpen
  \bibfield  {author} {\bibinfo {author} {\bibfnamefont {L.}~\bibnamefont
  {Balents}}, \bibinfo {author} {\bibfnamefont {M.~P.~A.}\ \bibnamefont
  {Fisher}}, \ and\ \bibinfo {author} {\bibfnamefont {S.~M.}\ \bibnamefont
  {Girvin}},\ }\href@noop {} {\bibfield  {journal} {\bibinfo  {journal} {Phys.
  Rev. B}\ }\textbf {\bibinfo {volume} {65}},\ \bibinfo {pages} {224412}
  (\bibinfo {year} {2002})},\ \Eprint {http://arxiv.org/abs/cond-mat/0110005}
  {cond-mat/0110005} \BibitemShut {NoStop}%
\bibitem [{\citenamefont {Kitaev}(2006)}]{Kita06}%
  \BibitemOpen
  \bibfield  {author} {\bibinfo {author} {\bibfnamefont {A.}~\bibnamefont
  {Kitaev}},\ }\href@noop {} {\bibfield  {journal} {\bibinfo  {journal} {Ann.
  Phys. (NY)}\ }\textbf {\bibinfo {volume} {321}},\ \bibinfo {pages} {2}
  (\bibinfo {year} {2006})},\ \Eprint {http://arxiv.org/abs/cond-mat/0506438}
  {cond-mat/0506438} \BibitemShut {NoStop}%
\bibitem [{\citenamefont {B{\'e}ri}\ and\ \citenamefont
  {Cooper}(2012)}]{BeCo12}%
  \BibitemOpen
  \bibfield  {author} {\bibinfo {author} {\bibfnamefont {B.}~\bibnamefont
  {B{\'e}ri}}\ and\ \bibinfo {author} {\bibfnamefont {N.~R.}\ \bibnamefont
  {Cooper}},\ }\href@noop {} {\bibfield  {journal} {\bibinfo  {journal} {Phys.
  Rev. Lett.}\ }\textbf {\bibinfo {volume} {109}},\ \bibinfo {pages} {156803}
  (\bibinfo {year} {2012})},\ \Eprint {http://arxiv.org/abs/1206.2224}
  {arXiv:1206.2224} \BibitemShut {NoStop}%
\bibitem [{\citenamefont {Altland}\ \emph {et~al.}(2013)\citenamefont
  {Altland}, \citenamefont {B{\'e}ri}, \citenamefont {Egger},\ and\
  \citenamefont {Tsvelik}}]{ABET13}%
  \BibitemOpen
  \bibfield  {author} {\bibinfo {author} {\bibfnamefont {A.}~\bibnamefont
  {Altland}}, \bibinfo {author} {\bibfnamefont {B.}~\bibnamefont {B{\'e}ri}},
  \bibinfo {author} {\bibfnamefont {R.}~\bibnamefont {Egger}}, \ and\ \bibinfo
  {author} {\bibfnamefont {A.~M.}\ \bibnamefont {Tsvelik}},\ }\href@noop {}
  {\bibfield  {journal} {\bibinfo  {journal} {preprint}\ } (\bibinfo {year}
  {2013})},\ \Eprint {http://arxiv.org/abs/1312.3802} {arXiv:1312.3802}
  \BibitemShut {NoStop}%
\bibitem [{\citenamefont {Altland}\ \emph {et~al.}(2014)\citenamefont
  {Altland}, \citenamefont {B{\'e}ri}, \citenamefont {Egger},\ and\
  \citenamefont {Tsvelik}}]{ABET14}%
  \BibitemOpen
  \bibfield  {author} {\bibinfo {author} {\bibfnamefont {A.}~\bibnamefont
  {Altland}}, \bibinfo {author} {\bibfnamefont {B.}~\bibnamefont {B{\'e}ri}},
  \bibinfo {author} {\bibfnamefont {R.}~\bibnamefont {Egger}}, \ and\ \bibinfo
  {author} {\bibfnamefont {A.~M.}\ \bibnamefont {Tsvelik}},\ }\href@noop {}
  {\bibfield  {journal} {\bibinfo  {journal} {J. Phys. A}\ }\textbf {\bibinfo
  {volume} {47}},\ \bibinfo {pages} {265001} (\bibinfo {year} {2014})},\
  \Eprint {http://arxiv.org/abs/1403.0113} {arXiv:1403.0113} \BibitemShut
  {NoStop}%
\bibitem [{\citenamefont {Kitaev}(2003)}]{Kita03}%
  \BibitemOpen
  \bibfield  {author} {\bibinfo {author} {\bibfnamefont {A.~{\relax Yu}.}\
  \bibnamefont {Kitaev}},\ }\href@noop {} {\bibfield  {journal} {\bibinfo
  {journal} {Ann. Phys. (NY)}\ }\textbf {\bibinfo {volume} {303}},\ \bibinfo
  {pages} {2} (\bibinfo {year} {2003})},\ \Eprint
  {http://arxiv.org/abs/quant-ph/9707021} {quant-ph/9707021} \BibitemShut
  {NoStop}%
\bibitem [{\citenamefont {Nayak}\ \emph {et~al.}(2008)\citenamefont {Nayak},
  \citenamefont {Simon}, \citenamefont {Stern}, \citenamefont {Freedman},\ and\
  \citenamefont {Sarma}}]{NSSF08}%
  \BibitemOpen
  \bibfield  {author} {\bibinfo {author} {\bibfnamefont {C.}~\bibnamefont
  {Nayak}}, \bibinfo {author} {\bibfnamefont {S.~H.}\ \bibnamefont {Simon}},
  \bibinfo {author} {\bibfnamefont {A.}~\bibnamefont {Stern}}, \bibinfo
  {author} {\bibfnamefont {M.}~\bibnamefont {Freedman}}, \ and\ \bibinfo
  {author} {\bibfnamefont {S.~D.}\ \bibnamefont {Sarma}},\ }\href@noop {}
  {\bibfield  {journal} {\bibinfo  {journal} {Rev. Mod. Phys.}\ }\textbf
  {\bibinfo {volume} {80}},\ \bibinfo {pages} {1083} (\bibinfo {year}
  {2008})},\ \Eprint {http://arxiv.org/abs/0707.1889} {arXiv:0707.1889}
  \BibitemShut {NoStop}%
\bibitem [{\citenamefont {Andrews}\ \emph {et~al.}(1984)\citenamefont
  {Andrews}, \citenamefont {Baxter},\ and\ \citenamefont {Forrester}}]{AnBF84}%
  \BibitemOpen
  \bibfield  {author} {\bibinfo {author} {\bibfnamefont {G.~E.}\ \bibnamefont
  {Andrews}}, \bibinfo {author} {\bibfnamefont {R.~J.}\ \bibnamefont {Baxter}},
  \ and\ \bibinfo {author} {\bibfnamefont {P.~J.}\ \bibnamefont {Forrester}},\
  }\href@noop {} {\bibfield  {journal} {\bibinfo  {journal} {J. Stat. Phys.}\
  }\textbf {\bibinfo {volume} {35}},\ \bibinfo {pages} {193} (\bibinfo {year}
  {1984})}\BibitemShut {NoStop}%
\bibitem [{\citenamefont {Friedan}\ \emph {et~al.}(1984)\citenamefont
  {Friedan}, \citenamefont {Qiu},\ and\ \citenamefont {Shenker}}]{FrQS84}%
  \BibitemOpen
  \bibfield  {author} {\bibinfo {author} {\bibfnamefont {D.}~\bibnamefont
  {Friedan}}, \bibinfo {author} {\bibfnamefont {Z.}~\bibnamefont {Qiu}}, \ and\
  \bibinfo {author} {\bibfnamefont {S.}~\bibnamefont {Shenker}},\ }\href@noop
  {} {\bibfield  {journal} {\bibinfo  {journal} {Phys. Rev. Lett.}\ }\textbf
  {\bibinfo {volume} {52}},\ \bibinfo {pages} {1575} (\bibinfo {year}
  {1984})}\BibitemShut {NoStop}%
\bibitem [{\citenamefont {Huse}(1984)}]{Huse84}%
  \BibitemOpen
  \bibfield  {author} {\bibinfo {author} {\bibfnamefont {D.~A.}\ \bibnamefont
  {Huse}},\ }\href@noop {} {\bibfield  {journal} {\bibinfo  {journal} {Phys.
  Rev. B}\ }\textbf {\bibinfo {volume} {30}},\ \bibinfo {pages} {3908}
  (\bibinfo {year} {1984})}\BibitemShut {NoStop}%
\bibitem [{\citenamefont {Feiguin}\ \emph {et~al.}(2007)\citenamefont
  {Feiguin}, \citenamefont {Trebst}, \citenamefont {Ludwig}, \citenamefont
  {Troyer}, \citenamefont {Kitaev}, \citenamefont {Wang},\ and\ \citenamefont
  {Freedman}}]{FTLT07}%
  \BibitemOpen
  \bibfield  {author} {\bibinfo {author} {\bibfnamefont {A.}~\bibnamefont
  {Feiguin}}, \bibinfo {author} {\bibfnamefont {S.}~\bibnamefont {Trebst}},
  \bibinfo {author} {\bibfnamefont {A.~W.~W.}\ \bibnamefont {Ludwig}}, \bibinfo
  {author} {\bibfnamefont {M.}~\bibnamefont {Troyer}}, \bibinfo {author}
  {\bibfnamefont {A.}~\bibnamefont {Kitaev}}, \bibinfo {author} {\bibfnamefont
  {Z.}~\bibnamefont {Wang}}, \ and\ \bibinfo {author} {\bibfnamefont {M.~H.}\
  \bibnamefont {Freedman}},\ }\href@noop {} {\bibfield  {journal} {\bibinfo
  {journal} {Phys. Rev. Lett.}\ }\textbf {\bibinfo {volume} {98}},\ \bibinfo
  {pages} {160409} (\bibinfo {year} {2007})},\ \Eprint
  {http://arxiv.org/abs/cond-mat/0612341} {cond-mat/0612341} \BibitemShut
  {NoStop}%
\bibitem [{\citenamefont {Trebst}\ \emph {et~al.}(2008)\citenamefont {Trebst},
  \citenamefont {Ardonne}, \citenamefont {Feiguin}, \citenamefont {Huse},
  \citenamefont {Ludwig},\ and\ \citenamefont {Troyer}}]{TAFH08}%
  \BibitemOpen
  \bibfield  {author} {\bibinfo {author} {\bibfnamefont {S.}~\bibnamefont
  {Trebst}}, \bibinfo {author} {\bibfnamefont {E.}~\bibnamefont {Ardonne}},
  \bibinfo {author} {\bibfnamefont {A.}~\bibnamefont {Feiguin}}, \bibinfo
  {author} {\bibfnamefont {D.~A.}\ \bibnamefont {Huse}}, \bibinfo {author}
  {\bibfnamefont {A.~W.~W.}\ \bibnamefont {Ludwig}}, \ and\ \bibinfo {author}
  {\bibfnamefont {M.}~\bibnamefont {Troyer}},\ }\href@noop {} {\bibfield
  {journal} {\bibinfo  {journal} {Phys. Rev. Lett.}\ }\textbf {\bibinfo
  {volume} {101}},\ \bibinfo {pages} {050401} (\bibinfo {year} {2008})},\
  \Eprint {http://arxiv.org/abs/0801.4602} {arXiv:0801.4602} \BibitemShut
  {NoStop}%
\bibitem [{\citenamefont {Gils}\ \emph {et~al.}(2009)\citenamefont {Gils},
  \citenamefont {Ardonne}, \citenamefont {Trebst}, \citenamefont {Ludwig},
  \citenamefont {Troyer},\ and\ \citenamefont {Wang}}]{GATL09}%
  \BibitemOpen
  \bibfield  {author} {\bibinfo {author} {\bibfnamefont {C.}~\bibnamefont
  {Gils}}, \bibinfo {author} {\bibfnamefont {E.}~\bibnamefont {Ardonne}},
  \bibinfo {author} {\bibfnamefont {S.}~\bibnamefont {Trebst}}, \bibinfo
  {author} {\bibfnamefont {A.~W.~W.}\ \bibnamefont {Ludwig}}, \bibinfo {author}
  {\bibfnamefont {M.}~\bibnamefont {Troyer}}, \ and\ \bibinfo {author}
  {\bibfnamefont {Z.}~\bibnamefont {Wang}},\ }\href@noop {} {\bibfield
  {journal} {\bibinfo  {journal} {Phys. Rev. Lett.}\ }\textbf {\bibinfo
  {volume} {103}},\ \bibinfo {pages} {070401} (\bibinfo {year} {2009})},\
  \Eprint {http://arxiv.org/abs/0810.2277} {arXiv:0810.2277} \BibitemShut
  {NoStop}%
\bibitem [{\citenamefont {Gils}\ \emph {et~al.}(2013)\citenamefont {Gils},
  \citenamefont {Ardonne}, \citenamefont {Trebst}, \citenamefont {Huse},
  \citenamefont {Ludwig}, \citenamefont {Troyer},\ and\ \citenamefont
  {Wang}}]{GATH13}%
  \BibitemOpen
  \bibfield  {author} {\bibinfo {author} {\bibfnamefont {C.}~\bibnamefont
  {Gils}}, \bibinfo {author} {\bibfnamefont {E.}~\bibnamefont {Ardonne}},
  \bibinfo {author} {\bibfnamefont {S.}~\bibnamefont {Trebst}}, \bibinfo
  {author} {\bibfnamefont {D.~A.}\ \bibnamefont {Huse}}, \bibinfo {author}
  {\bibfnamefont {A.~W.~W.}\ \bibnamefont {Ludwig}}, \bibinfo {author}
  {\bibfnamefont {M.}~\bibnamefont {Troyer}}, \ and\ \bibinfo {author}
  {\bibfnamefont {Z.}~\bibnamefont {Wang}},\ }\href@noop {} {\bibfield
  {journal} {\bibinfo  {journal} {Phys. Rev. B}\ }\textbf {\bibinfo {volume}
  {87}},\ \bibinfo {pages} {235120} (\bibinfo {year} {2013})},\ \Eprint
  {http://arxiv.org/abs/1303.4290} {arXiv:1303.4290} \BibitemShut {NoStop}%
\bibitem [{\citenamefont {Ludwig}\ \emph {et~al.}(2011)\citenamefont {Ludwig},
  \citenamefont {Poilblanc}, \citenamefont {Trebst},\ and\ \citenamefont
  {Troyer}}]{LPTT11}%
  \BibitemOpen
  \bibfield  {author} {\bibinfo {author} {\bibfnamefont {A.~W.~W.}\
  \bibnamefont {Ludwig}}, \bibinfo {author} {\bibfnamefont {D.}~\bibnamefont
  {Poilblanc}}, \bibinfo {author} {\bibfnamefont {S.}~\bibnamefont {Trebst}}, \
  and\ \bibinfo {author} {\bibfnamefont {M.}~\bibnamefont {Troyer}},\
  }\href@noop {} {\bibfield  {journal} {\bibinfo  {journal} {New J.Phys.}\
  }\textbf {\bibinfo {volume} {13}},\ \bibinfo {pages} {045014} (\bibinfo
  {year} {2011})},\ \Eprint {http://arxiv.org/abs/1003.3453} {arXiv:1003.3453}
  \BibitemShut {NoStop}%
\bibitem [{\citenamefont {Pfeifer}\ \emph {et~al.}(2012)\citenamefont
  {Pfeifer}, \citenamefont {Buerschaper}, \citenamefont {Trebst}, \citenamefont
  {Ludwig}, \citenamefont {Troyer},\ and\ \citenamefont {Vidal}}]{PBTL12}%
  \BibitemOpen
  \bibfield  {author} {\bibinfo {author} {\bibfnamefont {R.~N.~C.}\
  \bibnamefont {Pfeifer}}, \bibinfo {author} {\bibfnamefont {O.}~\bibnamefont
  {Buerschaper}}, \bibinfo {author} {\bibfnamefont {S.}~\bibnamefont {Trebst}},
  \bibinfo {author} {\bibfnamefont {A.~W.~W.}\ \bibnamefont {Ludwig}}, \bibinfo
  {author} {\bibfnamefont {M.}~\bibnamefont {Troyer}}, \ and\ \bibinfo {author}
  {\bibfnamefont {G.}~\bibnamefont {Vidal}},\ }\href@noop {} {\bibfield
  {journal} {\bibinfo  {journal} {Phys. Rev. B}\ }\textbf {\bibinfo {volume}
  {86}},\ \bibinfo {pages} {155111} (\bibinfo {year} {2012})},\ \Eprint
  {http://arxiv.org/abs/1005.5486} {arXiv:1005.5486} \BibitemShut {NoStop}%
\bibitem [{\citenamefont {Date}\ \emph {et~al.}(1986)\citenamefont {Date},
  \citenamefont {Jimbo}, \citenamefont {Miwa},\ and\ \citenamefont
  {Okado}}]{DJMO86}%
  \BibitemOpen
  \bibfield  {author} {\bibinfo {author} {\bibfnamefont {E.}~\bibnamefont
  {Date}}, \bibinfo {author} {\bibfnamefont {M.}~\bibnamefont {Jimbo}},
  \bibinfo {author} {\bibfnamefont {T.}~\bibnamefont {Miwa}}, \ and\ \bibinfo
  {author} {\bibfnamefont {M.}~\bibnamefont {Okado}},\ }\href@noop {}
  {\bibfield  {journal} {\bibinfo  {journal} {Lett. Math. Phys.}\ }\textbf
  {\bibinfo {volume} {12}},\ \bibinfo {pages} {209} (\bibinfo {year}
  {1986})}\BibitemShut {NoStop}%
\bibitem [{\citenamefont {Pasquier}(1988)}]{Pasq88}%
  \BibitemOpen
  \bibfield  {author} {\bibinfo {author} {\bibfnamefont {V.}~\bibnamefont
  {Pasquier}},\ }\href@noop {} {\bibfield  {journal} {\bibinfo  {journal}
  {Comm. Math. Phys.}\ }\textbf {\bibinfo {volume} {118}},\ \bibinfo {pages}
  {355} (\bibinfo {year} {1988})}\BibitemShut {NoStop}%
\bibitem [{\citenamefont {Roche}(1990)}]{Roch90}%
  \BibitemOpen
  \bibfield  {author} {\bibinfo {author} {\bibfnamefont {{\relax
  Ph}.}~\bibnamefont {Roche}},\ }\href@noop {} {\bibfield  {journal} {\bibinfo
  {journal} {Comm. Math. Phys.}\ }\textbf {\bibinfo {volume} {127}},\ \bibinfo
  {pages} {395} (\bibinfo {year} {1990})}\BibitemShut {NoStop}%
\bibitem [{\citenamefont {Finch}(2013)}]{Finch13}%
  \BibitemOpen
  \bibfield  {author} {\bibinfo {author} {\bibfnamefont {P.~E.}\ \bibnamefont
  {Finch}},\ }\href@noop {} {\bibfield  {journal} {\bibinfo  {journal} {J.
  Phys. A}\ }\textbf {\bibinfo {volume} {46}},\ \bibinfo {pages} {055305}
  (\bibinfo {year} {2013})},\ \Eprint {http://arxiv.org/abs/1201.4470}
  {arXiv:1201.4470} \BibitemShut {NoStop}%
\bibitem [{\citenamefont {Finch}\ and\ \citenamefont {Frahm}(2013)}]{FiFr13}%
  \BibitemOpen
  \bibfield  {author} {\bibinfo {author} {\bibfnamefont {P.~E.}\ \bibnamefont
  {Finch}}\ and\ \bibinfo {author} {\bibfnamefont {H.}~\bibnamefont {Frahm}},\
  }\href@noop {} {\bibfield  {journal} {\bibinfo  {journal} {New J. Phys.}\
  }\textbf {\bibinfo {volume} {15}},\ \bibinfo {pages} {053035} (\bibinfo
  {year} {2013})},\ \Eprint {http://arxiv.org/abs/1211.4449} {arXiv:1211.4449}
  \BibitemShut {NoStop}%
\bibitem [{\citenamefont {Pasquier}(1987)}]{Pasq87a}%
  \BibitemOpen
  \bibfield  {author} {\bibinfo {author} {\bibfnamefont {V.}~\bibnamefont
  {Pasquier}},\ }\href@noop {} {\bibfield  {journal} {\bibinfo  {journal}
  {Nucl. Phys. B}\ }\textbf {\bibinfo {volume} {285}},\ \bibinfo {pages} {162}
  (\bibinfo {year} {1987})}\BibitemShut {NoStop}%
\bibitem [{\citenamefont {Gepner}(1993)}]{Gepn93a}%
  \BibitemOpen
  \bibfield  {author} {\bibinfo {author} {\bibfnamefont {D.}~\bibnamefont
  {Gepner}},\ }\href@noop {} {\bibfield  {journal} {\bibinfo  {journal}
  {preprint}\ } (\bibinfo {year} {1993})},\ \Eprint
  {http://arxiv.org/abs/hep-th/9306143} {hep-th/9306143} \BibitemShut {NoStop}%
\bibitem [{\citenamefont {Bonderson}(2007)}]{Bonderson07}%
  \BibitemOpen
  \bibfield  {author} {\bibinfo {author} {\bibfnamefont {P.~H.}\ \bibnamefont
  {Bonderson}},\ }\emph {\bibinfo {title} {{N}on-{A}belian {A}nyons and
  {I}nterferometry}},\ \href@noop {} {Ph.D. thesis},\ \bibinfo  {school}
  {California Institute of Technology} (\bibinfo {year} {2007})\BibitemShut
  {NoStop}%
\bibitem [{\citenamefont {Preskill}(2004)}]{Preskill04}%
  \BibitemOpen
  \bibfield  {author} {\bibinfo {author} {\bibfnamefont {J.}~\bibnamefont
  {Preskill}},\ }\href
  {http://www.theory.caltech.edu/~preskill/ph219/topological.pdf} {\enquote
  {\bibinfo {title} {{T}opological {Q}uantum {C}omputation},}\ } (\bibinfo
  {year} {2004}),\ \bibinfo {note} {part of Lecture Notes to Physics
  219/Computer Science 219 "Quantum Computation"}\BibitemShut {NoStop}%
\bibitem [{\citenamefont {Moore}\ and\ \citenamefont {Seiberg}(1990)}]{MoSe89}%
  \BibitemOpen
  \bibfield  {author} {\bibinfo {author} {\bibfnamefont {G.}~\bibnamefont
  {Moore}}\ and\ \bibinfo {author} {\bibfnamefont {N.}~\bibnamefont
  {Seiberg}},\ }in\ \href
  {http://www.physics.rutgers.edu/~gmoore/LecturesRCFT.pdf} {\emph {\bibinfo
  {booktitle} {Physics, Geometry and Topology}}},\ \bibinfo {series} {NATO ASI
  Series.}, Vol.\ \bibinfo {volume} {238},\ \bibinfo {editor} {edited by\
  \bibinfo {editor} {\bibfnamefont {H.~C.}\ \bibnamefont {Lee}}}\ (\bibinfo
  {year} {1990})\ pp.\ \bibinfo {pages} {263--361}\BibitemShut {NoStop}%
\bibitem [{\citenamefont {Finch}\ \emph {et~al.}(2014)\citenamefont {Finch},
  \citenamefont {Frahm}, \citenamefont {Lewerenz}, \citenamefont {Milsted},\
  and\ \citenamefont {Osborne}}]{Finch.etal14}%
  \BibitemOpen
  \bibfield  {author} {\bibinfo {author} {\bibfnamefont {P.~E.}\ \bibnamefont
  {Finch}}, \bibinfo {author} {\bibfnamefont {H.}~\bibnamefont {Frahm}},
  \bibinfo {author} {\bibfnamefont {M.}~\bibnamefont {Lewerenz}}, \bibinfo
  {author} {\bibfnamefont {A.}~\bibnamefont {Milsted}}, \ and\ \bibinfo
  {author} {\bibfnamefont {T.~J.}\ \bibnamefont {Osborne}},\ }\href@noop {}
  {\bibfield  {journal} {\bibinfo  {journal} {Phys. Rev. B}\ }\textbf {\bibinfo
  {volume} {90}},\ \bibinfo {pages} {081111(R)} (\bibinfo {year} {2014})},\
  \Eprint {http://arxiv.org/abs/1404.2439} {arXiv:1404.2439} \BibitemShut
  {NoStop}%
\bibitem [{\citenamefont {Warnaar}\ and\ \citenamefont
  {Nienhuis}(1993)}]{WaNi93}%
  \BibitemOpen
  \bibfield  {author} {\bibinfo {author} {\bibfnamefont {S.~O.}\ \bibnamefont
  {Warnaar}}\ and\ \bibinfo {author} {\bibfnamefont {B.}~\bibnamefont
  {Nienhuis}},\ }\href@noop {} {\bibfield  {journal} {\bibinfo  {journal} {J.
  Phys. A}\ }\textbf {\bibinfo {volume} {26}},\ \bibinfo {pages} {2301}
  (\bibinfo {year} {1993})},\ \Eprint {http://arxiv.org/abs/hep-th/9301026}
  {hep-th/9301026} \BibitemShut {NoStop}%
\bibitem [{\citenamefont {Birman}\ and\ \citenamefont {Wenzl}(1989)}]{BiWe89}%
  \BibitemOpen
  \bibfield  {author} {\bibinfo {author} {\bibfnamefont {J.~S.}\ \bibnamefont
  {Birman}}\ and\ \bibinfo {author} {\bibfnamefont {H.}~\bibnamefont {Wenzl}},\
  }\href@noop {} {\bibfield  {journal} {\bibinfo  {journal} {Trans. AMS}\
  }\textbf {\bibinfo {volume} {313}},\ \bibinfo {pages} {249} (\bibinfo {year}
  {1989})}\BibitemShut {NoStop}%
\bibitem [{\citenamefont {Murakami}(1987)}]{Mura87}%
  \BibitemOpen
  \bibfield  {author} {\bibinfo {author} {\bibfnamefont {J.}~\bibnamefont
  {Murakami}},\ }\href@noop {} {\bibfield  {journal} {\bibinfo  {journal}
  {Osaka J. Math.}\ }\textbf {\bibinfo {volume} {24}},\ \bibinfo {pages} {745}
  (\bibinfo {year} {1987})}\BibitemShut {NoStop}%
\bibitem [{\citenamefont {Temperley}\ and\ \citenamefont
  {Lieb}(1971)}]{TeLi71}%
  \BibitemOpen
  \bibfield  {author} {\bibinfo {author} {\bibfnamefont {H.~N.~V.}\
  \bibnamefont {Temperley}}\ and\ \bibinfo {author} {\bibfnamefont {E.~H.}\
  \bibnamefont {Lieb}},\ }\href@noop {} {\bibfield  {journal} {\bibinfo
  {journal} {Proc. R. Soc. Lond. A}\ }\textbf {\bibinfo {volume} {332}},\
  \bibinfo {pages} {251} (\bibinfo {year} {1971})}\BibitemShut {NoStop}%
\bibitem [{\citenamefont {Cheng}\ \emph {et~al.}(1991)\citenamefont {Cheng},
  \citenamefont {Ge},\ and\ \citenamefont {Xue}}]{ChGx91}%
  \BibitemOpen
  \bibfield  {author} {\bibinfo {author} {\bibfnamefont {Y.}~\bibnamefont
  {Cheng}}, \bibinfo {author} {\bibfnamefont {M.~L.}\ \bibnamefont {Ge}}, \
  and\ \bibinfo {author} {\bibfnamefont {K.}~\bibnamefont {Xue}},\ }\href@noop
  {} {\bibfield  {journal} {\bibinfo  {journal} {Comm. Math. Phys.}\ }\textbf
  {\bibinfo {volume} {136}},\ \bibinfo {pages} {195} (\bibinfo {year}
  {1991})}\BibitemShut {NoStop}%
\bibitem [{\citenamefont {Cheng}\ \emph {et~al.}(1992)\citenamefont {Cheng},
  \citenamefont {Ge}, \citenamefont {Li},\ and\ \citenamefont {Xue}}]{CGLX92}%
  \BibitemOpen
  \bibfield  {author} {\bibinfo {author} {\bibfnamefont {Y.}~\bibnamefont
  {Cheng}}, \bibinfo {author} {\bibfnamefont {M.-L.}\ \bibnamefont {Ge}},
  \bibinfo {author} {\bibfnamefont {G.~C.}\ \bibnamefont {Li}}, \ and\ \bibinfo
  {author} {\bibfnamefont {K.}~\bibnamefont {Xue}},\ }\href@noop {} {\bibfield
  {journal} {\bibinfo  {journal} {J. Knot Theory Ramifications}\ }\textbf
  {\bibinfo {volume} {01}},\ \bibinfo {pages} {31} (\bibinfo {year}
  {1992})}\BibitemShut {NoStop}%
\bibitem [{\citenamefont {Grimm}(1994)}]{Grim94}%
  \BibitemOpen
  \bibfield  {author} {\bibinfo {author} {\bibfnamefont {U.}~\bibnamefont
  {Grimm}},\ }\href@noop {} {\bibfield  {journal} {\bibinfo  {journal} {J.
  Phys. A}\ }\textbf {\bibinfo {volume} {27}},\ \bibinfo {pages} {5897}
  (\bibinfo {year} {1994})},\ \Eprint {http://arxiv.org/abs/hep-th/9402076}
  {hep-th/9402076} \BibitemShut {NoStop}%
\bibitem [{\citenamefont {Owczarek}\ and\ \citenamefont
  {Baxter}(1987)}]{OwBa87}%
  \BibitemOpen
  \bibfield  {author} {\bibinfo {author} {\bibfnamefont {A.~L.}\ \bibnamefont
  {Owczarek}}\ and\ \bibinfo {author} {\bibfnamefont {R.~J.}\ \bibnamefont
  {Baxter}},\ }\href@noop {} {\bibfield  {journal} {\bibinfo  {journal} {J.
  Stat. Phys.}\ }\textbf {\bibinfo {volume} {49}},\ \bibinfo {pages} {1093}
  (\bibinfo {year} {1987})}\BibitemShut {NoStop}%
\bibitem [{\citenamefont {Aufgebauer}\ and\ \citenamefont
  {Kl{\"u}mper}(2010)}]{AuKl10}%
  \BibitemOpen
  \bibfield  {author} {\bibinfo {author} {\bibfnamefont {B.}~\bibnamefont
  {Aufgebauer}}\ and\ \bibinfo {author} {\bibfnamefont {A.}~\bibnamefont
  {Kl{\"u}mper}},\ }\href@noop {} {\bibfield  {journal} {\bibinfo  {journal}
  {J. Stat. Mech.}\ ,\ \bibinfo {pages} {P05018}} (\bibinfo {year} {2010})},\
  \Eprint {http://arxiv.org/abs/1003.1932} {arXiv:1003.1932} \BibitemShut
  {NoStop}%
\bibitem [{\citenamefont {Kulish}\ \emph {et~al.}(1981)\citenamefont {Kulish},
  \citenamefont {Reshetikhin},\ and\ \citenamefont {Sklyanin}}]{KuRS81}%
  \BibitemOpen
  \bibfield  {author} {\bibinfo {author} {\bibfnamefont {P.~P.}\ \bibnamefont
  {Kulish}}, \bibinfo {author} {\bibfnamefont {N.~{\relax Yu}.}\ \bibnamefont
  {Reshetikhin}}, \ and\ \bibinfo {author} {\bibfnamefont {E.~K.}\ \bibnamefont
  {Sklyanin}},\ }\href@noop {} {\bibfield  {journal} {\bibinfo  {journal}
  {Lett. Math. Phys.}\ }\textbf {\bibinfo {volume} {5}},\ \bibinfo {pages}
  {393} (\bibinfo {year} {1981})}\BibitemShut {NoStop}%
\bibitem [{\citenamefont {Bazhanov}\ and\ \citenamefont
  {Reshetikhin}(1989)}]{BaRe89}%
  \BibitemOpen
  \bibfield  {author} {\bibinfo {author} {\bibfnamefont {V.~V.}\ \bibnamefont
  {Bazhanov}}\ and\ \bibinfo {author} {\bibfnamefont {N.~{\relax Yu}.}\
  \bibnamefont {Reshetikhin}},\ }\href@noop {} {\bibfield  {journal} {\bibinfo
  {journal} {Int. J. Mod. Phys. A}\ }\textbf {\bibinfo {volume} {4}},\ \bibinfo
  {pages} {115} (\bibinfo {year} {1989})}\BibitemShut {NoStop}%
\bibitem [{\citenamefont {Pronko}(2000)}]{Pron00}%
  \BibitemOpen
  \bibfield  {author} {\bibinfo {author} {\bibfnamefont {G.~P.}\ \bibnamefont
  {Pronko}},\ }\href@noop {} {\bibfield  {journal} {\bibinfo  {journal} {Comm.
  Math. Phys.}\ }\textbf {\bibinfo {volume} {212}},\ \bibinfo {pages} {687}
  (\bibinfo {year} {2000})},\ \Eprint {http://arxiv.org/abs/hep-th/9908179}
  {hep-th/9908179} \BibitemShut {NoStop}%
\bibitem [{\citenamefont {Yang}\ \emph {et~al.}(2006)\citenamefont {Yang},
  \citenamefont {Nepomechie},\ and\ \citenamefont {Zhang}}]{YaNZ06}%
  \BibitemOpen
  \bibfield  {author} {\bibinfo {author} {\bibfnamefont {W.-L.}\ \bibnamefont
  {Yang}}, \bibinfo {author} {\bibfnamefont {R.~I.}\ \bibnamefont
  {Nepomechie}}, \ and\ \bibinfo {author} {\bibfnamefont {Y.-Z.}\ \bibnamefont
  {Zhang}},\ }\href@noop {} {\bibfield  {journal} {\bibinfo  {journal} {Phys.
  Lett. B}\ }\textbf {\bibinfo {volume} {633}},\ \bibinfo {pages} {664}
  (\bibinfo {year} {2006})},\ \Eprint {http://arxiv.org/abs/hep-th/0511134}
  {hep-th/0511134} \BibitemShut {NoStop}%
\bibitem [{\citenamefont {Walker}(1959)}]{Walk59}%
  \BibitemOpen
  \bibfield  {author} {\bibinfo {author} {\bibfnamefont {L.~R.}\ \bibnamefont
  {Walker}},\ }\href@noop {} {\bibfield  {journal} {\bibinfo  {journal} {Phys.
  Rev.}\ }\textbf {\bibinfo {volume} {116}},\ \bibinfo {pages} {1089} (\bibinfo
  {year} {1959})}\BibitemShut {NoStop}%
\bibitem [{\citenamefont {Johnson}\ \emph {et~al.}(1973)\citenamefont
  {Johnson}, \citenamefont {Krinsky},\ and\ \citenamefont {McCoy}}]{JoKM73}%
  \BibitemOpen
  \bibfield  {author} {\bibinfo {author} {\bibfnamefont {J.~D.}\ \bibnamefont
  {Johnson}}, \bibinfo {author} {\bibfnamefont {S.}~\bibnamefont {Krinsky}}, \
  and\ \bibinfo {author} {\bibfnamefont {B.~M.}\ \bibnamefont {McCoy}},\
  }\href@noop {} {\bibfield  {journal} {\bibinfo  {journal} {Phys. Rev. A}\
  }\textbf {\bibinfo {volume} {8}},\ \bibinfo {pages} {2526} (\bibinfo {year}
  {1973})}\BibitemShut {NoStop}%
\bibitem [{\citenamefont {Fateev}\ and\ \citenamefont
  {Zamolodchikov}(1982)}]{FaZa82}%
  \BibitemOpen
  \bibfield  {author} {\bibinfo {author} {\bibfnamefont {V.~A.}\ \bibnamefont
  {Fateev}}\ and\ \bibinfo {author} {\bibfnamefont {A.~B.}\ \bibnamefont
  {Zamolodchikov}},\ }\href@noop {} {\bibfield  {journal} {\bibinfo  {journal}
  {Phys. Lett. A}\ }\textbf {\bibinfo {volume} {92}},\ \bibinfo {pages} {37}
  (\bibinfo {year} {1982})}\BibitemShut {NoStop}%
\bibitem [{\citenamefont {Albertini}(1994)}]{Albe94}%
  \BibitemOpen
  \bibfield  {author} {\bibinfo {author} {\bibfnamefont {G.}~\bibnamefont
  {Albertini}},\ }\href@noop {} {\bibfield  {journal} {\bibinfo  {journal}
  {Int. J. Mod. Phys. A}\ }\textbf {\bibinfo {volume} {9}},\ \bibinfo {pages}
  {4921} (\bibinfo {year} {1994})},\ \Eprint
  {http://arxiv.org/abs/hep-th/9310133} {hep-th/9310133} \BibitemShut {NoStop}%
\bibitem [{\citenamefont {Bazhanov}\ and\ \citenamefont
  {Stroganov}(1990)}]{BaSt90}%
  \BibitemOpen
  \bibfield  {author} {\bibinfo {author} {\bibfnamefont {V.~V.}\ \bibnamefont
  {Bazhanov}}\ and\ \bibinfo {author} {\bibfnamefont {{\relax Yu}.~G.}\
  \bibnamefont {Stroganov}},\ }\href@noop {} {\bibfield  {journal} {\bibinfo
  {journal} {J. Stat. Phys.}\ }\textbf {\bibinfo {volume} {59}},\ \bibinfo
  {pages} {799} (\bibinfo {year} {1990})}\BibitemShut {NoStop}%
\bibitem [{\citenamefont {Yang}\ and\ \citenamefont {Yang}(1969)}]{YaYa69}%
  \BibitemOpen
  \bibfield  {author} {\bibinfo {author} {\bibfnamefont {C.~N.}\ \bibnamefont
  {Yang}}\ and\ \bibinfo {author} {\bibfnamefont {C.~P.}\ \bibnamefont
  {Yang}},\ }\href@noop {} {\bibfield  {journal} {\bibinfo  {journal} {J. Math.
  Phys.}\ }\textbf {\bibinfo {volume} {10}},\ \bibinfo {pages} {1115} (\bibinfo
  {year} {1969})}\BibitemShut {NoStop}%
\bibitem [{\citenamefont {Bl{\"o}te}\ \emph {et~al.}(1986)\citenamefont
  {Bl{\"o}te}, \citenamefont {Cardy},\ and\ \citenamefont
  {Nightingale}}]{BlCN86}%
  \BibitemOpen
  \bibfield  {author} {\bibinfo {author} {\bibfnamefont {H.~W.~J.}\
  \bibnamefont {Bl{\"o}te}}, \bibinfo {author} {\bibfnamefont {J.~L.}\
  \bibnamefont {Cardy}}, \ and\ \bibinfo {author} {\bibfnamefont {M.~P.}\
  \bibnamefont {Nightingale}},\ }\href@noop {} {\bibfield  {journal} {\bibinfo
  {journal} {Phys. Rev. Lett.}\ }\textbf {\bibinfo {volume} {56}},\ \bibinfo
  {pages} {742} (\bibinfo {year} {1986})}\BibitemShut {NoStop}%
\bibitem [{\citenamefont {Affleck}(1986)}]{Affl86}%
  \BibitemOpen
  \bibfield  {author} {\bibinfo {author} {\bibfnamefont {I.}~\bibnamefont
  {Affleck}},\ }\href@noop {} {\bibfield  {journal} {\bibinfo  {journal} {Phys.
  Rev. Lett.}\ }\textbf {\bibinfo {volume} {56}},\ \bibinfo {pages} {746}
  (\bibinfo {year} {1986})}\BibitemShut {NoStop}%
\bibitem [{\citenamefont {Mathur}\ \emph {et~al.}(1988)\citenamefont {Mathur},
  \citenamefont {Mukhi},\ and\ \citenamefont {Sen}}]{MaMS88}%
  \BibitemOpen
  \bibfield  {author} {\bibinfo {author} {\bibfnamefont {S.~D.}\ \bibnamefont
  {Mathur}}, \bibinfo {author} {\bibfnamefont {S.}~\bibnamefont {Mukhi}}, \
  and\ \bibinfo {author} {\bibfnamefont {A.}~\bibnamefont {Sen}},\ }\href@noop
  {} {\bibfield  {journal} {\bibinfo  {journal} {Phys. Lett. B}\ }\textbf
  {\bibinfo {volume} {213}},\ \bibinfo {pages} {303} (\bibinfo {year}
  {1988})}\BibitemShut {NoStop}%
\bibitem [{\citenamefont {Ginsparg}(1988)}]{Ginsparg87}%
  \BibitemOpen
  \bibfield  {author} {\bibinfo {author} {\bibfnamefont {P.~H.}\ \bibnamefont
  {Ginsparg}},\ }\href {\doibase 10.1016/0550-3213(88)90249-0} {\bibfield
  {journal} {\bibinfo  {journal} {Nucl. Phys. B}\ }\textbf {\bibinfo {volume}
  {295}},\ \bibinfo {pages} {153} (\bibinfo {year} {1988})}\BibitemShut
  {NoStop}%
\bibitem [{\citenamefont {Finch}\ and\ \citenamefont {Frahm}(2012)}]{FiFr12}%
  \BibitemOpen
  \bibfield  {author} {\bibinfo {author} {\bibfnamefont {P.~E.}\ \bibnamefont
  {Finch}}\ and\ \bibinfo {author} {\bibfnamefont {H.}~\bibnamefont {Frahm}},\
  }\href@noop {} {\bibfield  {journal} {\bibinfo  {journal} {J. Stat. Mech.}\
  ,\ \bibinfo {pages} {L05001}} (\bibinfo {year} {2012})},\ \Eprint
  {http://arxiv.org/abs/1108.3228} {arXiv:1108.3228} \BibitemShut {NoStop}%
\bibitem [{\citenamefont {Mong}\ \emph {et~al.}(2014)\citenamefont {Mong},
  \citenamefont {Clarke}, \citenamefont {Alicea}, \citenamefont {Lindner},\
  and\ \citenamefont {Fendley}}]{MCAL14}%
  \BibitemOpen
  \bibfield  {author} {\bibinfo {author} {\bibfnamefont {R.~S.~K.}\
  \bibnamefont {Mong}}, \bibinfo {author} {\bibfnamefont {D.~J.}\ \bibnamefont
  {Clarke}}, \bibinfo {author} {\bibfnamefont {J.}~\bibnamefont {Alicea}},
  \bibinfo {author} {\bibfnamefont {N.~H.}\ \bibnamefont {Lindner}}, \ and\
  \bibinfo {author} {\bibfnamefont {P.}~\bibnamefont {Fendley}},\ }\href@noop
  {} {\bibfield  {journal} {\bibinfo  {journal} {preprint}\ } (\bibinfo {year}
  {2014})},\ \Eprint {http://arxiv.org/abs/1406.0846} {arXiv:1406.0846}
  \BibitemShut {NoStop}%
\bibitem [{\citenamefont {Kedem}\ and\ \citenamefont {McCoy}(1993)}]{KeMc93}%
  \BibitemOpen
  \bibfield  {author} {\bibinfo {author} {\bibfnamefont {R.}~\bibnamefont
  {Kedem}}\ and\ \bibinfo {author} {\bibfnamefont {B.~M.}\ \bibnamefont
  {McCoy}},\ }\href@noop {} {\bibfield  {journal} {\bibinfo  {journal} {J.
  Stat. Phys.}\ }\textbf {\bibinfo {volume} {71}},\ \bibinfo {pages} {865}
  (\bibinfo {year} {1993})},\ \Eprint {http://arxiv.org/abs/hep-th/9210129}
  {hep-th/9210129} \BibitemShut {NoStop}%
\bibitem [{\citenamefont {Ardonne}\ \emph {et~al.}(2011)\citenamefont
  {Ardonne}, \citenamefont {Gukelberger}, \citenamefont {Ludwig}, \citenamefont
  {Trebst},\ and\ \citenamefont {Troyer}}]{AGLT11}%
  \BibitemOpen
  \bibfield  {author} {\bibinfo {author} {\bibfnamefont {E.}~\bibnamefont
  {Ardonne}}, \bibinfo {author} {\bibfnamefont {J.}~\bibnamefont
  {Gukelberger}}, \bibinfo {author} {\bibfnamefont {A.~W.~W.}\ \bibnamefont
  {Ludwig}}, \bibinfo {author} {\bibfnamefont {S.}~\bibnamefont {Trebst}}, \
  and\ \bibinfo {author} {\bibfnamefont {M.}~\bibnamefont {Troyer}},\
  }\href@noop {} {\bibfield  {journal} {\bibinfo  {journal} {New J. Phys.}\
  }\textbf {\bibinfo {volume} {13}},\ \bibinfo {pages} {045006} (\bibinfo
  {year} {2011})},\ \Eprint {http://arxiv.org/abs/1012.1080} {arXiv:1012.1080}
  \BibitemShut {NoStop}%
\bibitem [{\citenamefont {Figueroa-O'Farrill}(1990)}]{Figu90}%
  \BibitemOpen
  \bibfield  {author} {\bibinfo {author} {\bibfnamefont {J.~M.}\ \bibnamefont
  {Figueroa-O'Farrill}},\ }\href {\doibase 10.1016/0550-3213(90)90478-V}
  {\bibfield  {journal} {\bibinfo  {journal} {Nucl. Phys. B}\ }\textbf
  {\bibinfo {volume} {343}},\ \bibinfo {pages} {450} (\bibinfo {year}
  {1990})}\BibitemShut {NoStop}%
\bibitem [{\citenamefont {Lukyanov}\ and\ \citenamefont
  {Fateev}(1990)}]{LuFa90}%
  \BibitemOpen
  \bibfield  {author} {\bibinfo {author} {\bibfnamefont {S.~L.}\ \bibnamefont
  {Lukyanov}}\ and\ \bibinfo {author} {\bibfnamefont {V.}~\bibnamefont
  {Fateev}},\ }\href@noop {} {\bibfield  {journal} {\bibinfo  {journal} {Sov.
  J. Nucl. Phys.}\ }\textbf {\bibinfo {volume} {49}},\ \bibinfo {pages} {925}
  (\bibinfo {year} {1990})}\BibitemShut {NoStop}%
\bibitem [{\citenamefont {Bouwknegt}\ and\ \citenamefont
  {Schoutens}(1993)}]{BoSc93}%
  \BibitemOpen
  \bibfield  {author} {\bibinfo {author} {\bibfnamefont {P.}~\bibnamefont
  {Bouwknegt}}\ and\ \bibinfo {author} {\bibfnamefont {K.}~\bibnamefont
  {Schoutens}},\ }\href@noop {} {\bibfield  {journal} {\bibinfo  {journal}
  {Phys. Rep.}\ }\textbf {\bibinfo {volume} {223}},\ \bibinfo {pages} {183}
  (\bibinfo {year} {1993})}\BibitemShut {NoStop}%
\bibitem [{\citenamefont {Frenkel}\ \emph {et~al.}(1992)\citenamefont
  {Frenkel}, \citenamefont {Kac},\ and\ \citenamefont {Wakimoto}}]{FrKW92}%
  \BibitemOpen
  \bibfield  {author} {\bibinfo {author} {\bibfnamefont {E.}~\bibnamefont
  {Frenkel}}, \bibinfo {author} {\bibfnamefont {V.}~\bibnamefont {Kac}}, \ and\
  \bibinfo {author} {\bibfnamefont {M.}~\bibnamefont {Wakimoto}},\ }\href@noop
  {} {\bibfield  {journal} {\bibinfo  {journal} {Comm. Math. Phys.}\ }\textbf
  {\bibinfo {volume} {147}},\ \bibinfo {pages} {295} (\bibinfo {year}
  {1992})}\BibitemShut {NoStop}%
\bibitem [{\citenamefont {Blumenhagen}\ \emph {et~al.}(1995)\citenamefont
  {Blumenhagen}, \citenamefont {Eholzer}, \citenamefont {Honecker},
  \citenamefont {Hornfeck},\ and\ \citenamefont {Hubel}}]{BEHH95}%
  \BibitemOpen
  \bibfield  {author} {\bibinfo {author} {\bibfnamefont {R.}~\bibnamefont
  {Blumenhagen}}, \bibinfo {author} {\bibfnamefont {W.}~\bibnamefont
  {Eholzer}}, \bibinfo {author} {\bibfnamefont {A.}~\bibnamefont {Honecker}},
  \bibinfo {author} {\bibfnamefont {K.}~\bibnamefont {Hornfeck}}, \ and\
  \bibinfo {author} {\bibfnamefont {R.}~\bibnamefont {Hubel}},\ }\href
  {\doibase 10.1142/S0217751X95001157} {\bibfield  {journal} {\bibinfo
  {journal} {Int. J. Mod. Phys. A}\ }\textbf {\bibinfo {volume} {10}},\
  \bibinfo {pages} {2367} (\bibinfo {year} {1995})},\ \Eprint
  {http://arxiv.org/abs/hep-th/9406203} {hep-th/9406203} \BibitemShut {NoStop}%
\bibitem [{\citenamefont {Bouwknegt}\ and\ \citenamefont
  {Schoutens}(1995)}]{BoSc95}%
  \BibitemOpen
  \bibinfo {editor} {\bibfnamefont {P.}~\bibnamefont {Bouwknegt}}\ and\
  \bibinfo {editor} {\bibfnamefont {K.}~\bibnamefont {Schoutens}},\ eds.,\
  \href@noop {} {\emph {\bibinfo {title} {$\mathcal{W}$-{S}ymmetry}}},\
  \bibinfo {series} {Adv. Ser. Math. Phys.}, Vol.~\bibinfo {volume} {22}\
  (\bibinfo  {publisher} {World Scientific Publishing},\ \bibinfo {address}
  {Singapore},\ \bibinfo {year} {1995})\BibitemShut {NoStop}%
\bibitem [{\citenamefont {Zamolodchikov}\ and\ \citenamefont
  {Fateev}(1985)}]{ZaFa85}%
  \BibitemOpen
  \bibfield  {author} {\bibinfo {author} {\bibfnamefont {A.~B.}\ \bibnamefont
  {Zamolodchikov}}\ and\ \bibinfo {author} {\bibfnamefont {V.~A.}\ \bibnamefont
  {Fateev}},\ }\href@noop {} {\bibfield  {journal} {\bibinfo  {journal} {Sov.
  Phys. JETP}\ }\textbf {\bibinfo {volume} {62}},\ \bibinfo {pages} {215}
  (\bibinfo {year} {1985})}\BibitemShut {NoStop}%
\bibitem [{\citenamefont {Gepner}\ and\ \citenamefont {Qiu}(1987)}]{GeQi87}%
  \BibitemOpen
  \bibfield  {author} {\bibinfo {author} {\bibfnamefont {D.}~\bibnamefont
  {Gepner}}\ and\ \bibinfo {author} {\bibfnamefont {Z.}~\bibnamefont {Qiu}},\
  }\href@noop {} {\bibfield  {journal} {\bibinfo  {journal} {Nucl. Phys. B}\
  }\textbf {\bibinfo {volume} {285}},\ \bibinfo {pages} {423} (\bibinfo {year}
  {1987})}\BibitemShut {NoStop}%
\end{thebibliography}
%

\end{document}